# Multiple Mechanisms of Spiral Wave Breakup in a Model of Cardiac Electrical Activity

Flavio H. Fenton and Elizabeth M. Cherry

Center for Arrhythmia Research at Hofstra University and The Heart Institute, Beth Israel

Medical Center, New York, New York 10003

Harold M. Hastings

Center for Arrhythmia Research and Department of Physics, Hofstra University,

Hempstead, NY 11549

Steven J. Evans

Center for Arrhythmia Research at Hofstra University and The Heart Institute, Beth Israel

Medical Center, New York, New York 10003

It has become widely accepted that the most dangerous cardiac arrhythmias are due to re-entrant waves, i.e., electrical wave(s) that re-circulate repeatedly throughout the tissue at a higher frequency than the waves produced by the heart's natural pacemaker (sinoatrial node). However, the complicated structure of cardiac tissue, as well as the complex ionic currents in the cell, has made it extremely difficult to pinpoint the detailed mechanisms of these life-threatening reentrant arrhythmias. A simplified ionic model of the cardiac action potential (AP), which can be fitted to a wide variety of experimentally and numerically obtained mesoscopic characteristics of cardiac tissue such as AP shape and restitution of AP duration and conduction velocity, is used to explain many different mechanisms of spiral wave breakup which in principle can occur in cardiac tissue. Some, but not all, of these mechanisms have been observed before using other models; therefore, the purpose of this paper is to demonstrate them using just one framework model and to explain the different parameter regimes or physiological properties necessary for each mechanism (such as high or low excitability, corresponding to normal or ischemic tissue, spiral tip trajectory types, and tissue structures such as rotational anisotropy and periodic boundary conditions). Each mechanism is compared with data from other ionic models or experiments to illustrate that they are not model-specific phenomena. The fact that many different breakup mechanisms exist has important implications for antiarrhythmic drug design and for comparisons of fibrillation experiments using different species, electromechanical uncoupling drugs, and initiation protocols.

87.18.Hf, 87.18.Bb, 87.19.Hh, 05.45.-a

**Cardiovascular disease remains the most prevalent cause of death in the industrialized world. In the United States, an estimated 10 percent of all deaths are sudden and are mostly due to ventricular fibrillation, a fast-developing disturbance in heart rhythm that disrupts the coordinated contractions and renders the heart unable to pump blood effectively. After decades of research, researchers still are left**

**with an incomplete understanding of how arrhythmias like ventricular fibrillation initiate and evolve. Although the advent of implantable devices has yielded much success in avoiding arrhythmic episodes in patients diagnosed with a predisposition to ventricular arrhythmia, antiarrhythmic drug therapy is not always successful. In some cases drug use actually *increased* mortality, findings that underscore the lack of understanding of the mechanisms responsible for fibrillation. Experimental work has shown that some atrial and ventricular arrhythmias begin with the presence of scroll waves of electrical activation that rotate at higher frequencies than the heart's natural pacemaker, preventing normal function. In many cases, existing scroll waves have been found to break and to form new waves, which in turn yield further breakup and more waves. Many theories have been developed to explain how breakup occurs. In this paper, we use a single mathematical model of cellular electrical activity to demonstrate, categorize, and explain many of the mechanisms proposed here and elsewhere and to discuss the implications of the existence of multiple mechanisms.**

## I. Introduction

Cardiovascular disease is the most common cause of death in the industrialized world, with serious health and economic impacts. Nearly one million deaths annually are caused by cardiovascular disease in the United States alone, or over 40 percent of all deaths[1]. Almost 10 percent of these are victims of cardiac arrest[1], of whom an estimated 95 percent die before reaching the hospital[1]. Most of these are sudden cardiac deaths[2] attributed to ventricular fibrillation (VF), a fast-developing electrical disturbance in the heart's rhythm that renders it unable to pump blood. While implantable cardioverter defibrillators have been highly successful in terminating arrhythmic episodes in patients with diagnosed heart disease, studies indicate that more than half of sudden cardiac deaths arise in individuals with no previous symptoms[1]. Atrial fibrillation, although not as immediately life-threatening as ventricular fibrillation, causes an estimated 15 percent of all strokes due to clot formation from stagnation of blood in the atria[1]. Thus cardiac fibrillation remains a serious health problem.

Fibrillation arises when the heart's usual electrical rhythm is disturbed. Whereas the heart's muscular contraction usually is smooth and coordinated due to a single wave of electrical excitation that signals the cells to contract, during fibrillation the normal electrical signal is masked by higher frequency circulating activation waves, leading to small and out-of-phase localized contractions. Although some cases of fibrillation may be attributed to rapid formation of impulses arising from multiple spontaneous foci[3], which is known to occur experimentally[4] in cultured ventricular muscle and atrial tissue[5,6]. The most prevalent hypothesis is that during fibrillation, at least one[7,8,9,10] and possibly many[10,11,12,13,14,15,16] three-dimensional spirals or scroll waves of electrical activation are present, and these have been observed experimentally in many preparations. Therefore, the dynamical complexity that can be observed on the surface of the heart may be due to multiple scroll waves with short and long life spans due to collisions and various tissue heterogeneities, with the observed activity rendered more complex by the high degree of rotational anisotropy in the tissue.

Although the general processes that produce fibrillation have been characterized, the precise mechanisms responsible for its onset and maintenance are not yet understood.

The consequences of incomplete knowledge of arrhythmia mechanisms have been illustrated dramatically in the results of several antiarrhythmic drug trials, most notably the CAST[17 18] and SWORD[19] trials, in which mortality *increased* for post- myocardial infarction (MI) patients receiving pharmacotherapy compared to placebo. In the CAST trials, the drugs being tested reduced the occurrence of premature ventricular contractions (PVCs), phenomena that can trigger arrhythmias under the right conditions. By reducing the number of PVCs, it was hypothesized that the likelihood of arrhythmia initiation would be reduced as well. However, this did not turn out to be the case, as those PVCs that did occur were far more likely to induce fibrillation than without drugs. The CAST[17] trial of encainide and flecainide was stopped after a mean follow-up of 10 months, since at that time a total of 148 deaths, 106 in the antiarrhythmic drug group vs. 42 in the placebo group, had been registered out of the 1498 initial patients, producing a relative risk of 2.5. Similarly, the CAST II[18] trial using moricizine was terminated after studying the effects of therapy during a 14-day exposure period in which a total of 20 deaths, 17 with antiarrhythmic therapy vs. 3 with placebo, had occurred out of 665 in the study group, yielding a relative risk of 5.6.

In the SWORD trial, the pure potassium channel blocker d-sotalol was tested on the assumption that prolongation of the action potential duration might be protective and reduce all-cause mortality in high-risk patients with MI and left ventricular dysfunction, a hypothesis based on studies using amiodarone, a class III antiarrhythmic drug which showed potential for improving survival rate. Previous experiments also had shown antiarrhythmic effects for d-sotalol, such as non-inducibility of VT in rabbit preparations[20]. The SWORD trial was terminated after a mean follow-up of 148 days because of an increase in mortality in patients taking d-sotalol (78 deaths vs. 48 for placebo out of 3121 patients, with a relative risk of 1.65). From the results of these trials, it is clear that a detailed understanding of arrhythmogenic mechanisms is needed so that the experiences of the CAST and SWORD trials are not repeated. In addition, the failure of the PVC suppression hypothesis, as evidenced in the CAST trial results, suggests that more than one mechanism can produce fibrillation[21,22], in which case trying to suppress one mechanism possibly can enhance another.

Much of our understanding of arrhythmia mechanisms has come from experiments in different preparations, most of which have been conducted in single cells or in small tissue preparations using surface arrays of electrodes. Over the last decade, optical mapping techniques[23] that allow visualization of the surface potentials at all sites have advanced substantially, increasing our understanding of arrhythmia dynamics, but intramural dynamics have remained largely hidden. Recently developed transmural recording techniques such as transillumination[24] and optical fibers[25], both of which use voltage-sensitive dyes, and fiberglass needle electrodes[26] may address this limitation by allowing further investigation of the intramural dynamics and illuminating some of the mechanisms underlying fibrillation.

On the other hand, for over half a century, computer simulations have complemented traditional animal experiments and have contributed to the understanding of arrhythmias. In the 1940s, Winer and Rosenblueth[27] began the field of computational cardiac electrophysiology by using a simple cellular automata model to describe action potential dynamics and to investigate the conditions under which arrhythmias could develop. Despite the simplicity of their models, they explained how reentrant waves could circulate around obstacles and provided a theoretical framework for the circulating waves observed in cardiac tissue experiments by Mines[28] in 1913 and Garrey in 1914[29]. [McWilliams[30] first suggested the concept of reentry as one of the mechanisms responsible for the onset of cardiac arrhythmias in 1887.] In the 1960s, Moe[31] and collaborators[32] expanded the simple cellular automata model of Winer and Rosenblueth by adding more states and introducing a degree of randomness in the refractory period duration. Their contributions produced a dispersion of refractoriness and under certain conditions allowed reentrant excitations (spiral waves) that did not require an anchoring obstacle to rotate and that could break into multiple wavelets reminiscent of experimentally observed fibrillatory patterns. By the 1960s and early 1970s, continuous models of general excitable media were introduced and were used to demonstrate the existence of spiral waves in isotropic tissue. With increased computer power by the early 1990s, many mechanisms of spiral wave breakup were demonstrated in two dimensions [33 34 35 36 37]. By the late 1990s and early 2000s, simulations have advanced to examine three-dimensional effects[38], including rotational anisotropy[39 40], and are beginning to investigate the contributions of anatomical structures to arrhythmogenesis[41 42 43]. Together, the simulation results of the last several decades have advanced spiral wave theory by proposing various mechanisms for the onset and evolution of spiral wave breakup.

In this paper, we discuss a number of mechanisms of spiral wave breakup that have been hypothesized to contribute to arrhythmia initiation and maintenance in homogeneous cardiac tissue. While some of these mechanisms (but not all) have been described before, each using one particular ionic model and thus requiring the use of several different models to show all the mechanisms, we intend to demonstrate all of them using one single model, for two reasons. First, illustrating all the mechanisms using only one model shows that they are model-independent phenomena. Second, analyzing how these different mechanisms relate to parameter regimes corresponding to different electrophysiological properties within the context of one model facilitates comparison of the requirements for each mechanism that can underlie the transition from tachycardia to fibrillation.

The rest of this paper is organized as follows. In section II we review the restitution properties of action potential duration and conduction velocity, which are mesoscopic characteristics that can help in describing cardiac tissue, along with the different trajectories and dynamics of stable spiral waves. In section III we discuss the simplified ionic model used and its parameters. Section IV describes six different mechanisms for spiral wave breakup in 2D. In section V we show a quasi-3D case where boundary effects can become important in some regimes. Section VI describes three different mechanisms that can occur in 3D. In section VII we discuss the ten mechanisms described earlier and the implications of our results. Section VIII presents conclusions and summarizes further prospects. Finally, in the appendix, the equations and model parameters for all simulations are presented.

## II. Review

In this section we describe two mesoscopic characteristics found in cardiac tissue, namely action potential duration restitution and conduction velocity restitution. These restitution properties incorporate in a functional form the ionic complexity underlying the relationship of pulse duration and velocity with respect to the previous dynamical state. In addition, we describe many possible types of spiral wave tip trajectories in cardiac models.

**1. Restitution of action potential duration and conduction velocity**

When the heart beats at a faster rate than normal, such as during increased physical activity, the relative durations of systole and diastole are adjusted to ensure that both filling of the chambers and ejection of blood occur efficiently. An increase in frequency without a change in the systole duration would lead to a disproportionate decrease in the corresponding diastole duration, and at high frequencies, the ventricles would not be filled before contracting. For this reason, if a second action potential is initiated soon after the first, when not all ionic processes have recovered fully to their rest states, the duration of the second action potential is shorter than the first because the transmembrane current will be reduced. It follows that an APD is a function of both the previous APDs and the time between excitations, also known as the recovery time or diastolic interval (DI). The function depends on the different characteristics of all the ionic currents found in cardiac cells; however, the mesoscopic dynamics of the depolarization wave front and repolarization wave back, as well as the interaction between these two fronts, can be obtained by using simple experimental curves relating only APDs to DIs. These curves commonly are referred to as APD restitution curves (for example, see Figure 4 and other figures throughout the text). Note that distinguishing the APD from the DI requires defining a voltage cutoff during repolarization. In general, a percentage of the voltage repolarization (e.g., 80 percent) is used as a threshold to separate the APD from the DI, and the cutoff is indicated by writing the percentage as a subscript (e.g., $APD_{80}$). Throughout this manuscript we use the 80 percent cutoff ($APD_{80}$) when calculating restitutions.

A second fundamental mesoscopic property of excitable media and thus cardiac tissue is the restitution of conduction velocity. When a sequence of propagating pulses is produced, the influence of the preceding pulse on the subsequent one is reflected not only in its action potential duration but also in its propagation speed. Conduction velocity restitution, therefore, is the direct analog of APD restitution in the sense that it relates the speed of a pulse at a given site to the recovery time at that site or its preceding DI (for example, see Figure 5 and other figures throughout the text). The CV depends on the orientation of the wave front with respect to the fiber axis of the cells. However, in a continuous medium with straight, parallel fibers, we can arbitrarily choose to define the fiber axis parallel to a coordinate axis. Conduction generally is slower in partially recovered tissue, so that the CV decreases with decreasing DI to a minimum velocity greater than zero. This means that propagation fails for any DI shorter than the minimum DI.

The APD and CV restitution curves are fundamental to characterizing the wave dynamics of the system, since as a first approximation they reflect the mesoscopic effects of changes that occur in ionic currents and concentrations at the cellular level. The importance of restitution in understanding cardiac dynamics has been emphasized in numerous numerical[33,34,44,45,46,47,48] and experimental studies[49,50,51,52,53,54,55,56].

**2. Spiral Wave Tip Trajectories**

Spiral waves in cardiac tissue models and experiments can follow different types of paths[8,15,57,58,59,60,61,62], from circles to linear trajectories with sharp turns. In between there is a wide range of different trajectories that follow a series of winding, loop-like bends that turn upon themselves and are known as meandering trajectories (see Figure 1). Although useful for analyzing the stability of spiral waves, to date no theory explaining spiral tip trajectories has been developed in terms of the APD and CV restitution curves, partly because tip trajectories depend not only on the wavelength[62] and the excitable gap[63], both of which can be obtained from the restitution curves, but also on the critical radius of curvature[64,65] (i.e., tissue diffusion and excitability given by the rate of rise and the threshold for excitation) and to some extent on electrotonic effects[66] (produced by the shape of AP and tissue diffusion). These last two elements in principle may be obtained by knowing the APD and CV restitutions at all possible cutoffs or thresholds (i.e., APD restitution curves formed using $APD_{90}$, $APD_{80}$, $APD_{70}$, etc.) to allow the effects of the shape of the action potential and interactions with neighboring cells to become apparent. Nevertheless, predicting a spiral wave's period of rotation and tip trajectory as a function of model parameters is complex, and at present there is no analytical theory for spiral wave motion except in the weakly excitable limit[67,68].

Despite the lack of a general theory, some insights into which parameters can cause tip trajectories to change qualitatively among the different patterns of motion have been developed. The transition from circular to meandering motion originates by a supercritical Hopf bifurcation[61,69,70], which introduces a second frequency that adds modulated waves and whose onset has been explained rigorously in the large core regime by Hakim and Karma[68]. Within the meander regime, the transition from epicycloidal to hypocycloidal occurs as the second frequency grows larger than the original one[61], with cycloidal occurring in between when both frequencies are the same. The transition to linear core has been argued to occur[62] when the radius of rotation $r$ produced by the faster frequency is much smaller than the spiral wavelength $\lambda$, with a rotation of the linear core at a slow rate proportional to $r/\lambda$.

In Figure 1, we show examples of these trajectories obtained using the model described in the next section. Transitions such as those depicted can be obtained by varying either the sodium[57,62,65] (excitability) or the calcium and/or potassium dynamics (wavelength)[62,71]. It is important to note that several of these trajectories have been observed experimentally in cardiac tissue[8,59,72,73], particularly linear cores (such as the ones shown in Figure 1 E-F), which apparently occur predominantly in normal cardiac tissue with a line of block approximately 1-2 cm long (see, for example, Figure 7 in Ref. 58 , Figure 6 in Ref. 74, and Figures 2-5 in Ref. 15).

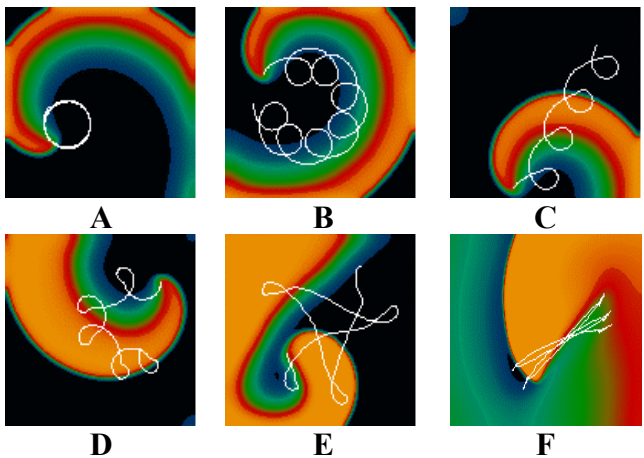

**Figure 1. Possible varieties of spiral wave tip trajectories. Shown are (A) circular, (B) epicycloidal (also known as meander with inward petals), (C) cycloidal, (D) hypocycloidal (also known as meander with outward petals), (E) hypermeandering, and (F) linear trajectories. Spirals are obtained with the model described in the appendix by using parameter set 1 for A-E with progressively increasing excitability (A, $\tau_d$=0.41; B, $\tau_d$=0.39; C, $\tau_d$=0.381; D,**

τd=0.35; E, τd=0.25) and by using parameter set 2 for F. The voltage field is colored ranging from orange and red (excited) to green and blue (refractory) to black (quiescent), and the levels can be compared to the voltage values in **Figure 3**. Tissue size is 4.75x4.75 cm, with Δx=0.019 cm and Δt=0.3 ms.

To trace the tip trajectory, different methods have been used. For example, the so-called pivot method[75] tracks the spatially discretized grid points that did not cross a given subthreshold level of voltage in a previous period. This method only defines an *unexcited* region during a former period[38] and only works when the spiral wave moves very slowly and the tip trajectory is almost circular. Biktashev et al.[38] used a method that consisted of identifying the spatially discretized grid cell with neighboring nodes found in three different states that depended on a selected constant voltage. A similar method[76] differentiates between four states of the cells (exciting, excited, refractory, and rest) and identifies points that have neighbors in each of the four states. However, the accuracy of these two methods depends on the spatial resolution. Barkley et al.[77] used a method that is less sensitive to the effect of spatial resolution, which consists of finding the intersection between the contours of two variables. However, this method may only work for model variables like the FHN model, since one of the contours depends on the slow gate variable. Another method used efficiently for FHN-type models with two variables is to find the point of maximum cross product for the gradients of the two variables[78]. We should mention that some of these methods do not work in three dimensions, especially when spiral waves are not perpendicular to the *x-y* plane.

Throughout this manuscript, we identify spiral wave tips as points with zero normal velocity at an arbitrarily chosen isopotential line that defines the boundary between the depolarization and the repolarization wave back. This method, defined in Ref. 40, can be used at any given time with an interpolation routine, so that the spiral tip trajectory is found accurately and independently of the spatial resolution as a continuous trajectory in space as a function of time. With this method the trajectory of a scroll wave (3D spiral wave) can be obtained accurately, along with information about its curvature, twist, and torsion, which are important characteristics when analyzing scroll wave dynamics. Two other commonly used methods for determining spiral tip trajectories work equally well. The first method, which is the only method that has been used so far in both experiments[16] and numerical simulations[16,79], uses a phase map for each recorded site as a function of time. The phase map assigns values between 0 and $2\pi$ depending on the location in phase space, or equivalently as a function of voltage as well as its time derivative, thereby avoiding the duplicate values of voltage found during the upstroke and during repolarization. Points in the tip trajectory correspond to phase singularities, which are the points where the wave front meets the back, thus producing singularity in the phase. This method has been shown to give the same results as the zero-normal velocity technique[80]. The second method follows the spiral wave trajectory by finding the points with null curvature over a given isopotential[65,81]. The point with zero curvature is not very sensitive to slight changes in wave shape as compared to other characteristic points, such as the point of maximum curvature[65], so that an accurate trajectory also may be obtained using this method.

Independently of the method used, the precision in positioning the spiral wave tip in cardiac tissue becomes a function of the spatially diffuse wave back because of the much slower repolarization period compared to the AP depolarization, as shown in Figure 2. This unavoidable problem slightly changes the size of the trajectory depending on the isopotential or percentage of repolarization used. Nevertheless, as illustrated in Figure 2, the basic morphology of the trajectory (e.g., linear, circular, or meandering) remains independent of the cutoff chosen. For models such as the FHN, where the wave front and wave back are thin boundary layers of width ε, this problem is not present because the trajectories obtained with different isopotential values differ by less than ε.

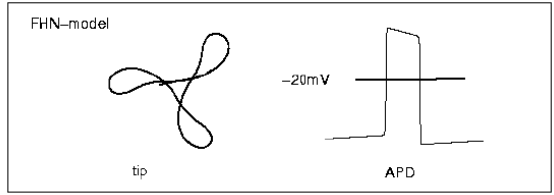

A

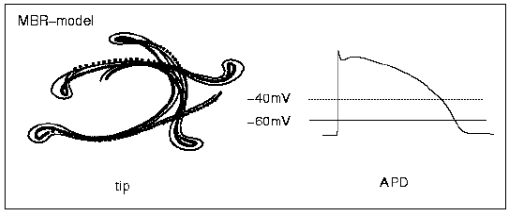

B

**Figure 2.** Spiral tip trajectories found using the zero normal velocity method. (A) Trajectory found using an isopotential of –20 mV for the Fitzhugh-Nagumo model[82]. Since the depolarization and repolarization times are almost the same, it does not matter which isopotential value is chosen, as they all give the same result. (B) Tip trajectory of the 8-variable Beeler-Reuter model[83] with a speedup in calcium dynamics by a factor of 2 (MBR model), using isopotentials at –60 mV (solid) and –40 mV (dashed). The dashed line is smaller because the repolarization wave gives a shorter APD at –40mV than at –60mV. Although the specific trajectory changes slightly at different isopotential values, the overall shape of the trajectory remains the same.

## III. Ionic Model.

Throughout this paper, we use a previously described[40 84 85] simplified mathematical ionic model for the cardiac action potential, whose purpose is not to replicate faithfully the micro-scale ionic complexity of cardiac cells, but rather to reproduce the action potential dynamics at a meso-scale level where restitution properties can be measured. The model was constructed so as to incorporate only the minimum set of ionic membrane currents necessary to reproduce generic restitution curves, and thus consists of three independent ionic currents. These currents can be thought as effective sodium, calcium, and potassium currents; however, we refer to them as fast and slow inward and slow outward currents as a reminder that they do not represent quantitatively measured currents. Despite their simplicity, these currents retain enough minimal structure of the basic currents involved in cardiac excitation that their parameters can be varied to reproduce the restitution curves of more complex ionic models[40,66,43] as well as those obtained from experimental data[40,86]. Furthermore, Figure 3 shows examples of the model with parameters fitted to reproduce the action potential shapes and APD and CV restitution curves of two different ionic models as well as one set of experimental data.

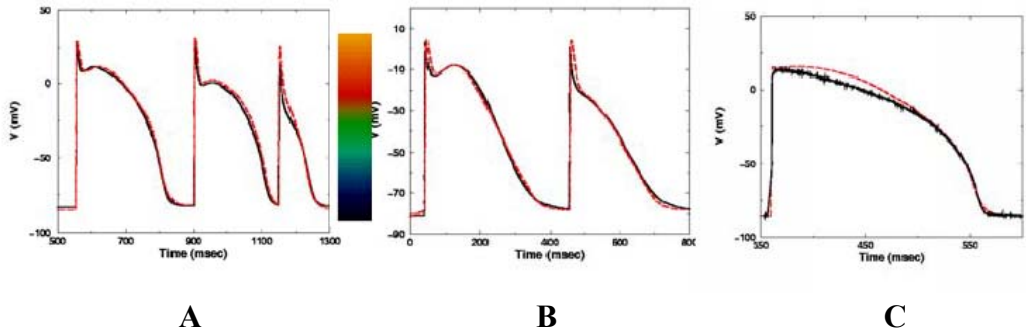

**A**  **B**  **C**

**Figure 3. Comparison of action potentials obtained using a variation of the model described in the appendix with other models and experiments. (A) Three consecutive activations using the 8-variable Beeler-Reuter model[83] (black) and the simplified model[66] (red). (B) Two consecutive activations using the 19-variable Courtemanche et al. atrial model[87] (black) paced at a 300 ms cycle length and the simplified model fitted to it (red)[43]. (C) Experimental optical AP from rabbit ventricle using Cytochalasin-D (uncoupling drug ) and the corresponding model fit[86]. The color bar in A indicates the voltage color scale used in all plots throughout this paper.**

The model consists of three variables: the membrane voltage $V$, a fast ionic gate $v$, and a slow ionic gate $w$. Therefore, we refer to the model as the 3V-SIM (three-variable simplified ionic model)[40,84,85]. The 3 variables are used to produce a total membrane current $I_m = I_{fi}(V;v) + I_{so}(V) + I_{si}(V;w)$ given by the sum of the three independent phenomenological currents. The current $I_{fi}(V;v)$ is a fast inward inactivation current used to depolarize the membrane when an excitation above threshold is induced. It depends on the inactivation gate variable $v$ and on a fast activation gate modeled by a Heaviside step function. The current $I_{so}(V)$ is a slow, time-independent rectifying outward current used to repolarize the membrane back to the resting potential. The current $I_{si}(V;w)$ is a slow inward inactivation current used to balance $I_{so}(V)$ and to produce the observed plateau in the action potential. It depends on the inactivation gate variable $w$ and on a very fast activation gate variable $d$, which has been replaced by its steady-state function $d_\infty(V)$. Using the steady state function keeps the model as simple as possible; nevertheless, the variable $d$ can be used instead of the steady-state function in order to reproduce a specific AP shape[43] (see Figure 3) and to include electrotonic effects in the model dynamics[66].

In the model, the slow currents $I_{so}(V)$ and $I_{si}(V;w)$ are independent of the fast current $I_{fi}(V;v)$, allowing the APD restitution curve and the CV restitution curve to be set independently. [The APD and CV restitution curves are found to be coupled at small diastolic intervals for small sodium conductances (low excitability), but as the excitability increases, the two curves decouple[85].] Six model parameters are needed to fit an arbitrary CV restitution curve: four time constants, $\tau_v^+$, $\tau_{v1}^-$, $\tau_{v2}^-$, and $\tau_d$, corresponding to the opening and closing time scales for the fast variable $v$ and the depolarization time, and two voltage thresholds, $V_c$ and $V_m$, which are used to define the range of the membrane potential. Seven model parameters are used to fit an arbitrary APD restitution curve: five time constants, $\tau_w^-$, $\tau_w^+$, $\tau_r$, $\tau_{si}$, and $\tau_0$, corresponding to the opening and closing time constants for the slow $w$ gate variable and the time constants for the slow current; the threshold potential $V_c^{si}$; and the activation width parameter $k$. The complete set of equations that describe the 3V-SIM in cardiac tissue is described further and summarized in the appendix.

All simulations were performed by discretizing the equation in the appendix on a uniform mesh using a semi-implicit Crank-Nicholson scheme. Except where noted otherwise, Neumann (no-flux) boundary conditions were used to ensure no current leaks at the boundaries. The integration time step $\Delta t$ and grid spacing $\Delta x$ were chosen depending on the particular parameters of the model, especially the excitability, so as to give numerically resolved solutions. See Appendix for further details.

## IV. Mechanisms of Spiral Wave Breakup in 2D.

The transition from a single spiral wave to multiple waves is accomplished via wave break, which in the strictest sense is due, at least in 2D, to conduction blocks that form whenever a wave encounters tissue that is absolutely refractory and fails to propagate. However, it is possible to differentiate among breakup processes based on electrophysiological and structural tissue properties and to classify them into different mechanisms in terms of the APD and CV restitution curves, tissue excitability, spiral period, and spiral tip trajectory, as well as initial conditions. In this section, we describe six mechanisms that lead to spiral wave breakup and analyze the source of the conduction blocks. We note that the numbering of mechanisms here and in other sections is made only to facilitate clarity in referencing and discussion within this paper and not to indicate prevalence or importance.

**Steep APD Restitution**

Although the breakup of spiral waves has been known to occur since the early numerical experiments of Winer and Rosenblueth[27] and Moe[31], it was not until 1991 when the first 2D simulations using ionic cell models were performed[88 89 90]. Those simulations showed that spiral breakup was possible even in a uniform medium without heterogeneities. Courtemanche and Winfree[88] showed that speeding up the dynamics of the calcium current by a factor of at least two in the Beeler-Reuter (BR) model[83] could prevent spiral breakup. This modification of the BR model is denoted as the MBR model. Courtemanche[33] later explained the breakup seen in the BR model in terms of the steepness of the APD restitution curve.

Since the pioneering paper of Nolasco and Dahlen[91], it has been known that steep APD restitution curves (those having a region with slope greater than one) can produce oscillations in APD via a Hopf bifurcation[44 47 92] that can result in spiral wave breakup in 2D[34 33 93]. When the slope of the restitution curve is greater than one, small changes in DI are magnified into larger changes in APD, whereas changes in APD due to changes in DI are damped out at smaller slopes. Because the minimum DI is greater than zero, for a fixed period, a long APD produced by the oscillations can demand a DI below the minimum and cause conduction block, as shown in shown in Figure 4 [for hands-on examples we refer to the JAVA applets in Ref. 94]. In the BR model, the slope of the APD restitution curve becomes greater than one at a period of about 310 ms (corresponding to a DI and APD of approximately 100 ms and 210 ms, respectively). Figure 4 shows transient oscillations leading to steady state when periodically pacing at a period of 320 ms (solid line), which is slightly above the point at which the slope exceeds one, using an initial DI of 205 ms. The dashed trajectory illustrates the transient oscillations and subsequent conduction block (denoted by the arrow) for a constant period of 295 ms, which is below 310 ms and thus within the region of the curve where the slope is greater than one. (A slightly different initial DI of 200 ms, rather than the value of 205 ms used for the stable case, was used here to facilitate visualization of the trajectories.) In categorizing spiral wave breakup mechanisms, we distinguish between two types of breakup within this regime of steep APD restitution, one occurring close to the tip and within the first few rotations of the spiral[33 88 90], and another occurring farther from the tip and in some cases requiring many oscillations to develop[34 93 95].

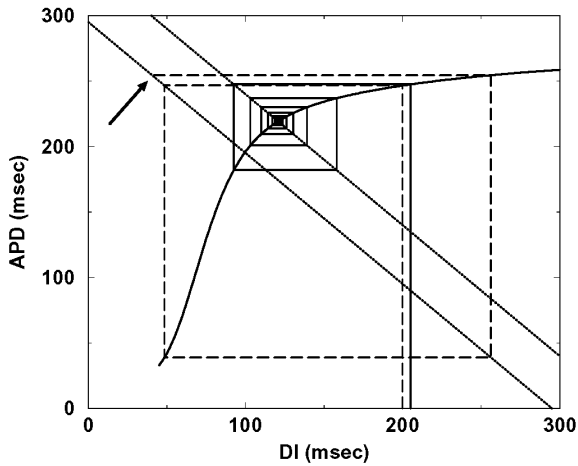

Figure 4. Beeler-Reuter APD restitution curve obtained from the 3V-SIM illustrating the geometrical argument of Nolasco and Dahlen[91] for oscillations of APD and conduction block. We show two cases in which a constant periodic stimulation is applied following an initial DI. Because the constant period of stimulation is the sum of APD and DI, the period T can be determined from the points at which a line at –45° intersects the axes, at (0,T) and (T,0), as shown. Two examples are shown: steady state obtained by pacing at T=320 ms, and conduction block obtained when pacing at T=290ms. The dynamics is displayed following a cobweb similar to a logistic map (see text for more details). Because the 3V-SIM parameters are fitted to accurately represent the BR restitution curves using set 3, the curve shown here is identical[40 85 66] to that obtained using the full BR model. The region with slope greater than one occurs for periods below about 310 ms (DI≈100, APD≈210), and the minimum DI possible before reaching conduction block is $DI_{min} \approx 43$ ms.

*Mechanism 1: Steep APD Restitution with Breakup Close to the Tip*

Courtemanche showed using a delay equation[33] that when the slope of the restitution curve is greater than one, abrupt changes in pulse duration affect the relationship between recovery and excitation, thereby modifying the speed of the wave back or "recovery front." In other words, the steepness of the restitution curve decreases the speed of the wave back in the presence of recovery gradients, so that the steeper the restitution curve in a given region of DIs, the slower the velocity of the wave back in that region. Once the DI is in the region where the APD restitution slope is greater than one, the front and back velocities begin to separate, with the velocity of the back slowing considerably compared to the velocity of the front (see Figure 5 and Figure 6). Therefore, the wave back under these conditions propagates more slowly than the wave front, resulting in what Courtemanche called slow recovery fronts (SRFs). The divergence of velocities of the wave front and the wave back in the region where the slope of the APD restitution curve is greater than one (DI<100 ms) can be observed by plotting the CV restitution for both the front and the back, as shown in Figure 5. The wave back CV restitution can be obtained either by delay equation[44][96] once the APD and wave front CV restitutions are known or directly from numerical simulation.

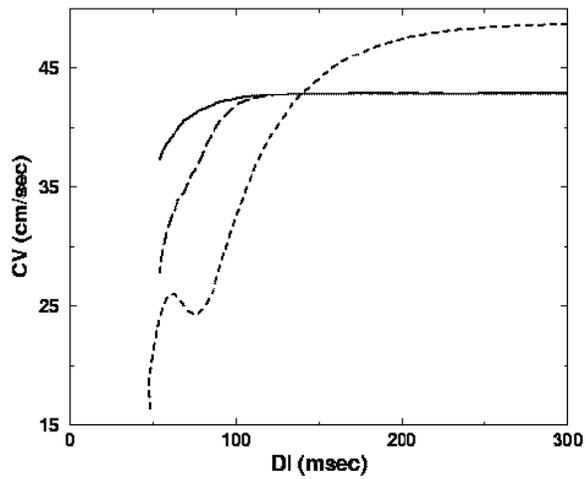

**Figure 5.** CV restitution curve for the wave front (solid) and wave back (long dashes) using parameter set 3. At large DIs, (slope less than one) the front and back velocities are the same, but they diverge at small DIs corresponding to the region where the slope of the APD restitution curve is greater than one (DI<≈100 ms). The CV is measured at the center of a 1D cable 4 cm long. Although the velocity of the wave back is a useful theoretical tool, its measurement is not straightforward. When the wave back velocity is measured in a shorter cable (1.5 cm), electrotonic effects induced by the boundaries can change the curve substantially (short dashes).

However, it is important to mention that the determination of the restitution curve is not as straightforward for the wave back as it is for the wave front. While the front corresponds to the steep, rapid upstroke (on the order of 1 ms), the wave back experiences a much more gradual change in voltage (on the order of tens or hundreds of ms). Therefore, the wave back can become distorted[33,48,97] as it moves due to increased sensitivity to electrotonic effects, which makes characterization of its velocity more difficult. An example of how electrotonic effects can alter the CV restitution of the wave back can be seen in Figure 5, where the short-dashed curve was calculated using a shorter cable than the one used to obtain the long-dashed curve, and in which strong electrotonic effects with the boundaries are introduced.

An example of propagating waves with SRFs and further development of conduction block in a one-dimensional cable can be seen in Figure 6. In this case, waves are generated using the BR model at pacing cycle lengths in the region where the APD restitution curve is steep (slope greater than one). Therefore the velocity of the wave back (recovery front) becomes slower than that of the wave front. Figure 6A-E shows a pulse introduced at the left edge of the cable 60 ms after a previous wave already had passed through the cable (i.e., a DI of 60 ms). Because the slope of the APD restitution curve at this small DI is greater than one, the wave front of the second pulse moves faster than its back, producing a slow recovery front (SRF) and a concomitant change in the wave shape. Figure 6F shows the wave fronts at different times of this pulse superimposed so that it is easier to observe that the front moves faster than the back. When a SRF exists, there is a minimum period $T_{min}$ of pacing in a one-dimensional cable for which it is impossible to produce a train of waves because the SRF has not moved enough and effectively blocks the following wave. This is shown in Figure 6G-L, where the back of a previous wave moves slowly enough to block propagation of the following excitation. In Figure 6, the period of excitation is T=75 ms, which is much less than the minimum period $T_{min}$≈296 ms, thus resulting in conduction block. Note that even though the period was short, it was still above the minimum period corresponding to $DI_{min}$ since it was able to produce an excitation, but the wave was blocked toward the center of the tissue due to the SRF.

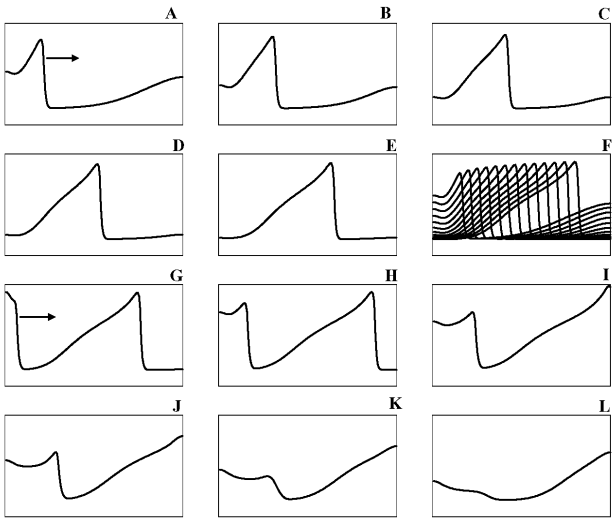

Figure 6. Slow recovery front in the BR model in 1D. The x-axis represents the length of the cable (L=3.3 cm) and the y-axis the voltage (-90-10 mV). Frames A through E show a wave induced at the left end of the cable at a DI of 60 ms. The front moves at a speed of 18.5 cm/s while the wave back moves at 11 cm/s, which makes the shape of the APD change in time. Frame F shows a superposition of wave profiles 5 ms apart to highlight the different speeds of the front and back. Frames G through L show a newly initiated wave front interacting with the back of the previous wave. Because the front moves faster than the slow back of the previous wave, it gets blocked and fails to propagate as it reaches the minimum DI.

Although steep APD restitution produces oscillations of APD even in a single cell or cable, the effects of the oscillations can become more complex in 2D and 3D systems, where the APDs vary throughout the tissue as fronts propagate. For large DIs, the front and back velocities of a wave are the same. However, once the DI is in the regime where the APD restitution slope is greater than one, the front and back velocities begin to separate, with the back slowing considerably compared to the front.

The divergence of the velocities of the front and back is manifested in 2D by the appearance of scalloping along the wave back, as shown in Figure 7. The scallop does not occur at the tip itself due to the high curvature, slower velocity, and smaller difference between front and back velocities there. However, the scallop forms relatively close to the tip, typically within the first arm of the spiral, because of the short DI produced as the wave quickly rotates. The longer APD and slower wave back of the scallop delay the recovery of the tissue's excitability. When the spiral turns and tries to invade the region of the scallop, it encounters refractory tissue, which causes conduction block and wave break. The break generally occurs within the first arm of the spiral because the block tends to occur soon after the spiral turns. In addition, since the waves break before one rotation is completed, no spirals exist with more than one arm.

An example of the evolution of wave break close to the spiral tip due to scalloping is shown in Figure 7. After the formation of a scallop close to the tip of the spiral wave (A), the spiral encounters refractory tissue and propagates along the existing wave back until it has enough excitable tissue available to turn (B-C). By this time a second scallop has formed closer to the tip. As the spiral continues to rotate (D-G), it turns back toward the location of the second scallop and finally collides with it (H-I), pinching off a new front that forms two counter-rotating spirals (J-K). The right spiral is unable to turn, as it continues to encounter refractory tissue, but the left spiral turns and propagates toward the scallop on the original broken front (L), where it again encounters refractory tissue and breaks (not shown).

Figure 7 (M) and (N) show complex states that emerge as the system continues to evolve through wave breaks. Ultimately, however, the breakup is transient, as all wave fronts move off the tissue and leave behind only wave backs (last frame). This transience is not uncommon, and in fact the breakup resulting from this mechanism often is not sustained due to the large variation in wavelengths that eventually can leave no quiescent areas to support continued propagation[88 90]. As the tissue size is increased, the system can support breakup for longer periods of time[98].

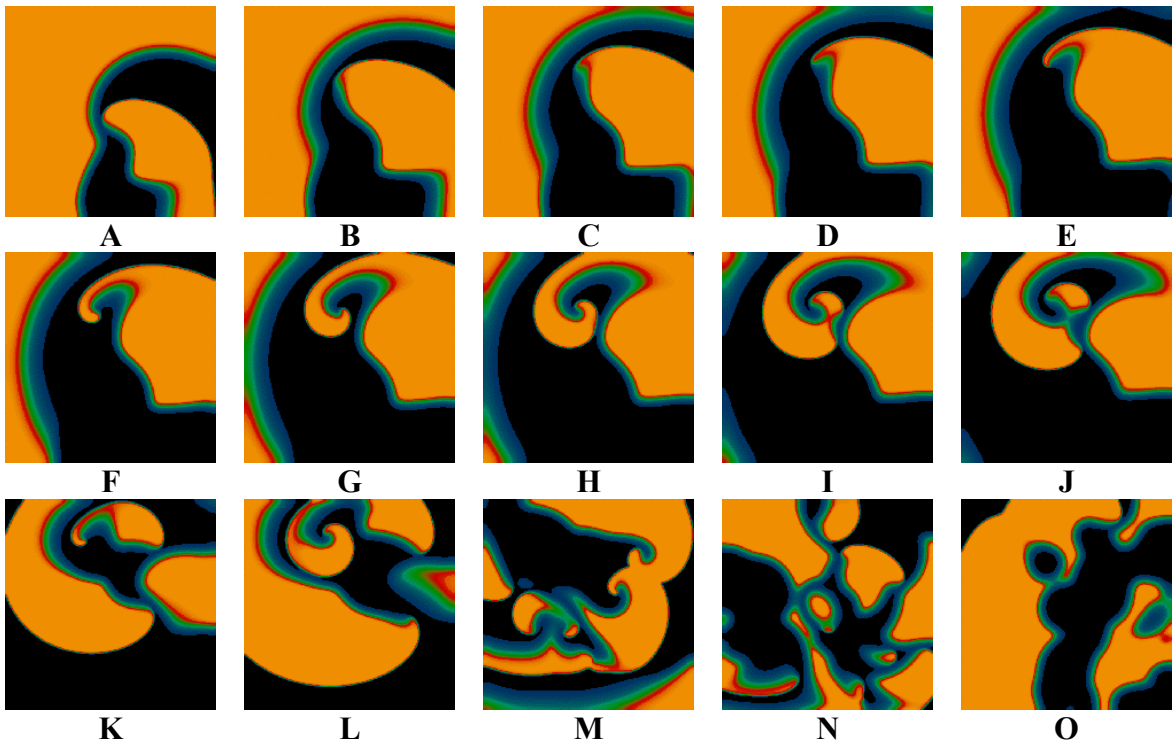

**Figure 7. Evolution of breakup close to the tip due to abruptly steep APD restitution using parameter set 3 in a 12.5x12.5 cm domain. Wave fronts are blocked by refractory regions due to scallops that have formed along wave backs because of the divergence of the front and back conduction velocities. Ultimately, the breakup is transient. The numerical parameters $\Delta x$ and $\Delta t$ are set to 0.025 cm and 0.25 ms, respectively.**

Breakup close to the tip produced by SRFs leading to scalloping can occur in models with abruptly steep APD restitution curves for spirals with periods in the slope greater than one regime. Examples of such models are the Beeler-Reuter model, as described in Refs. 33, 85, and 88; the four-variable Noble model[99], as seen in Refs. 85, 90, and 100; and the Luo-Rudy-I[101] model with altered parameter settings, such as in Ref. 102, where the maximum conductances of the calcium current and the time-dependent potassium current were changed.

*Mechanism 2: Steep APD Restitution with Breakup Far from the Tip*

For APD restitution curves having a region with slope greater than one, a range of periods can exist for which stable oscillations of APD can be sustained before conduction block is produced. The range of periods depends on the steepness of the APD curve, i.e., the width of the range of DIs for which the slope of the restitution curve is greater than one, and on how large the oscillations can become before reaching a DI smaller than $DI_{min}$, which produces conduction block. For example, in the 3V-SIM using parameter set 3, corresponding to the BR model (Figure 4), this region is very narrow (for periods between 285 and 315 ms)[48 85] in comparison to the region shown in Figure 8, where parameter set 4 (see Appendix for parameters) was used to produce a restitution curve with a slower rate of change while still retaining a period of slope greater than one for DIs $\approx\leq$ 110 ms (the region of periods with slope greater than one having stable alternans is about 80 ms, between 180 and 260 ms, as shown in Figure 11).

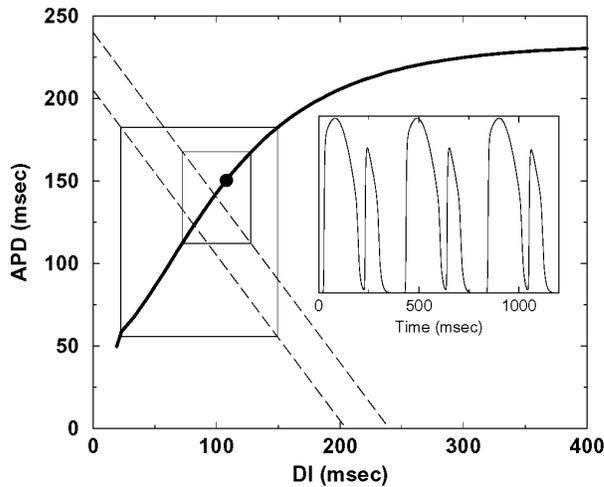

**Figure 8.** APD restitution curve with slope greater than one for a model with a larger range of DIs in the slope greater than one region (due to reduced steepness and decreased $DI_{min}$) compared to the curve shown in Figure 4, allowing alternans to develop over a wider range of periods before reaching conduction block. The circle indicates the point on the curve where the slope is one, with a period of about 260 ms (DI≈110 ms, APD≈150 ms). Periods below 260 ms display alternans; two examples with periods of 240 ms and 205 ms are shown. The dashed lines indicate the period of constant stimulation as the *x*- and *y*-intercepts and the rectangular regions illustrate the two alternating APD solutions. The inset shows the AP alternation resulting from periodic stimulation at 205 ms. Note that while the activations are produced at a constant period, the APD alternates between 180 and 52 ms and therefore the DI alternates between 25 and 153 ms. Note that the range of DIs that give stable alternans here is much larger than in Figure 8. Parameter set 4 is used. See text for further details.

While the cobweb diagram of Figure 8 reflects only the dynamics of a single cell, it has been shown in spatially extended systems that APD oscillations can vary throughout the tissue. When the entire medium exhibits the same long-short pattern in APD (a long APD produced on one beat and a short APD produced on the next beat, with the pattern repeating), the alternans is called concordant. However, due to initial conditions[48] or conduction velocity restitution[48,103], concordant alternans can further develop into discordant alternans, which occurs when some part of a medium oscillates long-short, while another area oscillates short-long—that is, on a given beat, there is a gradient of APD from long to short as the wave propagates through the medium, and on the subsequent beat, the gradient is reversed and becomes short to long. In this case, the APD produced at one site varies along the tissue and the alternans is out of phase. Both concordant and discordant APD alternans are known to occur in cardiac tissue and have been observed in experiments when waves circulate along small paths or rings of tissue[104] or, equivalently, when high-frequency stimulation is applied periodically to linear strands of cardiac tissue[95] or isolated hearts[105][106]. Clinically, beat-to-beat oscillations of APD correlate to T-wave alternans in the ECG, which often has been observed as a precursor to ventricular fibrillation and sudden death[107].

Figure 9 A-D shows the spatial distribution of APDs during discordant alternans produced in a 1D cable when pacing at a cycle length of 230 ms using parameter set 4. The APDs of all points along the cable on odd (even) beats are shown as a solid (dashed) black line, and the lines represent the steady state reached after prolonged pacing. Between every two out-of-phase regions is a node, where the APD does not oscillate on successive beats but remains the same. Previously[48] we demonstrated that along with cycle length, tissue size affects how many regions of opposite alternans phase can exist: for a given frequency, the longer the tissue, the larger the number of nodes. The positions and number of nodes depend on the APD and CV restitution curves[48] as well as on electrotonic effects[108][109] and tissue heterogeneities[48]. Figure 9 E-F show, for example, the effect of CV restitution on the node distribution, which can be analyzed by varying the parameter $\tau_w^-$ in the 3V-SIM to obtain different slopes in the CV restitution curve[85]. Figure 9F shows in black the CV restitution curve obtained with the original parameter set 4, while blue and green indicate the curves obtained by setting $\tau_w^-$ from 15.6 to 5 and 80, respectively. Figure 9E shows that when the CV restitution varies over a wider range of DIs, the node positions occur closer to the stimulus site, as demonstrated previously using iterative maps[48], and the width of the alternans regions decreases, allowing more nodes to be formed (observe three nodes, just about three nodes, and only two nodes are distributed as the CV restitution is flattened out).

It is important to mention that nodes are not always stationary. In some cases they can migrate to the pacing site[95 109], and as they disappear new nodes are formed at the other end [the two variable Karma model[34] is a numerical model that exhibits the migration of nodes with its original parameter settings]. Although the behavior in shorter cables is similar to a truncated version in longer cables, there can be differences. For instance, a short tissue may not contain a node, even though a node can occur in a longer tissue at a location within the length of the short tissue[48]. For example, compare panels A and B in Figure 9, where electrotonic effects in the short tissue produce concordant alternans whereas discordant is produced in a longer cable. More dramatic examples can be observed depending on the electrotonic effects of the model[48].

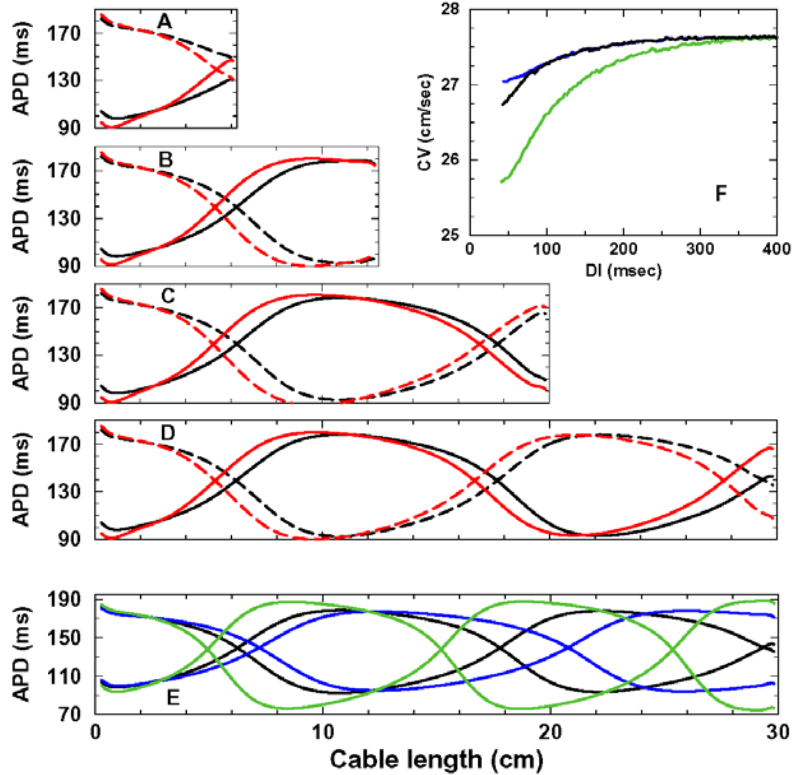

**Figure 9.** (A-D) Discordant alternans in a 1D cable with (black) and without (red) curvature. The steady state spatial distributions of APDs over a cable for even (dashed) and odd (solid line) beats at a basic cycle length of 230 ms are shown for four different cable sizes (6.25, 12.5, 20, and 30 cm) ($\Delta x$=0.0266 cm, $\Delta t$=0.1 ms). Pacing site is at the left edge. Note how curvature shifts the node locations closer to the pacing site. (E) Distribution of nodes in a cable without curvature as a function of CV restitution, while keeping the APD restitution unaltered. Black shows the original distribution from (D), green shows a more packed distribution of nodes obtained when the CV restitution varies over a wider range of DI's, and blue shows a less packed distribution from a flatter CV restitution. (F) CV restitution curves corresponding to the plots in E, where the parameter values were varied from set 4 by changing $\tau_v^-$ from 15.6 to 5 (blue) and 80 (green).

Wave front curvature also can affect the position and distribution of nodes. Although Kay and Gray[110] found that curvature could alter the APD restitution curve itself, our numerical simulations show less than five percent difference in APD restitution curves when using planar vs. curved wave fronts. Comtois and Vinet[111] showed that loading effects due to curvature in an activation wave can increase its duration substantially (more than 40 percent in their simulations using the BR model), but only at very high curvatures corresponding to circles of radius 0.25 to 1 mm), which can be close to the critical radius of propagation[63] and thus difficult to capture in a restitution protocol of extended systems. A more important effect of curvature on wave front propagation is the decrease in conduction velocity as a function of curvature[112] (see the eikonal equation[63 113]). This additional variation in velocity changes the position of the nodes, as shown in Figure 9 A-D, where the APD distribution for curved fronts is displayed in red in a 1D cable reduction of a target wave simulation in which the distance from the pacing site (left edge) is the radius.

Although longer lengths often can support more nodes, this is not always the case. At high frequencies, the shorter APD can be blocked as a node is being formed because the oscillations can become so large that the wave no longer can propagate. Where the short APD becomes blocked, only the fronts corresponding to the long impulses far from the stimulation point and produced every other beat are sustained, and the alternans progresses to what is known as 2:1 block far from the stimulation site. Figure 10 shows how conduction block can form at a given frequency (here, T=212 ms) as the cable length is increased, with longer cables developing block sooner (if block is generated at all). Only concordant alternans is present for the 6.25 cm cable, while discordant alternans develops in the 12.5 cm cable, and conduction block occurs in cables 20 and 30 cm long. The conduction block forms sooner in the 30 cm long cable and takes longer to develop in the 20 cm cable. In both cases, the block is produced far from the stimulation site, as shown experimentally[95]. Although we have shown that, CV restitution aside, electrotonic effects plays a role in the development of the block, further analysis is needed in order to quantify its onset. We should note that the cable sizes used in these examples are unphysiologically large, since the model parameters were tuned to facilitate explanation (the values of $\tau_d$ used produced very slow CV compared to healthy tissue, thus scaling up the size). Nevertheless, the dynamics is generic, as the distribution and position of nodes as well as the range of periods that support alternans depend on the APD and CV restitution curves and electrotonic effects[66 109].

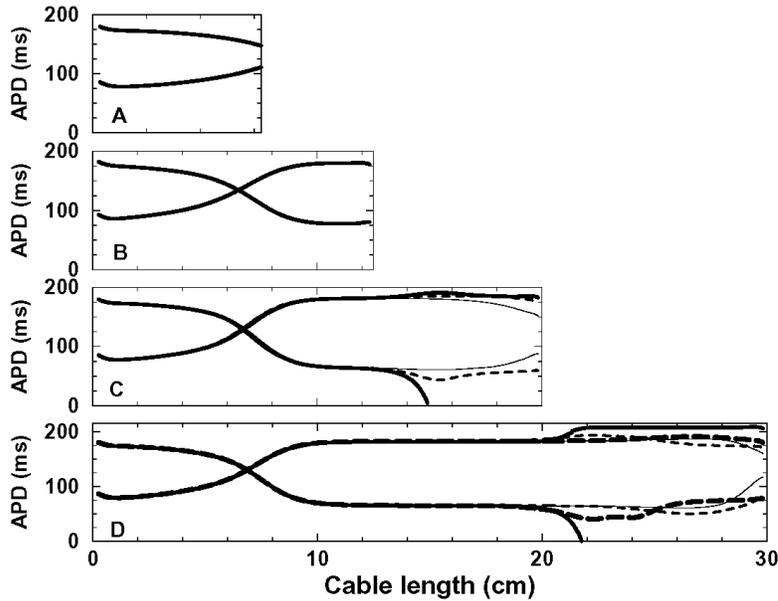

**Figure 10. Size-induced conduction block in a cable with discordant alternans using parameter set 4. The steady state is shown for alternate beats as solid lines. Previous beats, before steady state is reached, are shown as dashed and thinner lines. Although one node forms at 12.5 cm , longer tissues do not support multiple nodes because conduction block occurs. Note that only one wave every two beats propagates along the whole cable, leading to what is called 2:1 block. The cable lengths are the same as in Figure 9, but the period of stimulation is 212 ms instead of 235 ms.**

The relationship between cable length, number of nodes, and frequency of stimulation for a given model can be observed in a parameter space diagram[48 114]. Figure 11 shows the different node regimes for the 3V-SIM fitted to parameter set 4. The diagram on the left corresponds to plane waves and the one on the right to 1D reductions of target wave patterns where curvature effects are included. Dark gray indicates conduction block, while white below 260 ms indicates concordant alternans. In between are three shades of gray indicating the progression from one to two to three nodes as the shade of gray deepens. Curvature effects do not alter the transitions between regions substantially, with the predominant effect being an increase of the 2:1 block region for curved fronts at smaller periods. Far from the pacing site the distribution for plane waves and circular waves coincide as the curvature of the circular waves decreases. Figure 11 shows that at long cycle lengths (T >≈260 ms), only stable non-oscillating solutions are obtained, while at short cycle lengths, conduction block occurs in much the same way as in Figure 4 and alternans solutions lie between these two regimes. For a given pacing cycle length (constant frequency of stimulation), concordant alternans occurs only in short cables, while discordant alternans only appears in cables greater than a minimum length, with multiple nodes emerging as the cable length increases[48][95]. Note that any number of nodes can be present before conduction block is formed. It is important to note, as mentioned earlier, that the different region sizes and distributions shown in Figure 11 are a function of both the APD and the CV restitution curves. The wider the range of DIs where the slope of the APD restitution is greater than one, the larger the region for alternans. Similarly, the wider the range of DIs where the CV restitution varies, the more densely packed the node regions in space (see Figure 9E-F).

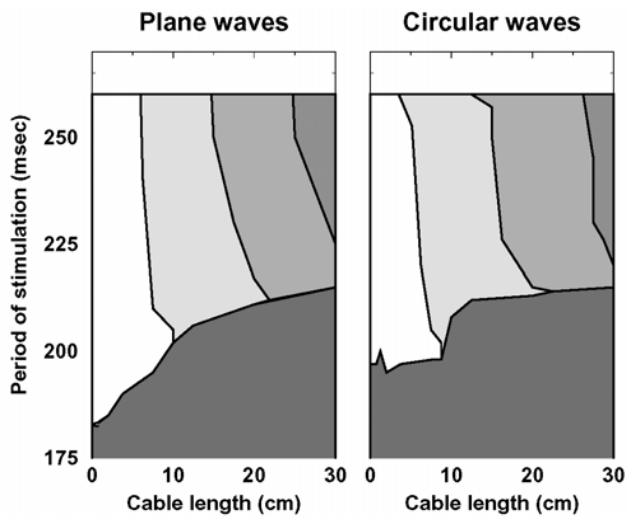

**Figure 11. Behavior of cables of different lengths and cycle lengths. The first figure illustrates the locations of regimes in a 1D cable without curvature, while the second includes curvature effects. Above the top horizontal line, stable, non-oscillatory behavior is observed. In the dark gray regime at**

the bottom, conduction block occurs. In between are regimes of concordant alternans (white) and discordant alternans with one (light gray), two (medium-light gray), and three (medium gray) nodes. For shorter cables, alternans is concordant, while at longer lengths only discordant alternans is observed, with the number of nodes increasing with size. The figures are similar with and without curvature effects, with curvature mainly changing the regimes by inducing block for larger cycle lengths in short cables.

Because of the spatial heterogeneity generated in APD and the eventual possibility of 2:1 conduction block, discordant alternans can lead to wave breaks and further generation of reentrant (spiral) waves when present in two and three dimensions, as has been shown numerically[34 93 103] and experimentally[105]. Since a spiral wave acts as a source of periodically paced waves, breakup can occur if its period of rotation falls within the alternans region. Figure 11 shows that this breakup is a function of both frequency and tissue size. That is, depending on its frequency, a spiral wave can be unstable and break into multiple waves in a large domain while remaining stable in a smaller domain[34]. Figure 12 illustrates how discordant alternans can lead to the breakup of a spiral wave. Parameter set 4, used through this subsection, was designed to produce spiral waves with circular tip trajectories and an APD restitution curve with slope greater than one over a broad range of DIs to allow a large region for alternans. Varying the size of the circular core can change the period of rotation and can allow the transition from stable waves to alternans and finally to conduction blocks. The tissue size is 30x12.5 cm (as mentioned earlier, the size is unphysiologically large, but the purpose here is simply explanatory). The first panel of Figure 12 shows a spiral wave with a period of 265 ms ($\tau_d$=0.415), which remains stable since the period is above the largest alternans period of 260 ms (see Figure 11). As the excitability is slowly increased by setting $\tau_d$ to 0.41, the spiral wave period decreases to 220 ms and the resulting APD oscillations develop into discordant alternans. Early oscillations (B) grow in magnitude until the difference in the long and short APDs is great (C), with one APD much longer than the next. After two more rotations (D), conduction block is almost formed far from the tip (stimulation site) as in Figure 10, and two rotations later (E) conduction block and spiral breakup occurs. Panels E-G show consecutive snapshots of the wave break as it forms. Once initiated, the breakup process continues (H-J), but the core and the three inner arms of the spiral remain unaffected.

The location of the initial wave break relative to the spiral core can be changed by further decreasing the period of rotation (i.e., stimulation period), as shown in Figure 11. Figure 12 K and L show breakup occurring closer to the core, with only two spiral arms remaining intact, as the period is further decreased by increasing excitability ($\tau_d = 0.409$). An additional decrease in period ($\tau_d = 0.405$) leave only one arm intact (M and N), and finally at a period of 160 ms ($\tau_d = 0.4$) no full arms remain (O and P), with breakup pervading the entire medium and only a small spiral wave circulating around a small core that eventually disappears by collision with another wave.

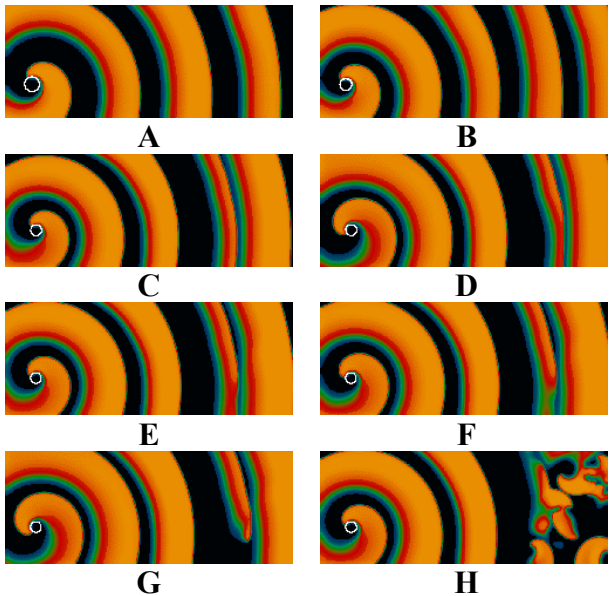

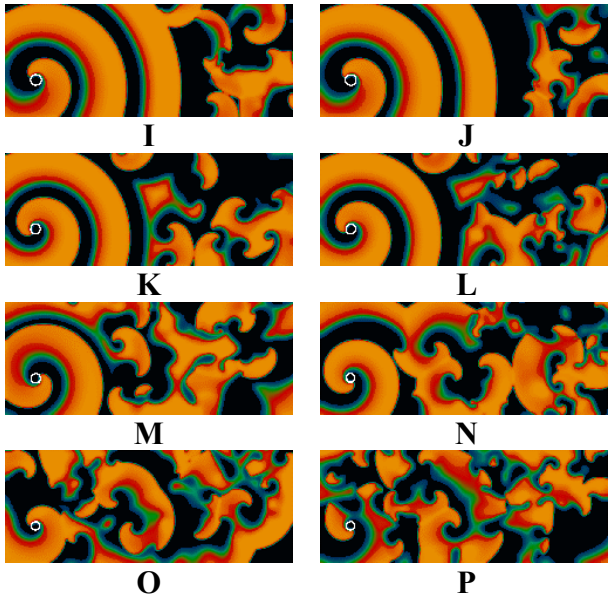

I J
K L
M N
O P

**Figure 12. Discordant alternans-induced breakup far from the tip in a 30x12.5 cm domain. (A) Stable spiral wave with a period of 265 ms. (B-J) After decreasing the period to 220 ms by slightly increasing the excitability, APD oscillations develop and grow until a short APD following a long APD cannot continue to propagate and breaks, as in Figure 10. Additional breakup occurs, but only far from the spiral tip, while the tip itself and the three innermost spiral arms remains protected. (K-L) By changing $\tau_d$ further, the period of rotation is shortened and the distance beyond which wave break occurs is moved progressively closer to the tip of the spiral leaving only two arms intact. (M-N) Increasing the excitability again decreases the period further, so that only one spiral arm remains unbroken. (O-P) A final decrease of the period to 160 ms leaves less than one full spiral arm intact, and eventually the entire medium is filled with broken waves. Note that frames (K-P) are not part of a time series but instead are created by modifying parameters (spiral period). Without these modifications, the breakup remains restricted to a certain region of the tissue far from the spiral tip (source), as shown in Figure 10 and Figure 11. Simulations were done using $\Delta x=0.025$ cm, $\Delta t=0.1$ ms. and $\tau_d$ varies from 0.415 to 0.4 as explained in the text.**

Spiral wave breakup far from the tip due to discordant alternans can occur in any model having a steep APD restitution curve and a spiral period in the slope greater than one regime, provided that the spiral tip follows a circular or low meander trajectory, or perhaps if it is pinned to a scar or inhomogeneity. Models that can break in this manner are the Karma model[34] and the 1962 Noble model [99]. Discordant-alternans-induced breakup also has been observed experimentally in rapidly paced tissue preparations[95,106]. Recently, stable spiral waves surrounded by multiple breaking waves as in Figure 12 H-N have been shown to occur experimentally and in simulations[10]; however, their occurrence was attributed to different densities in the background current between the right and left ventricle. Nevertheless, is interesting to note that such patterns can be obtained in homogenous tissue.

***APD Restitution Curve with One Region of Slope Greater Than One And Two Regions of Slope Less Than One***

An interesting special case of the steep APD restitution curve slope mechanisms can occur when two regions with slope less than one are present on either side of the steep part of the curve. Physiologically, a second region with slope less than one can occur in ionic action potential models where the currents responsible for the AP plateau (mostly calcium) do not activate fully at very short DIs. These alterations in cell dynamics lead to action potentials with very small plateaus at high frequencies of stimulation. Therefore, the APDs vary little at the shortest DIs, producing a second region in the restitution curve with slope less than one. The 1962 Noble model[99] for Purkinje fibers exemplifies this phenomenon. The slope of its APD restitution curve is less than one for periods greater than 256 ms and for periods less than 123 ms, with a region of slope greater than one at periods between 123 and 256 ms[85]. It is important to note that the 1962 Noble model does not include a calcium current in its formulation, since the calcium current was not discovered until a few years later, and instead one of its potassium currents is responsible for the plateau. Another model showing two regions with slope less than one, which was developed more recently and includes a greater number of ionic currents, is the canine model by Fox et al.[115]. In that model, the slope is greater than one for periods between 145 and 210 ms and is less than one for all other periods, down to a minimum of about 85 ms.

Using parameter set 5, the 3V-SIM produces an APD restitution curve with two regions having slope less than one, for periods greater than 375 ms and for periods less than 150 ms. Figure 13 shows the maximum and minimum APD oscillations as a function of cycle length obtained by simulating propagating waves in 1D rings of various sizes[85]. For periods of stimulation greater than 375 ms and below 150 ms, it is possible to obtain stable non-oscillating waves (see top right and left plots), while for periods in between, oscillations of APD (see top two center plots) can be obtained along with a region of conduction block, shown in gray, where no propagation is possible on a ring. Because of the second region with slope <1, the conduction velocities of the wave front and back not only diverge, as described earlier, but also merge back together at short DIs, as shown in Figure 14, where the inset indicates the difference in the front and back velocities as a function of DI.

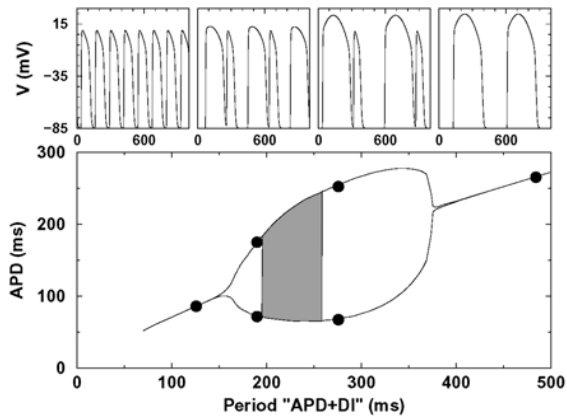

Figure 13. APD vs. cycle length (APD+DI) for the 3V-SIM using parameter set 5, which produces two regions with APD restitution curve slope less than one and consequently two Hopf bifurcations, one at 375 ms and the other at 150 ms. The small figures on top show the voltage as a function of time for the four cycle lengths indicated by the large dots (from left to right, 125, 188, 276, and 485 ms). Stable behavior is observed for the smallest and largest cycle lengths, while APD alternans occurs for the intermediate cycle lengths. The gray area indicates periods for which conduction block occurs in a 1D ring.

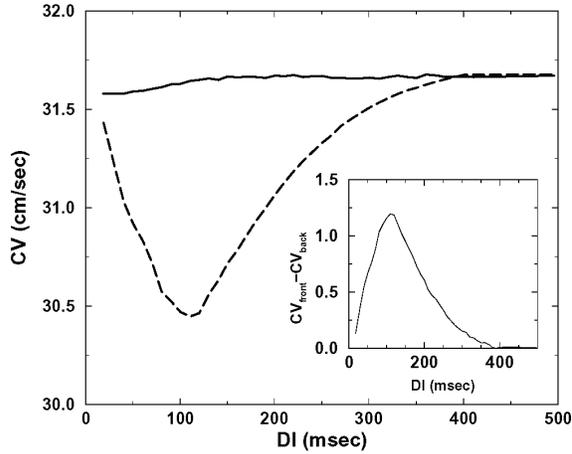

Figure 14. CV restitution curves for the wave front (solid) and wave back (dashed) corresponding to the APD restitution curve with two regions of slope less than one. While the front and back velocities initially diverge as the DI is decreased, the difference in velocities decreases as the second region with slope less than one is approached. Therefore, stable pulses can occur at short as well as long DIs, as shown in the top left inset of Figure 13.

Under these conditions, stable spiral waves can be supported if their period of rotation is large and falls in the region where the APD restitution curve has slope less than one, while breakup will occur if the period lies in the region with slope greater than one, since there can be SRFs or discordant alternans, as in breakup mechanisms 1 and 2. However, if the period is small and falls in the second region with slope less than one, then the solution can be either a stable spiral wave or breakup, depending on the initial conditions[85].

Panels A and B in Figure 15 show the two possible solutions obtained when the period of rotation is 75 ms, using parameter set 5 with $\tau_d$=0.355. In A, a spiral wave was initiated from a broken pulse[88] propagating into quiescent tissue. As the spiral rotated, SRFs led to breakup and complex spatiotemporal dynamics in the same way as observed in the Noble model[90,100] and similarly to Figure 7. In B, a stable spiral wave was achieved also from a broken pulse but following a train of impulses at short intervals so that the spiral never reached the periods between 150 and 375 ms that can lead to conduction block and breakup. The spiral actually was initiated in a smaller tissue (to assure no breakup), following which the size was increased gradually until reaching the same size as in panel A. Two aspects of these spiral wave dynamics are noteworthy. First, at short periods two solutions are possible depending on initial conditions, either a stable spiral wave or spiral breakup and complex spatiotemporal dynamics. Second, a stable spiral wave formed at a short period will remain stable in any tissue size, unlike mechanism 2 where there may be a limit in tissue size before breakup starts. However, in this case, if the period of a stable spiral is slowly increased, breakup will occur as in mechanism 2 when the period becomes larger than that corresponding to the Hopf bifurcation (approximately 160 ms for this model). Panels C-F in Figure 7 show the evolution of the stable spiral wave in panel B as the period is increased from 75 ms to 180 ms (by slowly changing $\tau_d$ from 0.3547 to 0.359 (C), 0.368 (D) and finally to 0.375 in (E))). In E the period is 180 ms (greater than the period for the first Hopf bifurcation ~ 160 ms), at which point discordant alternans begins to develop, as shown in the bottom left corner. The oscillations eventually grow and lead to breakup and multiple spirals (F), following mechanism 2. Note that the density of waves is different between A and F, due to the different sizes of their tip trajectories, and that a denser set of waves can be obtained at higher excitabilities (smaller values of $\tau_d$) as the tip trajectories can change to epicycloidal and hypocycloidal (not shown).

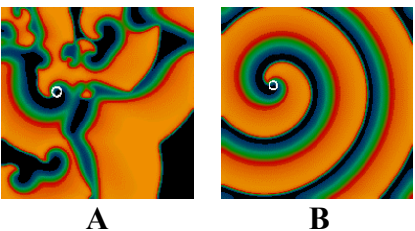

A  B

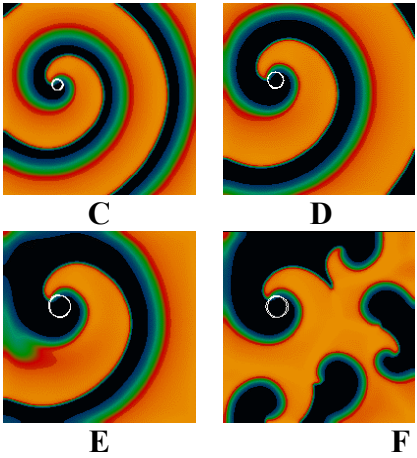

Figure 15. Range of possible dynamics of the 3V-SIM using parameter set 6, which leads to an APD restitution curve that has two regions with slope less than one ($\Delta x=0.068$ cm, $\Delta t=.42$ ms). Panel A and B show two different dynamics, spiral wave breakup and a stable spiral wave, obtained with the same parameters but using different initial conditions. (C-E) The spiral wave from panel B remains stable until the period is increased to 180 ms (E). (E-F) Breakup by discordant alternans leading to various spiral waves.

**Mechanism 3: Bistability, Hysteresis, and 2:1 Block**

Bistability in cardiac tissue is a phenomenon where a given frequency of stimulation results in one of two possible APDs, depending on the initial conditions. The shorter APD is obtained naturally when pacing at a constant frequency, with every pacing beat producing an activation (1:1 response), while the longer APD occurs when only one activation is produced for every two pacing beats (2:1 response). This scenario can be visualized by plotting the APD as a function of the pacing cycle length (period) instead of the DI. Figure 16A shows an example of this bistability in APD using parameter set 6, with the inset showing the normal APD restitution curve as a function of DI ($DI_{min}$ = 11ms, $APD_{min}$ = 61 ms, and thus the minimum period $T_{min}$ is 72 ms). Note that for clarity of exposition and to avoid confusion among different mechanisms, parameter set 6 was designed to produce bistability using an APD restitution curve with slope less than one for all DIs. Nevertheless, bistability can occur with or without steep APD restitution and alternans[116].

The main plot in Figure 16 shows the 1:1 response branch (solid line) as a function of the pacing cycle length. As the pacing cycle length is decreased gradually below $T_{min}$, conduction block occurs, so that only one activation is induced for every two pacing beats (thus jumping into the 2:1 branch, shown by the dashed line). Because the period of stimulation T is fixed, the following APD can be obtained from the APD restitution curve (inset) using a DI given by $DI_{min}$ + T. Since this DI is large, the resulting APD also is large. As the cycle length is decreased further, the APDs for the 2:1 response branch can be obtained, as shown in Figure 16 as a dashed line. If the cycle length now is gradually increased above $T_{min}$, the system remains on the 2:1 branch rather than dropping immediately to the available 1:1 branch because the APD produced is large and every other stimulus comes too early in the refractory period to generate a response, thereby creating a region of bistability. The system shifts back to the 1:1 branch only when the period of stimulation is larger than $DI_{min}$ + APD from the 2:1 branch, again allowing an activation to be produced once for every beat. The difference in dynamics depending on whether the cycle length is being increased or decreased is an example of hysteresis, an effect found in many nonlinear systems. Bistability and hysteresis have been observed in a variety of cardiac experimental preparations including sheep Purkinje fibers and papillary muscles[117], sheep atria[118], guinea pig ventricular cells[119], and frog[116] and rabbit[120] ventricular tissue. The range of cycle lengths in the hysteresis region depends on the slope of the restitution, but mostly on the size of the $DI_{min}$ and $APD_{min}$, and has been shown experimentally to last from several milliseconds to as long as hundreds of milliseconds[116].

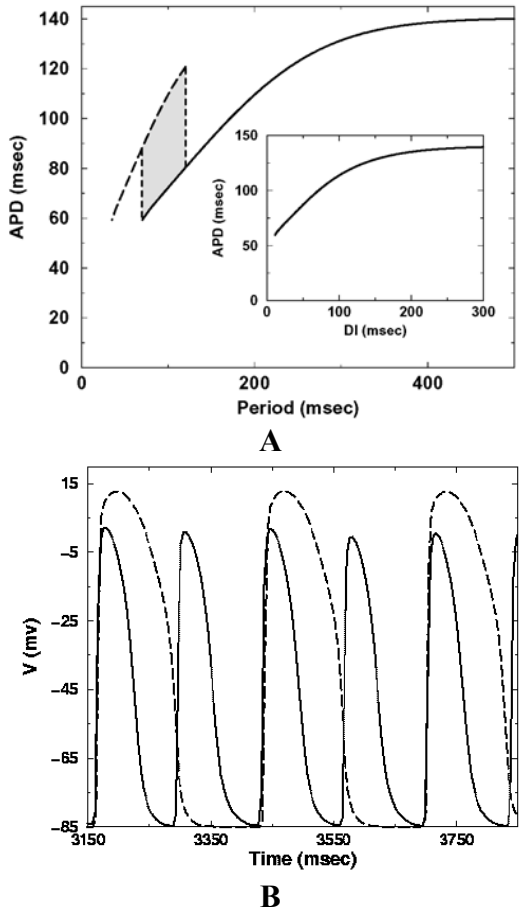

**Figure 16. Bistability and hysteresis of APD. (A) The solid and dashed lines show the 1:1 and 2:1 branches, respectively. The width of the region of periods that can access two stable solutions is about 70 ms in this case. (B) Voltage traces obtained at two different sites from the simulation shown in Figure 17. Close to the core (solid line), every rotation of the spiral is propagated, but farther away (dashed line), every other impulse is blocked, resulting in 2:1 conduction. The 2:1 trace has been shifted in time so that the upstrokes of the 2:1 trace coincide with the upstrokes in the 1:1 trace to facilitate visualization and comparison of the resulting APDs. Note that the APD is sufficiently long in the 2:1 region to prevent propagation of every other impulse.**

In a spatially-extended system, it is possible for some regions of the tissue to experience 2:1 block while the rest of the tissue conducts every impulse. This situation can arise when a target or spiral wave stimulus encounters a gradient of recovery region (DIs) produced by a previous wave[121]. Figure 17 shows a spiral wave induced by a broken wave that demonstrates this effect. The spiral was initiated at the top of the domain, and its period fell within the range of periods with both 1:1 and 2:1 solutions available. Near the tip of the spiral, the small DIs produced small APDs and waves of short wavelength, all of which propagated successfully. However, slightly farther away from the spiral tip, the tissue was quiescent longer before the first arm of the spiral arrived, producing larger DIs that resulted in long APDs on the 2:1 solution branch and longer wavelengths. The second arm of the spiral then was blocked when it reached these sites (Figure 17C-E). Therefore, away from the tip, only every other spiral arm propagated successfully. Figure 17 shows snapshots during two rotations to illustrate that far from the spiral wave, only one activation propagated for every two spiral rotations. Panels A and B show the spiral generating a new arm that was blocked and broke in C-F. The spiral continued to rotate (G-I), and after the longer DIs occurred far from the tip, the next impulse propagated successfully through the entire tissue (J-L). In this example, the broken end of the wave did not generate a reentrant wave, since the entire tissue surrounding the tip was in the region of 2:1 block and since the boundaries were close to the spiral tip. Similar 2:1 conduction block during the rotation of a spiral wave has been demonstrated using a variation of the LR-I model[65] and a cellular automata model[121].

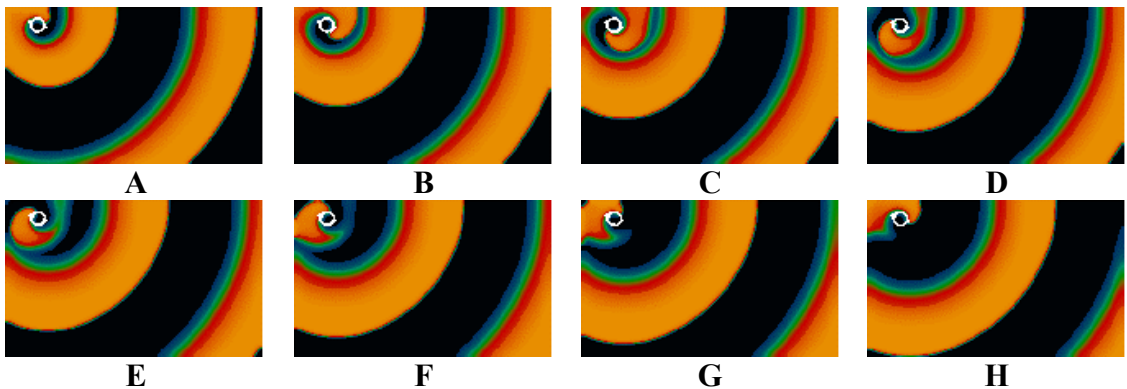

A  B  C  D

E  F  G  H

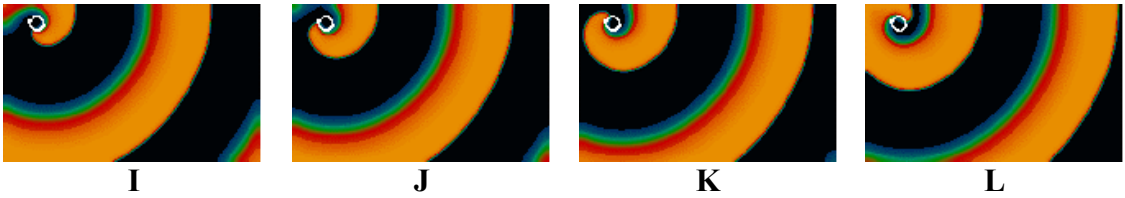

| I | J | K | L |

**Figure 17. Spiral wave with 2:1 conduction block far from the tip. Tissue size is 16.25x10cm (dx=0.025 cm, dt=.62 ms). Because the period of the spiral wave lies in the window of APD bistability, both 1:1 and 2:1 solutions exist. After spiral wave initiation following a plane wave propagating from left to right, sites near the tip conduct at the same frequency as the spiral. Farther from the tip, however, longer DIs occurred before the first arm of the newly created spiral arrived, resulting in longer APDs. When the subsequent spiral arm arrived at these distant sites, the tissue had not yet recovered from the long APDs. As a result, conduction block occurred, and a 2:1 conduction pattern was initiated. Figure 16B shows voltage traces for one site in the 1:1 region and for another site in the 2:1 region.**

**Mechanism 4: Bistability and Doppler Shift by Tip Trajectories**

In the previous section, we showed for a given set of parameter values how a spiral with a very short rotation period can be on the 1:1 branch close to the tip and on the 2:1 branch far from the tip. Although the 2:1 block away from the tip in that case was achieved by using a recovery gradient as the initial condition, rotating spiral waves can access the same state dynamically as they turn and, after multiple wave breaks, turbulence can arise. Figure 18 shows an example of how 2:1 block can develop for a spiral wave following a circular core using the same parameter set 6 without the need for specific initial conditions. As the spiral is formed (A-B) and the tip turns tightly (because of its short period of rotation), the spiral encounters a DI smaller than $DI_{min}$ and is blocked. However, the tip continues to propagate because it remains on the 1:1 branch even while adjacent parts of the arm are in the 2:1 branch (C-E). Eventually, as more tissue becomes available for excitation, the two broken ends find larger DIs that cause them to leave the 2:1 branch and return to the 1:1 branch (F), now forming new spirals because they are able to propagate (G). While the tip of the original spiral remains on the 1:1 branch as it continuously rotates around a small stationary circular core (H-L), the newly generated spirals evolve and repeat the initial breakup process. Because of the wide distribution of refractory periods throughout the tissue, which can correspond to either the 1:1 or 2:1 regime, the dynamics becomes increasingly complicated as new wave breaks continue to form. Notice that unlike mechanisms 1 and 2, in this case the slope of the APD restitution curve is never greater than one, as shown in Figure 16A.

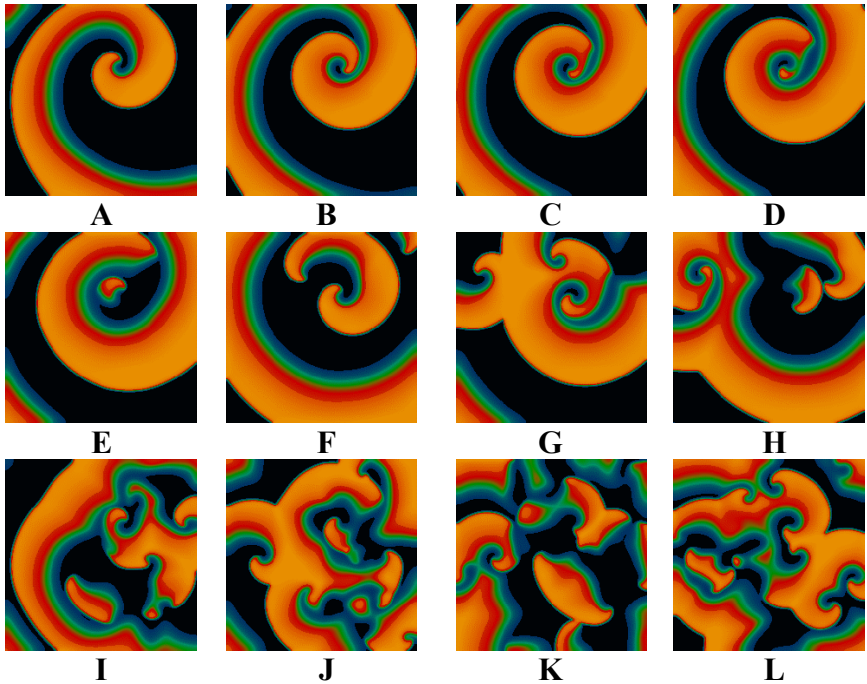

**Figure 18. Breakup of a spiral wave with a circular core due to the formation of 2:1 block.** The tip follows a tight circular trajectory (A) as its rotation period is short, which causes the spiral to encounter a DI smaller than $DI_{min}$ as it rotates (B). The wave cannot continue to propagate outward at this point, so the wave breaks (C-D) and this region of tissue necessarily jumps to the 2:1 branch. Despite the 2:1 block, the original spiral tip finds enough recovered tissue close to the core to sustain itself (E). Once the original spiral has made a full rotation and the 2:1 region is excitable again, the two broken waves are able to propagate and form new spirals (F). The breakup process repeats itself, forming increasingly disorganized states (G-L). Note that the original spiral tip has not disappeared even when surrounded by breakup.

When bistability is present in a model, spiral wave breakup still can occur even when the tip trajectory is not as tight as it is in the case shown in Figure 18. Transitions between the 1:1 and 2:1 branches can occur instead due to Doppler shift in the spiral's frequency induced by meander in the tip trajectory. Essentially, the rotating and meandering spiral functions as a moving source of periodic waves. Because of the Doppler effect, the frequency is higher in the direction in which the spiral is moving. Figure 19 shows an example of the Doppler effect produced by a moving spiral wave. The cycloidal tip trajectory causes the spiral arms to pack more closely together in the direction in which the spiral is moving (to the left), resulting in a higher frequency along the left side of the tissue and a lower frequency along the right. This stable spiral with a cycloidal trajectory is obtained using the same parameters as in Figure 18 but with a slight increase in its excitability ($\tau_d$=0.35) that causes the different trajectory. Similar behavior has been observed experimentally in cardiac preparations.[122]

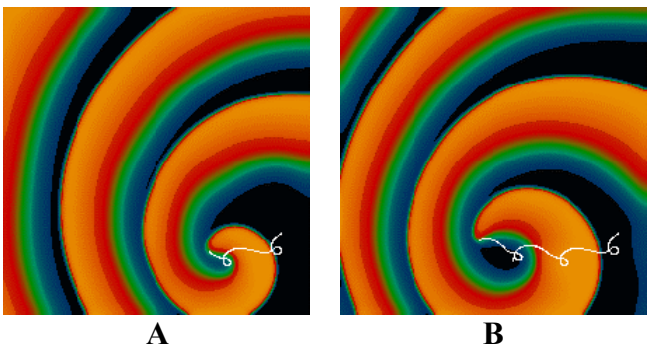

**Figure 19. Stable spiral wave with cycloidal trajectory obtained using parameter set 6 with $\tau_d$=0.35. The wave fronts become more closely packed on the left side of the domain because the spiral is moving in that direction. The Doppler shift in frequency also decreases the period of waves to the left, resulting in the observed shorter wavelength.**

In the case shown in Figure 19, the smallest period produced by the Doppler effect still is greater than the minimum period allowed for propagation (the period obtained at $DI_{min}$), so that no conduction blocks occur and no transitions to the 2:1 branch are present, resulting in a stable (but nonstationary) spiral wave. However, certain trajectories can impose larger Doppler shifts on the period and bring it below the minimum period for propagation, which results in conduction block.. Figure 20 shows an example of this type of conduction block occurring for a spiral wave following an epicycloidal trajectory. Each time the spiral makes an inner loop (inward petal, A), the shift in frequency is enough to produce conduction block somewhere in the arm and initiate reentry (B-C). Continued evolution yields additional wave breaks and spiral formation (D-H). The parameters are the same as in Figure 18 and Figure 19 except for a change in $\tau_d$ to 0.3, which brings the spiral into the regime of epicycloidal trajectories. The broken waves in Figure 20 do not fill the domain as densely as those in Figure 18 because of the difference in the sizes of the epicycloidal trajectory and the small circular core of Figure 18, which allowed waves to break more often and to pack more closely together than in the present case. Breakup resulting from Doppler shift-induced 2:1 block also can be obtained in some of the hypocycloidal trajectories when the outer loops are able to produce a large frequency shift, as in the case obtained using $\tau_d$=0.25 (not shown). Because of the large trajectories in the epi- and hypocycloidal cases compared to the tissue size, the breakup was transient as in the example of mechanism 1, lasting only a few seconds before all spirals eventually extinguished from collisions with other waves and with the boundaries. However, in the circular core case (Figure 18), breakup was continuous during the full 10 seconds simulated.

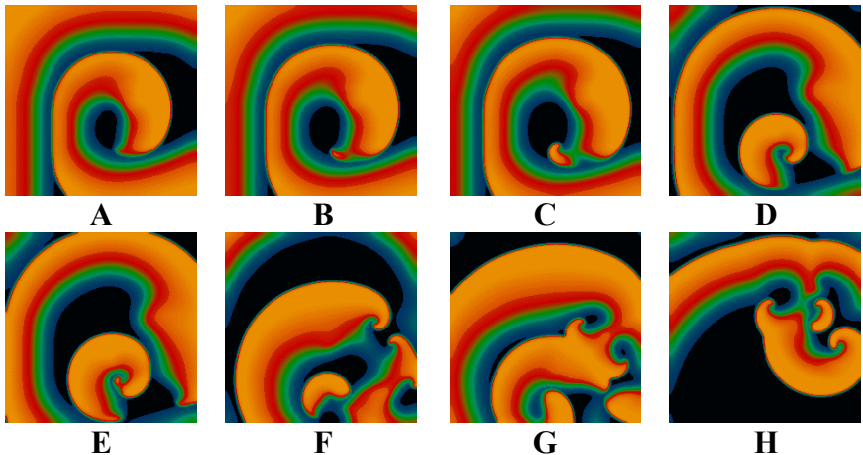

**Figure 20. Breakup of a spiral wave whose tip follows an epicycloidal trajectory using parameter set 6 with $\tau_d$ =0.3. The Doppler shift induced by the meandering tip causes the period to fall below the minimum period for propagation, causing 2:1 block and wave break (A-C). As the wave continues to meander, regions with 2:1 conduction block continue to develop (D-H). Breakup also can be obtained in the hypocycloidal regime (e.g., using $\tau_d$=0.25).**

Meandering trajectories are not the only types that can produce a Doppler shift in a spiral wave leading to breakup. In the linear core regime, where spiral waves follow long lines of block (for models with very flat restitutions, the linear core size is approximately $CV_{max} \times APD/2$)[85] with sharp turns, Doppler shift can cause 2:1 conduction blocks and wave breaks. Figure 21 shows such an effect when the model is brought into the linear core regime by setting $\tau_d$ to 0.115. The first frame (A) shows the spiral at the beginning of its second rotation after initiation from a broken pulse. As the spiral tip makes a sharp turn (B-C), part of the arm gets too close to the adjacent spiral arm due to the Doppler shift and breaks as it falls below the minimum period for propagation (D-F). The two broken ends eventually recombine (H-J), but on the next rotation the conduction block is large enough (K-M) to prevent the two arms from reconnecting, and two new spiral waves are formed (N). Eventually new spirals break in the same way, resulting in turbulence (O-P).

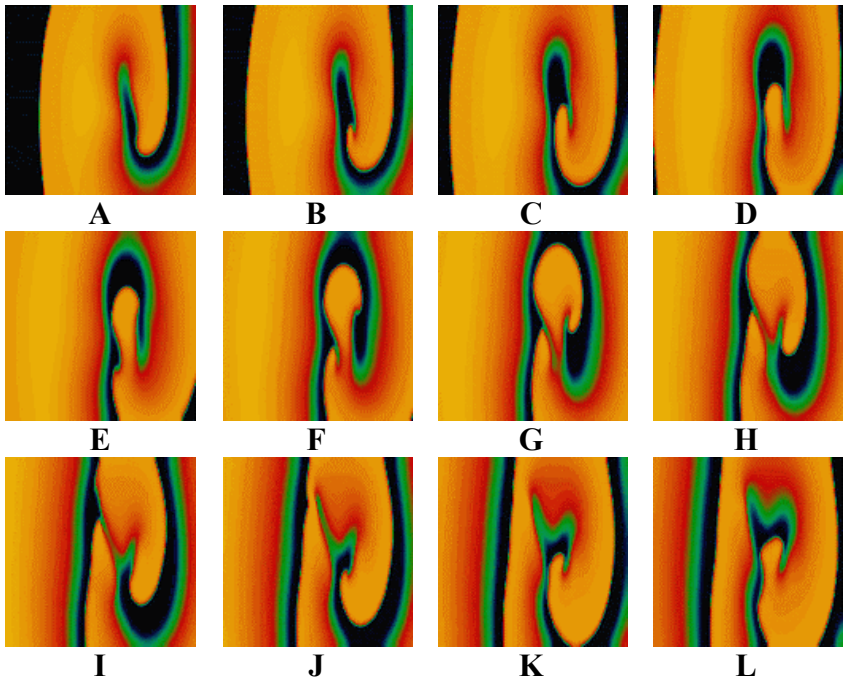

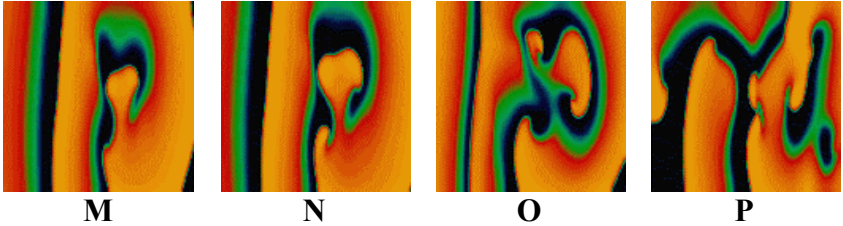

| M | N | O | P |

**Figure 21.** Breakup of a spiral wave following a linear core due to the Doppler effect. As the spiral tip turns sharply (A-C), part of it gets too close to a previously generated wave. The Doppler effect reduces the period below the minimum period for propagation, and wave break ensues (D-F). The break mends (H-J), but when the break occurs again during the next rotation (K-M) new spirals are formed (N). As the breakup evolves, the dynamics become increasingly complex (O-P). Note that the linear core causes the tissue to appear anisotropic, when in fact it is isotropic. Parameter set 6 with $\tau_d$ =0.115 was used.

An interesting case arises when the Doppler shift occurs at the spiral tip itself rather than nearby. Sharp turns in the tip trajectory can bring the tip of a spiral wave below the minimum period and halt its propagation. Then a *secondary* wave of depolarization can evolve and continue the spiral's rotation, as shown in Figure 22. Leon et al.[123] previously observed this behavior using a modified version of the BR model in an anisotropic domain. They referred to this phenomenon as a *secondary wave of repolarization* instead, because a pronounced repolarization region formed between the stopped original tip and the newly formed one. In the example of Figure 22, only the new wave is observed without the repolarization island due to the very high excitability of the system and almost flat restitution, so that we use the term secondary wave of depolarization in this case. Leon et al[123] modified the BR model by increasing the sodium conductance and eliminating the *j* gate, thereby decreasing the minimum DI from 43 ms to 25 ms and increasing the minimum APD. Enhanced by the anisotropy of the system, these changes created a window of bistability that allowed the cessation of the tip's motion due to Doppler shift and the formation of a secondary wave of depolarization.

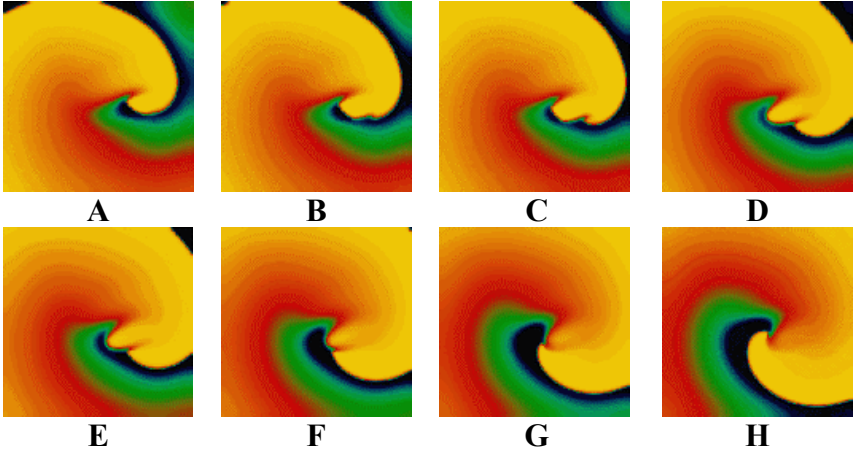

**Figure 22. Secondary waves of depolarization due to Doppler shift occurring at the spiral tip itself using parameter set 7. The period at the tip falls below the minimum necessary for propagation, causing the tip to halt its motion (A). However, a second wave of depolarization develops (B-C), moves around the stalled and now repolarizing original tip (D-F), and continues the motion of the tip (G-H).**

Doppler shift due to spiral wave drift has been observed in cardiac preparations arising from either experimentally induced[122] or naturally occurring[8,124] electrophysiological heterogeneity. In simulations, breakup by Doppler shift was first observed and described by Bär et al.[35,125] using a two-dimensional simplified model for CO oxidation where breakup is produced by meander. However, in some cases additional breakup in their model is produced by a *backfiring* effect that allows new waves to be generated in the wake of previous ones, which makes characterization of the breakup more complicated. The Luo-Rudy-I model[101] also exhibits breakup by Doppler shift with its original parameter values and with calcium dynamics speeded up by as much as a factor of 2.8[126]. It is important to indicate that this breakup mechanism shown in different regimes corresponding to various tip trajectories does not require steep APD restitution. In fact, both the LR-I and the Bär models have relatively flat restitution curves, where the slope is always less than one.

**Mechanism 5: Biphasic APD Restitution Curve**

So far only APD restitution curves that are monotonically decreasing functions of DI have been considered. However, some studies[127 128 129] have found APD restitution curves with a range of DIs for which the APD prolongs to a local maximum as DI decreases. These *biphasic* APD restitutions have been shown to lead to complex dynamics in 1D maps[130] and to spatiotemporal chaos in 1D rings[131]. However, the precise ionic mechanisms responsible for the supernormal phase still are not understood fully; furthermore, some experiments have shown that their existence may depend on the protocol used to measure the restitution curve[129 132]. Nevertheless, we include a brief discussion for completeness.

Experiments have shown different possible shapes and slopes for biphasic APD restitution curves. In addition, as mentioned above, the physiological phenomena that produce the biphasic restitution curves observed experimentally have not yet been identified. Therefore in the context of the 3V-SIM, we obtain a biphasic restitution curve by adding an extra current that deactivates at short DIs (see Appendix and set 8). We note that our intent here is only to replicate the mesoscopic restitution characteristics observed in experiments and not to reproduce the ionic basis. The shape of the restitution presented here is based on those obtained in rabbit ventricle preparations[127]; however, biphasic restitutions also have been measured in humans[129].

When the supernormal part of the restitution curve has a region with slope greater than one in magnitude, the complex dynamics[130 131] naturally would lead to spiral breakup in 2D. However, even when the magnitude of the restitution curve never exceeds one, the biphasic shape of the restitution curve can produce small variations in recovery (i.e., values of DIs) that can lead to conduction block, as shown in Figure 23.

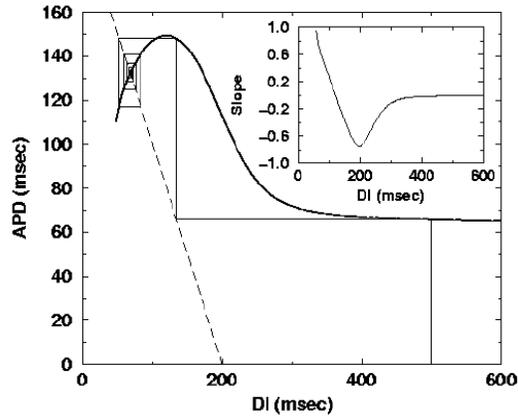

A

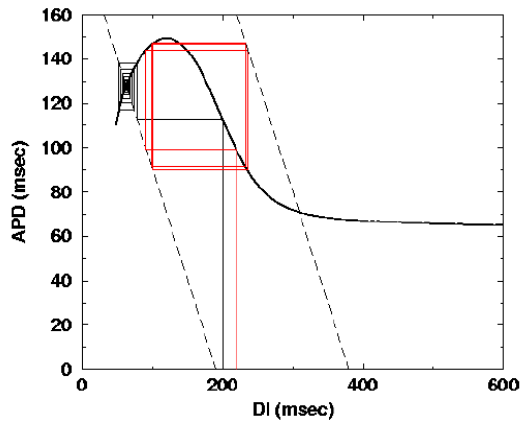

B

Figure 23. Oscillations due to a biphasic APD restitution curve. The shape of the restitution curve is based on observed restitution curves obtained from ventricular rabbit preparations[127]. Even when the slope of the restitution curve never exceeds one in magnitude (see inset), conduction block can form and lead to breakup. (A) For a period of 200 ms, a stable fixed point can be obtained even when using an initial condition with a large DI. (B) For the shorter period of 190 ms (dashed line shows the period and twice the period), two different initial conditions can result in either a stable solution (black line) or in 2:1 continuous block (red line).

In 2D, the conduction block generated by the biphasic portion of the APD restitution curve at given DIs can cause spiral wave breakup. Figure 24 shows an example of how breakup can develop under these conditions. An initially uniform spiral (A) develops a thicker wavelength as it turns and finds smaller DIs (B) and eventually breaks as a result of 2:1 block (C-D), as described in Figure 23. The local maximum in APD due to the biphasic restitution can lead to different wavelengths, large depressions in the wave back, and subsequent occurrences of 2:1 block (E-L).

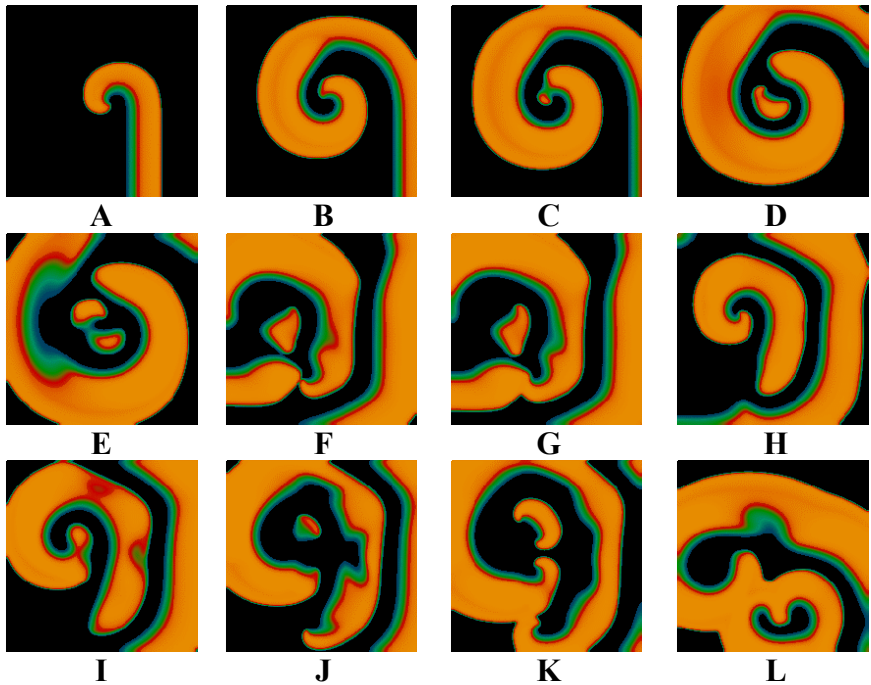

**Figure 24. Spiral wave breakup due to a biphasic APD restitution curve using parameter set 8. Although the spiral appears to initiate uniformly (A), its wavelength increases as it turns and encounters smaller DIs (B). Eventually, 2:1 block forms and causes breakup (C-D). As the breakup evolves (E-L), large variations in wavelengths, wide depressions in the wave back, and subsequent occurrences of 2:1 block (E-L) can occur.**

**Mechanism 6: Supernormal Conduction Velocity**

Like the APD restitution curve, the conduction velocity restitution curve also has been found in some experiments not to be a monotonically decreasing function. In the cardiology literature, such CV restitution curves are called supernormal (rather than biphasic). Supernormal CV also can be expressed as supernormal excitability, which is commonly represented as the dependence of excitation threshold on diastolic interval[133], Experiments have demonstrated supernormal conduction in the His-Purkinje system[133], papillary muscles[134], and the outflow tract of the right ventricle[135 136,] as well as in a related excitable system, the 1,4-cyclohexanedione Belousov-Zhabotinsky (BZ) reaction[137 138]. Simulations using supernormality have produced chaotic dynamics in 1D maps[133] and spiral wave breakup in a modified FHN model[37]. Although measurement protocols often are indirect and its existence in most regions of the heart is questionable[139,] we include supernormality as a possible arrhythmogenic mechanism, as in the case of biphasic resitutions, mostly for completeness.

Supernormal conduction, if it exists in cardiac tissue, may be arrhythmogenic by producing conduction blocks due to rapidly moving waves that collide and stack together at small DIs, as shown in the BZ reaction[137]. The solid curve in Figure 25A shows an example of a supernormal CV restitution obtained using the 3V-SIM with parameter set 9, while the dashed line illustrates a normal curve for comparison (using $\tau_v^-$ =15). Figure 25B shows the corresponding APD restitution curve, whose slope never exceeds one. The rapid waves generated by supernormality rush toward relatively slow-moving wave backs and practically slam into them, breaking and generating new waves, as shown in Figure 26. The breaks tend to develop fairly near the spiral tip because the small DIs at the tip produce waves that conduct supernormally. Scalloping also develops and leads to wave break, but in this case the scallops form because the supernormal conduction velocity generates heterogeneity of refractoriness, rather than from the steep APD restitution mechanism, since the slope of the APD restitution curve never exceeds one (see Figure 25B). In addition, because DI_min is very small in this case, some second waves of repolarization as in Figure 22 can occur once breakup has started (see Figure 26C-D). To demonstrate that the breakup is due to the supernormality, we show in Figure 26 panel I a stable spiral wave obtained using the same parameters but excluding supernormality (see dashed lines in Figure 25).

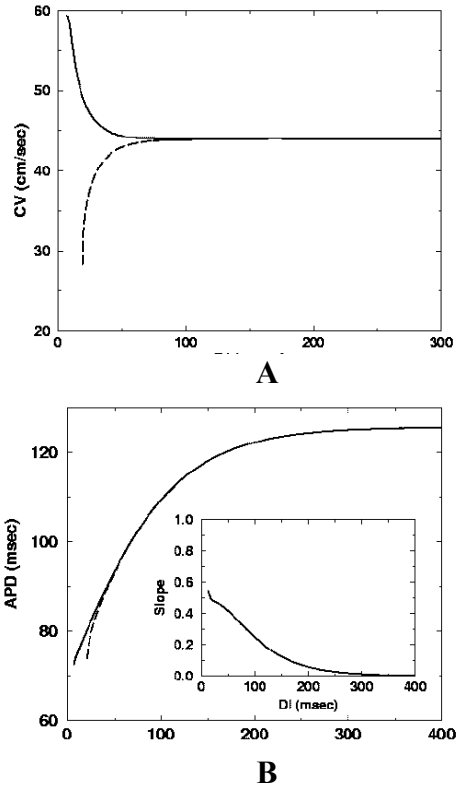

Figure 25. CV (A) and APD (B) restitution curves for parameter set 9, which produces supernormal conduction velocity. Solid lines indicate the restitution curves used with supernormal conduction velocity, while the dashed lines are associated with a more usual CV restitution curve (by using parameter set 9 but with $\tau_v-$ =15). The slope of the APD restitution curve never exceeds one, as shown in the inset of B.

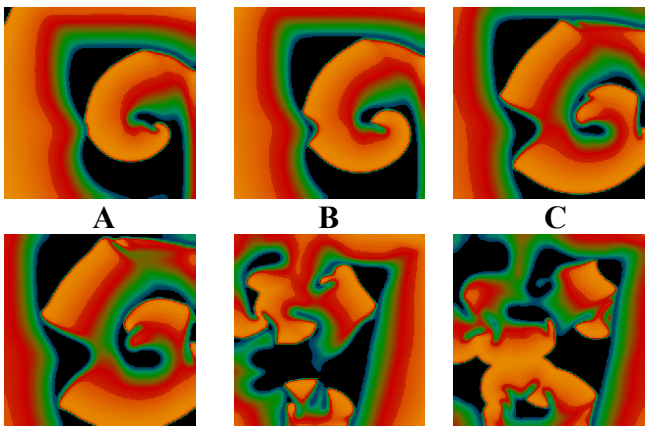

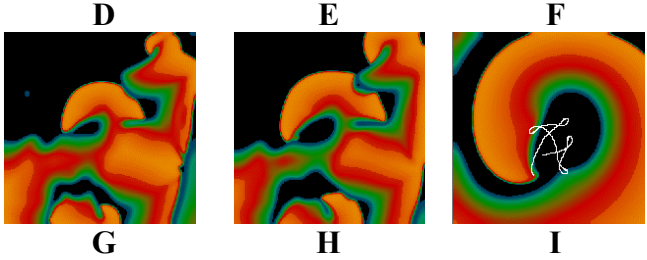

Figure 26. Breakup due to supernormal conduction velocity. (A-D) Breakup occurs as the spiral speeds toward the wave back in several locations, including at the left, where the spiral catches up to and collides with the previous wave back, and at the right of the medium, where boundary effects also play a role. Additional wave breaks occur as the medium evolves. (E-F) The wave in the upper right speeds up and collides with the back of the previous wave, resulting in a wave break. (G-H) As the wave in the upper right central portion of the domain speeds toward the scalloped back of another wave, it begins to break. (I) Stable hypermeandering spiral wave with no breakup obtained when the supernormal component of the CV restitution curve is excluded (dashed line in Figure 25A).

## V. Mechanism of Spiral Wave Breakup in Quasi-3D

**Mechanism 7: Periodic Boundary Conditions with Hypermeandering Tip Trajectories**

To this point, only rectangular shapes with no-flux boundary conditions have been used in all the simulations and no anatomical structure has been considered. However, periodicity is an important feature of cardiac structure and can affect the stability of spiral waves. Using periodic boundary conditions along two parallel edges of a 2D surface essentially forms a cylinder, which represents to a first degree a simplified geometry for some of the regions in the heart (such as the area between a valve rim and a blood vessel, or a ventricle and the septum), while still retaining the computational simplicity of a 2D plane. Figure 27 illustrates how a 2D plane with periodic boundary conditions on the left and right can be wrapped into a cylinder. Because such a cylinder is constructed from periodic conditions on a plane and not from cylindrical coordinates in space, surface curvature effects are not included. Although experimental results from chemical spiral waves on spherical surfaces[140] have shown that spiral tip dynamics are not affected by the curvature of the surface, curvature may induce breakup due to loading effects in some situations[141].

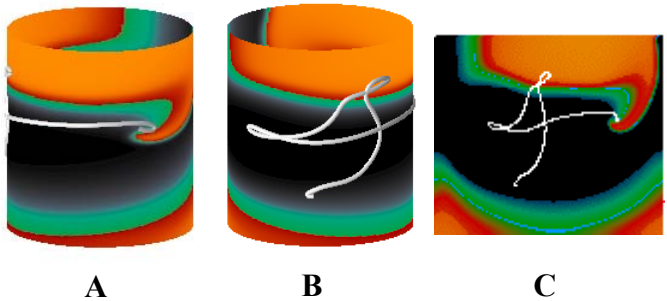

**Figure 27. Periodic boundary conditions along two parallel edges of a 2D sheet transform it to the topological equivalent of a cylinder. The voltage plot shown with its tip trajectory on the cylinder in A and rotated by $120^0$ in B depicts the same data as shown in the rectangular domain in C. Hereafter, the domains are shown only as 2D sheets. The height of the cylinders was reduced by 40 percent in the figures to aid in visualization.**

The dynamics of a spiral wave on a cylinder depend on the size of the tissue relative to the wavelength of the spiral. If the perimeter of the cylinder (i.e., the distance between the edges with periodic boundary conditions) is large compared to the spiral wave tip trajectory, the boundaries play no role in the tip dynamics and the trajectory is the same for both periodic and no-flux boundary conditions, as shown in Figure 28 A and B using parameter set 1 (same spiral as in Figure 1E). However, the interaction between colliding fronts (coming from both sides due to periodicity) produces some regions with different patterns of refractoriness using periodic boundary conditions compared to no-flux (Figure 28A-B). The dispersion of refractoriness changes as the perimeter is decreased further (Figure 28 C-D), but the spiral wave trajectory remains protected and unperturbed until the perimeter becomes comparable to the wavelength of the spiral (Figure 28 E), at which point the spiral tip becomes perturbed by self-generated incoming waves. In related previous work, Yermakova et al.[142] studied the dynamics of spiral waves periodically perturbed with plane waves and showed that when the pacing frequency is higher than that of the spiral wave, a drift is induced on the spiral wave tip. The same results are obtained using periodic boundary conditions when the cylinder perimeter is comparable to the wavelength of the spiral[97 143], since the wave front generated by the spiral propagates outward and is forced to collide with the spiral periodically as the spiral rotates, with the collision period dependent on the conduction velocity and the cylinder's perimeter.

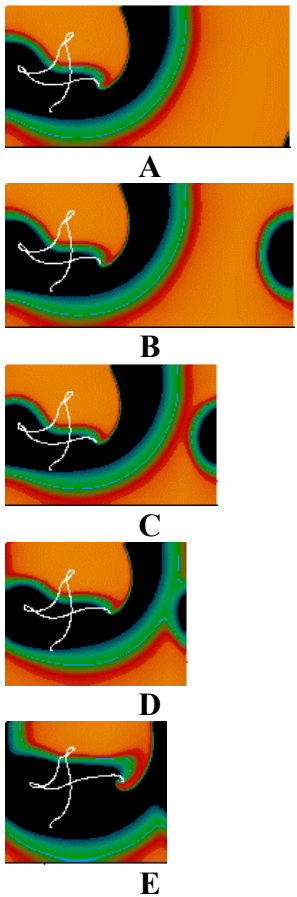

A

B

C

D

E

**Figure 28. Effects of periodic boundary conditions on domains of decreasing length using the 3V-SIM with parameter set 1. (A) Tip trajectory and voltage image with no-flux boundary conditions imposed on all four boundaries (size 12.64cm × 6.32 cm). (B) Tip trajectory in a tissue of the same size as (A) using periodic boundary conditions at the left and right edges and no-flux boundary conditions on the top and bottom. (C-E) Tip trajectories using periodic boundary conditions in progressively shorter domains (perimeter decreased from 12.64 cm to 9.5, 7.9 and, 7.3 cm, respectively). Although the trajectories are the same for B and C, differences in repolarization are apparent. Note that for the shortest perimeters (D and E) the tips and their trajectories begin to diverge from the originals and start to drift.**

Figure 29 shows how periodic boundary conditions and the size of the domain affect the evolution of a spiral wave following a circular core. The two plots in Figure 29A and B show four snapshots during one rotation for two domains that are identical except that B has periodic boundary conditions at the left and right edges; all other boundary conditions are no-flux, and parameter set 1 is used with $\tau_d$=0.403. Because the spiral wave interacts with itself under periodic boundary conditions, differently shaped quiescent regions form at the right side of the domain depending on the boundary conditions. When periodic boundary conditions are used, the quiescent region develops sooner because it was stimulated earlier by an encroaching wave that passed across the periodic boundary. As the perimeter of the cylinder is decreased from 4.74 cm to 4.58 cm in Figure 29C, the spiral wave tip is perturbed by incoming waves (since the period of rotation is smaller than the time required to travel once around the cylinder) and a drift in the tip trajectory develops. This drift is purely an effect of the boundaries and is independent of the type of tip trajectory (i.e., circular or meandering, see Figure 29D) and of the model used[85]. The only requirement is that the cylinder perimeter is smaller than the spiral wavelength, so that the spiral tip can interact with the waves it generates.

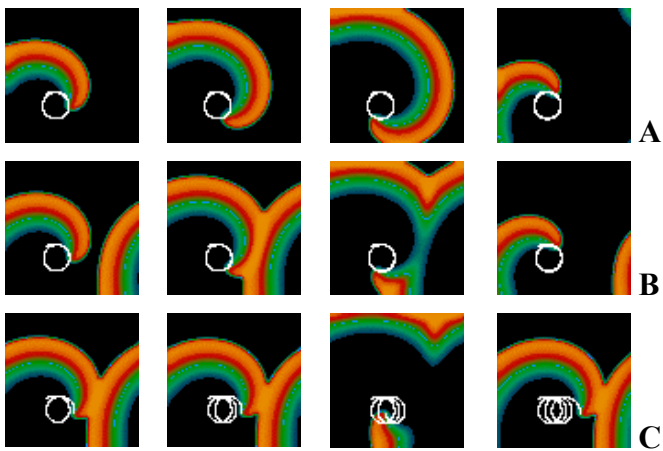

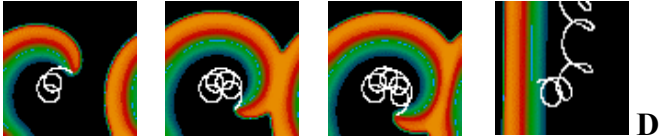D

Figure 29. Effects of periodic boundary conditions on spiral tip trajectories. (A) Four snapshots during one rotation of a spiral wave with circular core in a square domain with no-flux boundary conditions. (B) Same as in A, but with periodic boundary conditions on the left and right edges. (C) Drift induced by incoming waves in the periodic case once the length is decreased from 4.74 cm to 4.58 cm and the period of rotation is larger than the time required to travel along the entire length. Under these conditions, waves interact with the spiral tip and produce drift. (D) The first three panels show an unperturbed meandering spiral wave following an epicycloidal trajectory. When the length is decreased slightly, the spiral drifts, which can cause it to vanish at the boundary (as shown by the trajectory remaining in the last panel) and leave a rotating stable wave front that circulates around the cylinder indefinitely.

Once drift is present, a further decrease in the cylinder's perimeter makes the spiral wave encounter its own incoming waves sooner, equivalent to pacing a spiral wave at a frequency much faster than its own and resulting in an increase in the drift velocity[142]. Once the perimeter is smaller than the tip trajectory, no spiral wave activity can be sustained. However, for spiral waves in the hypermeandering linear regime, there is a window of cylinder perimeters between drift and termination for which spiral waves will break[143][97]. The breakup is produced because the hypermeandering wave tip repolarizes some regions along the cylinder unevenly[85]. Therefore, the incoming waves produced by the spiral itself can block the tip trajectory and form new spiral waves. Activations can disappear when the spirals annihilate with the no-flux boundaries. Figure 30 shows a sequence of voltage plots of spiral wave breakup on a small cylinder with perimeter 7.11cm using parameter set 1. The spiral wave shown in A collides with a portion of itself moving to the right across the periodic boundary in B. The new and old fronts merge (C), but interaction with the wave back causes only a small fragment of the original spiral to remain (D). As the spiral continues to turn (E-F), it again encounters the wave back and fragments into multiple waves (G-I). These two fronts merge again (J), but further breakup occurs as the fronts continue to interact with the wave backs (K-P).

Ultimately, the no-flux boundary conditions, at the top and bottom of the cylinder, absorb all the wave fronts (Q), leaving behind only quiescent tissue. The last frame of Figure 30 shows how the same initial condition in the same size domain but with all no-flux boundary conditions results in a stable hypermeandering spiral wave. Similar breakup due to periodic boundary conditions has been obtained using the FHN model with linear cores[57][58] on a cylinder with a perimeter of 280 grid points and using the BR model with the speed of calcium dynamics increased by a factor of two (MBR), which generates stable spiral waves in 2D tissues, using no-flux boundary conditions[33][88] with a cylinder perimeter of about 4.5cm..

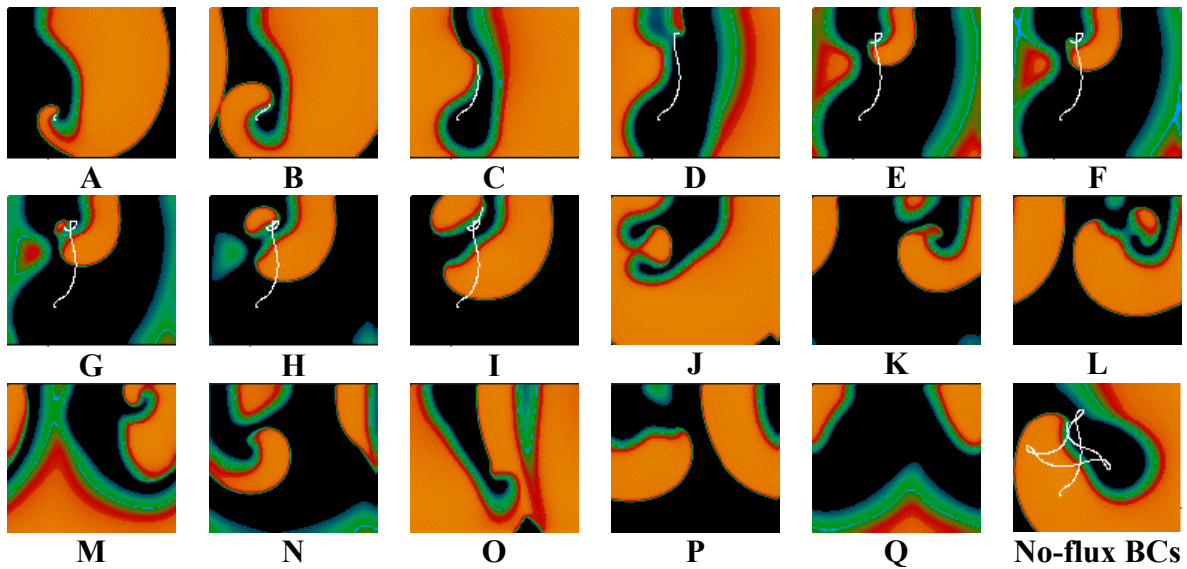

**Figure 30. Evolution of periodic boundary condition-induced breakup of a spiral wave with a hypermeandering tip trajectory. Because the tip meanders, regions are repolarized unevenly, allowing conduction blocks and subsequent wave breaks to develop, as described in the text. The tip trajectory is shown in frames A-I but is discontinued in the remaining frames because multiple spiral waves are present. Ultimately, the breakup is transient, as all the wave fronts eventually are absorbed by the no-flux boundaries (top and bottom) (Q). As shown in the last panel, no breakup occurs when no-flux boundary conditions are used on all edges in the same size domain.**

## VI. Mechanisms of Spiral Wave Breakup in 3D

While atrial tissue may be thin enough to be considered effectively two-dimensional (leaving aside its complex geometry, fiber structure, and regional variations in conduction velocity)[22 144], the ventricles are substantially thicker and three-dimensional effects may need to be considered. In particular, since the early experiments of Garrey[29] in 1914, who showed using canine hearts that fibrillation in the thinner right ventricle ceased when it was disconnected from the thicker left ventricle, the inclusion of a third dimension presents what has been since then a controversial open question: *is ventricular fibrillation purely a three-dimensional effect?*

The issue of whether ventricular fibrillation is intrinsically three-dimensional is a fundamental claim largely substantiated by the concept of a minimum mass necessary for fibrillation (more specifically perhaps, would be a minimum size in relation to wavelength) initiated with Garrey[29]. Among other things, he showed that pieces of ventricular muscle with a surface area of less than 4 $cm^2$ that had been shaved from the left ventricle stopped fibrillation, whereas the remaining portion of the left ventricle continued to fibrillate until 75 percent of the ventricle had been removed. Dillon et al.[145] and Kavanagh et al.[146] found comparable results also in canine ventricles when part of the ventricles were inactivated by a transmural infarct, leaving the rest of the tissue quasi-2D. Zipes et al.[147] suggested that fibrillation emanated only from the thicker left ventricle after they chemically depolarized the left ventricles of canine hearts and observed that fibrillation transitioned to sustained monomorphic tachycardia with a period of about 160 ms. Similarly, conversion from fibrillation to tachycardia in Langendorff-perfused rabbit hearts was obtained by Allessie et al.[148], Breithardt et al.[149], and Schalij et al.[150] when the intramural layers of the ventricles were eliminated by necrosis coagulation caused by freezing the endocardium with liquid nitrogen. This procedure left a surviving epicardial layer of about 1 mm thick[149] (quasi-2D), and in contrast to the infarction-induced thinning experiments, freezing seemed to preserve the electrophysiological properties[150] that otherwise potentially could alter the dynamics of electrical waves. More recently, two experiments of tissue reduction in porcine hearts, one[9] using freezing and the other[72] using sequential cuts of 2x4 cm portions of the fibrillating tissue parallel to the epicardium, also found evidence supporting the critical mass hypothesis, with a minimum mass of about 20 g needed to support fibrillation and a decreasing number of spiral waves as the tissue mass was decreased. The main conclusion of ventricular thinning experiments, emphasized by Winfree[151][152], is that hearts below a critical electrically active muscle thickness do not fibrillate, but instead support stable forms of tachycardia that can be associated with a single spiral wave. The minimum thickness, however, needs to be determined with respect to the wavelength[153] and perhaps the size of the tip trajectory[62].

In numerical simulations, the inclusion of a third dimension has been shown to widen the range of parameters that produce breakup for some of the 2D breakup mechanisms[85][154], although the parameter ranges increase only by five to ten percent. More importantly, however, simulations in3D have shown the existence of purely three-dimensional breakup mechanisms[38][39][40][84] that have no analogues in 2D. In the following subsections we discuss three mechanisms of 3D scroll wave breakup: negative tension in homogeneous tissue, twist instability in the presence of rotational anisotropy, and coarse discretization, also in the presence of rotational anisotropy.

**Mechanism 8: Negative Tension in the Low Excitability Regime**

In 3D, while a spiral waves becomes a scroll wave, the spiral tip expands from a single point to a one-dimensional line called a vortex line or filament. The simplest form of a scroll wave is a straight vortex obtained trivially when 2D spiral waves are stacked perpendicular to their plane of rotation over a finite thickness. Scroll waves like this both were first found in experiments with the Belousov-Zhabotinsky chemical reagent and were proposed to exist in myocardium by Winfree in 1973[155], and were first observed in ventricles extending upright 10-20 mm from endocardium to epicardium by Chen et al.[156] in 1988 and by Frazier et al.[157] in 1989. Vortex filaments, however, need not be straight lines, but instead can curve, bend, and twist[15,158,159,160] and sometimes form closed rings confined inside the medium without touching any boundaries[161,162]. Figure 31 shows an example of such a scroll ring. In the figure the wave front, shown in gold (over half the domain for clarity), forms a scroll-shaped surface. Notice how the spiral tip evolves from a point to a line (or vortex), which in this case closes to create a ring (vortex ring), shown in red. On the right, the 2D colored voltage image showing a spiral corresponds to what is seen throughout the tissue as a plane is rotated about the axis pictured in black.

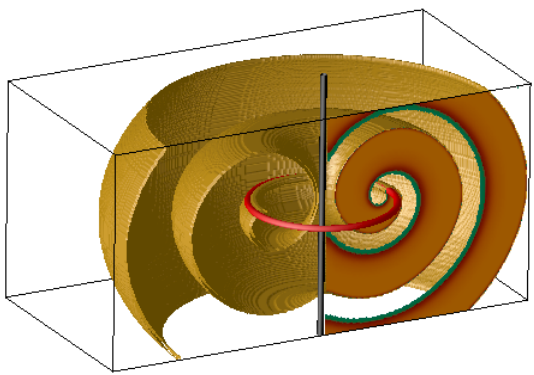

**Figure 31. Three-dimensional scroll wave. A 2D slice showing the voltage profile of a spiral wave is shown on the right. The gold scroll-shaped surface tracks the wave front in 3D, and the red circle shows the instantaneous location of all spiral wave tips in all slices, otherwise known as a filament or vortex line. Half the domain is shown for clarity.**

Numerical simulations of vortex rings have shown a wide range of dynamical behavior in various parameter regimes. While twisted and knotted scroll rings can be stable[163 164], untwisted scroll rings drift along the ring symmetry axis as they shrink or expand with velocities inversely proportional to the ring's radius[165 166]. They shrink, contract, and disappear when the excitability of the system is normal or high and expand when the excitability is low[165]. There are some small parameter regimes in which boundaries can have an effect in stabilizing or even reversing the initial drift direction[85 167 168 169]. 3D calculations of a scroll ring can be performed in 2D cylindrical coordinates ($\rho$ and z) whenever the ring symmetry axis is aligned with the z-axis since, due to the symmetry involved, all angular dependencies become zero ($\partial/\partial\phi = 0$). Examples of scroll ring drift using the 3V-SIM for the high and low excitability limits are shown in Figure 32, where two-dimensional cuts on the $\rho$-z plane similar to the 2D voltage color contour shown in Figure 31 are used. Figure 32A shows four snapshots during the contraction of a scroll wave in the normal-high excitable limit and in a regime where spiral waves follow hypocycloidal meander patterns. As the scroll radius (plotted on the horizontal axis) decreases (scroll contraction), the vortex ring follows the hypocycloidal trajectory and drifts along the *z*-axis. The last frame shows the moment before the ring collapses. Figure 32B and C show the evolution of two scroll waves in the opposite regime, the low excitability limit, where scroll waves expand. Panel B shows a scroll wave with an epicyclical meandering trajectory and panel C shows one with a circular trajectory. In both cases, the expanding rings also drift along the z-axis. It is important to note that, as mentioned in the introduction, tip trajectories can be modified by changing either the excitability or the wavelength, so that a tip trajectory does not necessarily indicate tension regime (for example, circular core regimes can exist for both the high and low excitability limits).

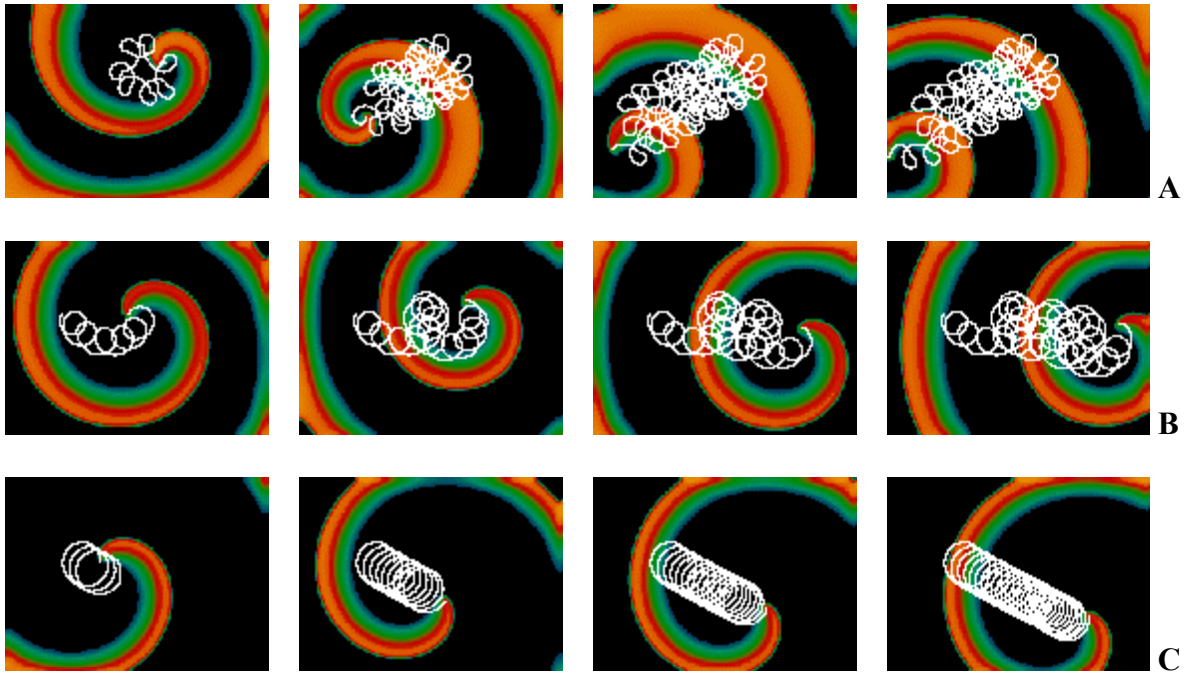

**Figure 32. Examples of contracting and expanding scroll rings. Each row shows four snapshots of a 2D plane (parallel to the scroll axis of symmetry as in Figure 31) showing the evolution of a scroll wave. The ring in (A) is in the negative tension regime and follows a hypocycloidal trajectory as it drifts along the z axis, contracts and eventually disappears (the left edge corresponds to zero radius). The rings in (B) and (C), which are in the negative tension regime, expand rather than contract as they also drift along the z-axis. Sometimes boundary effects can stabilize a scroll ring, so that a shrinking ring does not collapse but equilibrates with its mirror image in the z direction[85], and an expanding ring can change direction and shrink once it reaches the boundaries. Parameters correspond to set 1 using $\tau_d$=0.35, 0.39 and 0.416 in A, B and C, respectively.**

Biktashev et al.[38], based on Keener's asymptotic theory for reaction-diffusion systems[170][171], obtained a vortex filament evolution equation for the no-meander limit and introduced the concept of *filament tension* to describe the radial growth of an untwisted circular filament. The filament tension is the proportionality coefficient $\gamma$ in the local radial velocity of a circular filament ($V_r = dr/dt = -\gamma/r$). When $\gamma$ is positive (high excitable limit), scroll rings shrink and any perturbation to a straight filament tends to disappear. On the other hand, when $\gamma$ is negative (low excitability), scroll rings expand and any small perturbation to an initial straight filament grows exponentially[38]. Therefore, the term positive tension can be applied for $\gamma > 0$ where small perturbations decay and negative tension for $\gamma < 0$ where small perturbations grow.

When a straight scroll wave is induced in the negative tension regime, it remains straight unless a small perturbation is applied. Biktashev et al.[38] showed, using the FHN model in a low excitability regime, that if a perturbation is applied to a straight vortex filament, the vortex will elongate and curve until it collides with the boundaries, therefore breaking up and generating a second vortex. In addition, these vortices subsequently will produce new vortices by expanding and touching the boundaries or by breaks due to conduction blocks from collisions with other vortices. On the surface of the medium, the evolution of these vortices produces complex voltage activation patterns. Figure 33 shows an example of how a transmural filament evolves under these conditions, using the 3V-SIM with the parameters from Figure 32C. From the initial perturbation (A), the filament elongates (B-D) until it hits the boundary (E) and forms a second filament. Both filaments continue to elongate (F) and at times can disappear from both surfaces of the tissue and become intramural (G-H). Further evolution yields additional filaments, some by conduction blocks and some by collisions (J-K). Negative tension can substantially elongate a vortex (L-N) before it fragments again into multiple waves (O).

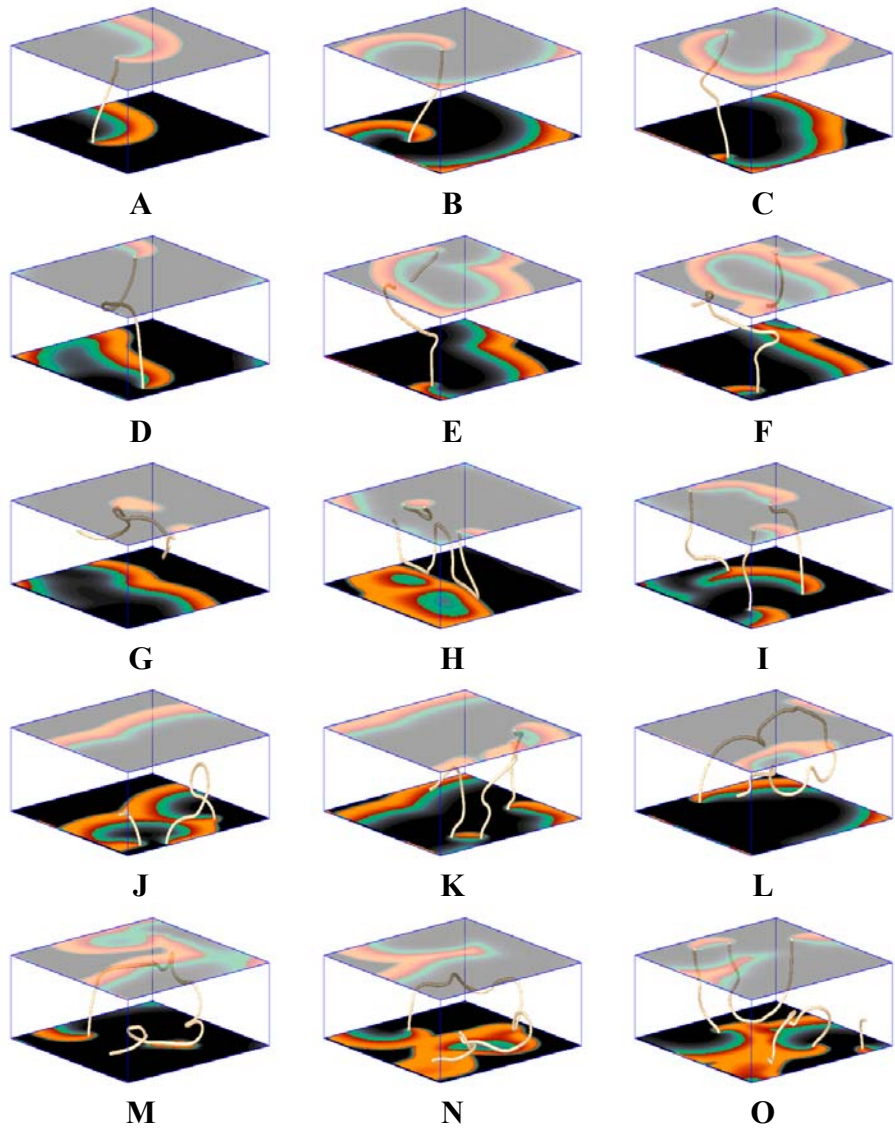

**Figure 33. Evolution and breakup of a three-dimensional vertical scroll wave due to negative tension in the low excitability limit. A small perturbation is applied to the initial filament. The voltage is shown for the bottom of the tissue and semi-transparently for the top to allow filament visualization throughout the tissue. Note that as the vortex elongates because of the negative tension, it touches the boundaries and forms multiple scroll waves. The vertical dimension has been stretched by a factor of 1.4 to allow easier visualization of the filaments.**

In cardiac tissue, the low excitability limit is reached when not enough oxygen is being supplied to the cells. This generally occurs during hypoxia or ischemia produced, for example, soon after a coronary occlusion.

**Mechanism 9: Fiber Rotation with Twist Instability**

Ventricular muscle is composed of elongated cells arranged roughly end-to-end to form fibers that conduct about three times faster along their axes than across. The fibers are arranged in sheets roughly parallel to the epicardial and endocardial surfaces, but their fiber axis rotates continuously from epicardium to endocardium[172 173]. Peskin's[174] derivation of the ventricular fiber architecture from mechanical principles predicts a total fiber rotation of about 180° between walls, which is in close agreement with what has been measured in dissection experiments[175]. According to experiments, this high angle of rotation seems to remain roughly constant for the right and left ventricles, despite the differences in thickness, and occurs in many mammalian species[151].

In previous numerical studies[40 84 85] of vortex dynamics in parallelepipedal slabs of ventricular muscle with various wall thicknesses and fiber rotation rates, it was found that rotational anisotropy can have a destabilizing effect on scroll waves in the high excitability limit. One of the effects of fiber rotation on scroll waves is the induction of a phase delay on the waves across the layers forming the thickness of the tissue. Figure 34 shows an example of this delay for a scroll wave in the circular core regime on a slab 0.22 cm thick and with a total of 26° in fiber rotation. Superimposed are the trajectories of the scroll wave produced at the simulated top (epicardium) and bottom (endocardium) portions of the slab during one period of rotation. The anisotropy (which is modeled by a conductivity tensor where propagation is about three times faster along the fibers axes than across and transmurally, see Ref. 40 for further details) transforms the circular trajectories into ellipses, which are not perfect because of diffusion in the *z* direction. A voltage contour (at 85% of repolarization) of the scroll wave (spiral wave) at each surface is also plotted at one instant in time. The figure shows that while the spiral wave on the epicardium has already completed the pivot turn on the curved section of the distorted elliptical trajectory, the spiral on the endocardium is just starting its pivot turn. Each time there is a pivot turn in the scroll trajectory (twice per period for a circular core that has been elongated), this phase lag induces a twist in the vortex line that is not uniformly distributed along its length, as in the case of sproing[75 176], but rather is highly localized[40 84 85]. The magnitude of the localized twist grows as the fiber rotation rate is increased, and the twist can produce elongation of the vortex filament as it travels along it[40 84 85].

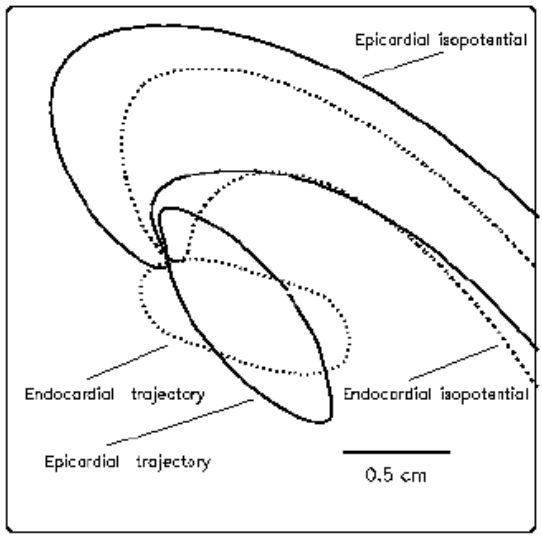

**Figure 34. Different timing of turns in 3D tissue with rotational anisotropy.** Elliptical trajectories (produced by anisotropy) of a scroll wave are shown for the top (solid line) and bottom (dashed line) of a slab with 26° in fiber rotation. While the spiral on the epicardium (solid line) has passed the pivot turn already, the spiral on the endocardium (dashed line) has not yet started its pivot turn. As a result, the vortex has developed a twist in its phase intramurally (as shown in Figure 35). Parameter set 1 is used with $\tau_d$=.416, using an anisotropy ratio of 5:1.

For scroll waves in the high excitability limit, where spiral waves tend to follow hypermeandering or linear core trajectories with high angle pivot turns, twist can produce large enough elongations to cause the filament to collide with boundaries and to produce new vortices. Figure 35 shows this formation, where twist is represented by plotting the normal vectors ($\mathbf{N}= \nabla V/|\nabla V|$, shown in blue) of the spiral wave tip equally spaced along the vortex line[40]. In the first panel the normal vectors at the top of the vortex are pointing to the right while those at the bottom are pointing to the left, indicating that the spiral at the top surface has already rotated around the pivot turn while the one at the bottom has not (as shown in Figure 34). The change in orientation occurs somewhere in between the top and bottom layers, where a highly localized transition denoted by a large degree of twist develops. This twisted section of the filament travels transmurally along the filament, as the lagging phase of the scroll wave finishes the pivot turn (B-C), resulting in elongation and curving of the filament that cause it to collide with a boundary and generate a new vortex (D). Although the breakup mechanism is different from the negative tension mechanism, the observed vortex elongation and collisions with boundaries are similar. This twist-induced destabilization of vortex filaments occurs above a critical thickness that depends on the tissue fiber rotation rate. In Refs. 40 and 84, the relation between fiber rotation rate, thickness, pivot turn angle of a spiral wave trajectory, and breakup was discussed in great detail. It was shown that while fiber rotation induces twist, the total amount of twist accumulated ultimately is determined by the trajectory of the spiral wave tip. For a given rotation rate, the sharper the pivot turn in the tip trajectory, the larger the twist induced by the lagging phase. Therefore, models that are in parameter regimes where their spiral waves show pronounced "petals," in which the trajectory makes a loop and crosses its own path as in the MBR model (which exhibits short petal distance ~2-3 mm, with a high angle of rotation ~180)[84], require a lower fiber rotation rate to induce breakup in 3D compared to models with faster sodium kinetics like the LR-I model (with the speed of calcium dynamics increased by a factor of three to produce stable spirals), whose trajectories also have sharp turns but have less pivot angle since they do not cross over[85 126]. Compare, for example, the trajectories in $^{\text{Figure 1}}$E and F, which are similar to the MBR and LR-I models with faster calcium dynamics[85 126], respectively.

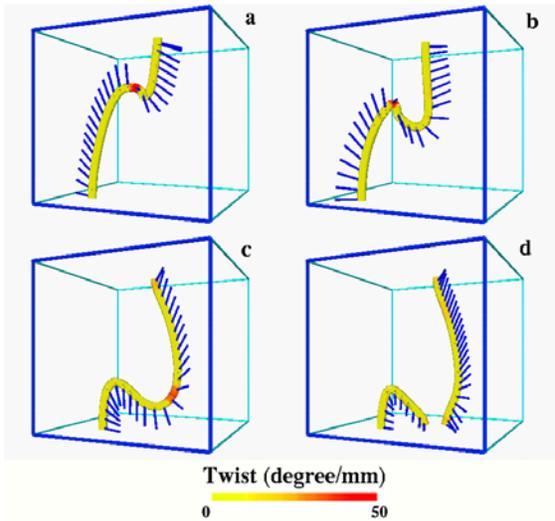

**Figure 35. Intramural twist and vortex elongation. Four snapshots of a transmural vortex using the 3V-SIM fitted to the MBR model (parameter set 1) in a slab 0.75 cm thick with a fiber rotation rate of 12°/mm[40]. The normal vectors shown in blue illustrate the direction of the spiral wave tips along the vortex and indicate an intramural highly localized twist (shown in red) produced by a phase difference (see text and Figure 34). As the twist propagates through the vortex line (at about 20 cm/s), it elongates and collides with a boundary, splitting the original filament into two segments, one that remains transmural and one corresponding to a half vortex ring.**

To emphasize that twist-induced breakup in tissue with fiber rotation is a function of the fiber rotation rate and the spiral wave trajectory[84 85 66 126] and not of the rotation rate and the slope of the APD restitution curve, as argued in Ref. 71, we illustrate this mechanism using parameter set 10, which produces spiral waves with linear core and has an APD restitution with slope less than one (see Figure 36). Figure 37 shows breakup of an otherwise stable 2D spiral wave (see Figure 36, inset) in a 3D slab (4.3 x 4.3 cm and 0.645 cm thick) with 180° degrees of total fiber rotation (28°/mm rotation rate). Because the pivot turn is not as pronounced as in the MBR model, a higher fiber rotation rate is needed for breakup compared to the MBR model[40 126]. A straight scroll wave is used as an initial condition (A). As the scroll wave rotates and evolves, the filament buckles as it elongates due to the traveling twiston[40 84] (B-F), forming a target pattern at the lower surface as the bent part of the filament approaches until it breaks into two vortices, one transmural and one half ring, when it touches the lower boundary (G). Half rings can expand further and break as they touch other boundaries (H-J). As time progresses the activity becomes increasingly irregular as more vortices are created and annihilated either by elongations or by conduction blocks (K-R). Note the creation of a single intramural scroll ring (M-N), which in an isotropic medium would collapse and disappear, but which here elongates and breaks at the boundaries due to the anisotropy, producing more vortices that sustain the fibrillatory-like behavior. Because the minimum APD produced in this model is larger than that of the MBR model, the density of scroll waves is smaller than in the MBR[126] and MBR-like simulations[40].

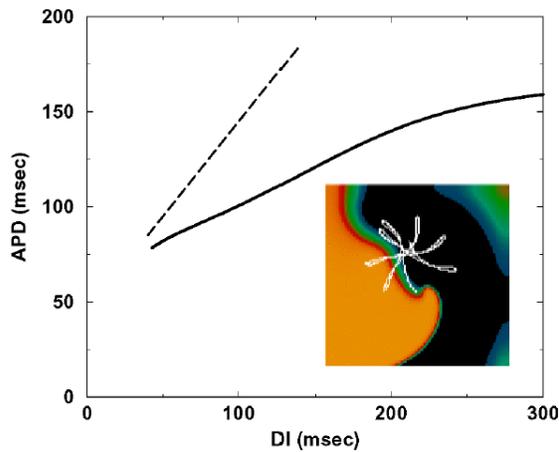

**Figure 36 APD restitution curve with slope less than one obtained using parameter set 10. The dotted line has slope one. The inset shows a stable 2D spiral wave and its tip trajectory using the same parameter values.**

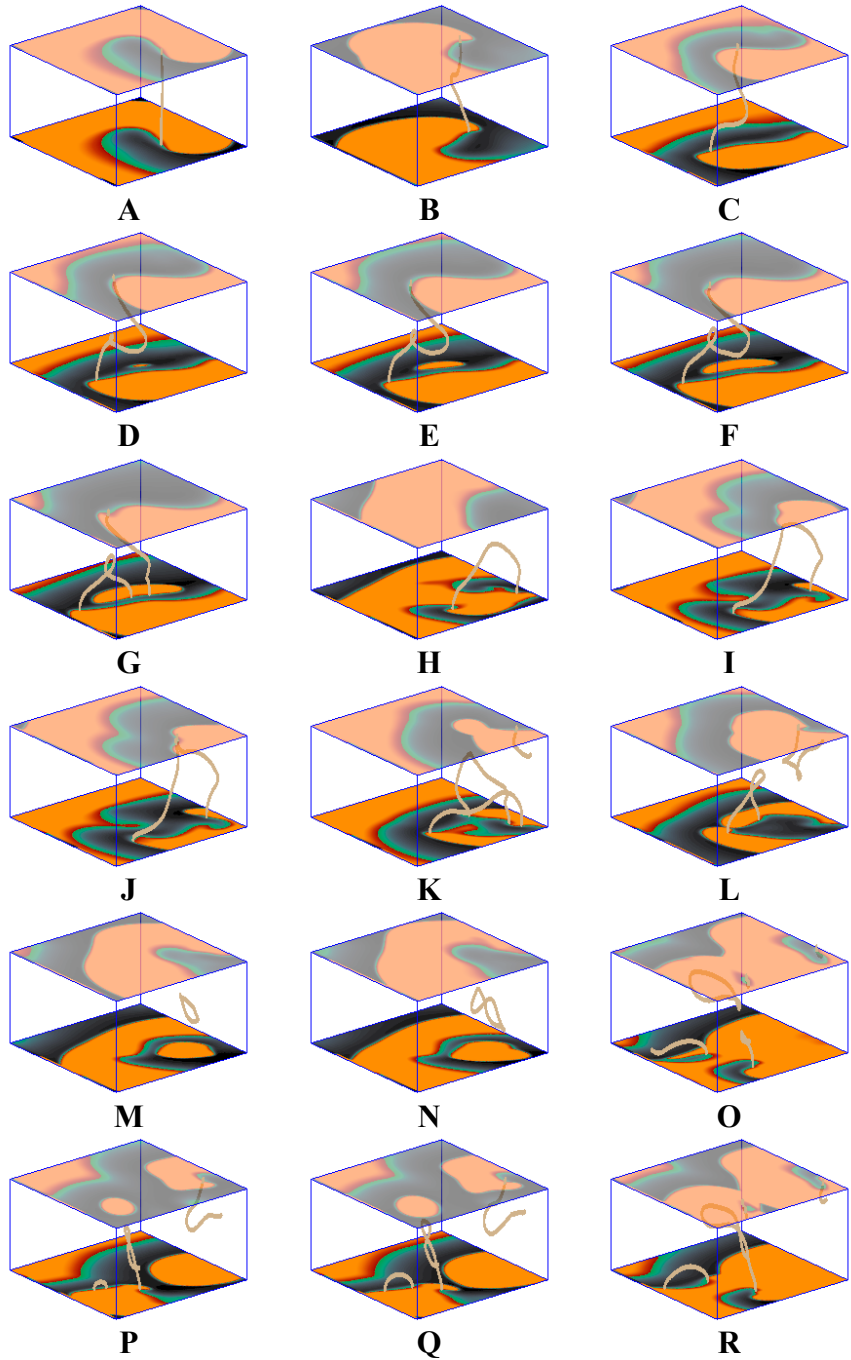

**Figure 37. Breakup and complex dynamics due to filament twist using parameter set 10 in a 3D slab of tissue (4.3 x 4.3 cm x 0.645 cm) with 180° degrees of total fiber rotation and using an anisotropy ratio of 5:1. An initially straight filament (A) rotates and evolves, accumulating twist and elongating (B-D). A target pattern appears at the lower surface as the bent part of the filament approaches it (E-F) until the filament touches the lower boundary and breaks into a transmural filament and one half ring (G). Later, a half ring expands and becomes two transmural filaments upon touching the boundary (H-J). Elongations and conduction blocks continue to occur (K-R), sustaining the fibrillation-like dynamics. Using the same parameter values in 2D produces a stable spiral wave (see Figure 36, inset). To facilitate visualization, the vertical dimension has been stretched by a factor of two.**

**Mechanism 10: Fiber Rotation with Coarse Discretization**

While so far we have considered cardiac tissue as a continuous medium, experimental evidence accumulating since the 1980s has suggested that the discrete nature of the cells and the anisotropic distribution of intracellular connections[177,178] can lead, in some cases, to discontinuous effects in propagating waves. In particular, it has been shown that cellular discreteness can affect the excitability and safety factor for propagation that can lead to reentry without the presence of spatial differences in refractory periods[50,179,180]. These types of discrete effects become stronger as tissue becomes ischemic[181,182].

Panfilov and Keener[39,183] made the first studies of scroll waves in parallelepipedal slabs with rotational anisotropy where discrete effects were considered. Using a piecewise linear FHN-model, they observed that rotational anisotropy could destabilize scroll waves and produce breakup[39]. The scroll wave instability resulted from the discrete anisotropic refractoriness, which can produce patchy propagation failure at coarse spatial resolutions[184,185,186]. This breakup mechanism, as the twist instability one, occurs as a function of tissue thickness, but otherwise the two mechanisms are quite different. Coarse discretization-induced propagation failure occurs preferentially in the transverse and transmural directions, an effect amplified by the rotational anisotropy. Therefore, the thickness required for breakup also depends on the spatial resolution used.

To show this effect, we use the same parameters as for the twist mechanism (set 10) in a slab of similar thickness as that shown in Figure 37, but with a much smaller rotation rate in which breakup due to twist cannot occur. In this case the slab size is 6.8x6.8 cm and 0.69 cm thick, with a total of 100° of fiber rotation. Therefore, the fiber rotation rate is 14.5°/mm, which is almost half of that used in the twist breakup example, and the spatial resolution used (Δx = 0.086 cm) is four times larger than in the twist example. (The spatial resolutions used in Ref. 39 for the FHN model breakup were 0.09 and 0.125 cm.) At this resolution and thickness, the slab contains nine layers, and Figure 38 shows every other layer of the slab as a column at one instant in time, with the top voltage contour plot corresponding to the top layer. The first column (A) shows the intramural scroll wave 20 ms after initiation. As the scroll wave evolves propagation failure begins to occur (B-D), and eventually leading to multiple waves (E-H). Because this breakup is due to discretization, the breakup disappears in the continuous limit: using a finer spatial resolution results in a stable scroll wave, since twist-induced breakup for this thickness and rotation rate (almost the same thickness and half the rotation rate as in Figure 37) does not occur for this parameter set.

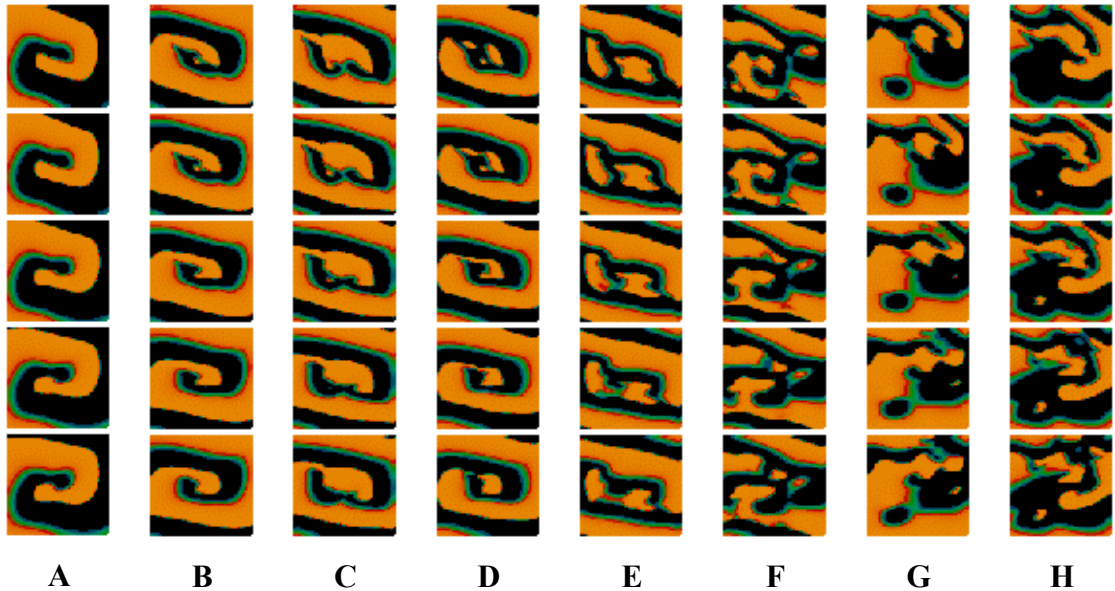

A    B    C    D    E    F    G    H

**Figure 38. Breakup in 3D due to coarse discretization with rotational anisotropy, using parameter set 10, in a 3D slab of tissue (6.8x6.8x0.69 cm) with 100° of total fiber rotation and using an anisotropy ratio of 5:1. Voltage plots of the top and bottom surfaces, along with three evenly spaced interior surfaces, are shown at eight different times (after 20, 250, 330, 410, 1000, 1300, 1600, and 2000 ms).**

**The initially straight scroll wave becomes distorted due to the rotational anisotropy, and the coarse discretization induces breakup. Note that breakup does not occur when the same simulation is performed with much finer spatial resolution, as discussed previously[183].**

It is important to note that when including discrete effects, either by discretizing directly the cardiac cells as in Ref. 180 or by using a coarse spatial resolution, caution is required when choosing the protocols and model parameters, since numerical artifacts leading to unrealistic results sometimes can be induced. As an example, Figure 39 shows how coarse discretization can lead to incorrect spiral trajectories due to lattice pinning. The spiral wave from the coarse simulation pins to the computational grid, producing an almost rectangular trajectory instead of the correct circular one and yielding a much longer period of rotation.

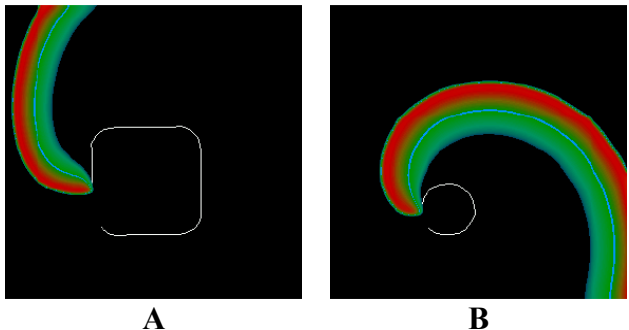

A                    B

**Figure 39. Lattice pinning resulting from coarse discretization in the low excitability regime. (A) Spiral wave trajectory pinning to the lattice. (B) Spiral wave trajectory obtained by decreasing the spatial resolution by 10%. The pinning occurs because the curvature at the spiral tip is close to the critical radius of curvature, so that only those adjacent cells on one side of the tip become excited (causing the straight line motion seen in A) until the spiral arm rotates enough for the tip to turn As a result, the trajectory is very different and the period is completely wrong.**

## VII. Discussion

### The Role of APD and CV Restitution Curves

Although the shape of restitution curves cannot solely predict spiral tip trajectories (e.g., circular cores can be obtained for both flat and steep APD restitution curves, as in <sup>Figure 1</sup>a and Figure 12, respectively), they are useful to explain and to determine some regimes in which certain conduction blocks can develop and produce spiral wave breakup. Throughout section IV six different mechanisms for breakup in 2D were described depending to some extent on their APD and CV restitutions.

*Steep APD Restitution Curves*

The first two mechanisms originate when the APD restitution curve has a slope greater than one over some range of DIs, which can produce alternans of APD and even conduction blocks at high frequencies. While in principle both mechanisms can be considered as one, we make a distinction based on two factors: the steepness of the APD restitution curve and the frequency of the source. If the APD restitution curve is abruptly steep (i.e., has a narrow range of DIs over which the restitution is much greater than one), then the region for possible periods with oscillating APDs that result in stable alternans rather than conduction block is narrow. Therefore, spiral waves will break close to the tip as they form, since conduction blocks will be present whenever a DI falls out of the stable alternans region, as shown in Figure 4 and Figure 7. We refer to this breakup as mechanism 1 or breakup by an abruptly steep APD restitution.

If instead the APD restitution curve is steep but not abruptly steep (i.e., has slope greater than one for a relatively large range of DIs), then the region for stable alternans is wider and there is a chance for a spiral wave to form without generating conduction blocks. Nevertheless, under these conditions, the conduction velocity restitution can affect the behavior of the spiral wave and produce breakup. It has been shown that when tissue is periodically stimulated at a frequency in the alternans region, CV restitution can induce discordant alternans[48,103] along the tissue, which in turn, can lead to conduction blocks[106 95]. The distribution of nodes that separates the discordant alternans regions becomes a function of the CV restitution[48]. In particular, nodes occur closer to the stimulus site and more densely packed as the CV restitution changes over a wide range, as shown in Figure 10E and F. Therefore, for a spiral wave whose period is in the alternans region, the CV restitution curve will dictate if discordant alternans and block can form in a specific tissue size[48 95]. In Figure 11 we show how, for a given frequency, conduction block depends on the distance from the source. This means that the stability of a spiral wave whose frequency is in the alternans region is a function of size[34], so that a spiral wave that is stable in a square domain of length $L$ may not necessarily be stable in a domain of length $r*L$ ($r>1$), as illustrated in Figure 12. Unlike mechanism 1, which occurs rather quickly and close to the tip, the breakup due to discordant alternans develops over time, requiring many beats to form and originating far from the source. Therefore, we distinguish it as a separate mechanism.

It is important to note that even when the APD restitution curve is abruptly steep, breakup by discordant alternans still can be produced (by pacing periodically at a constant frequency in the region of alternans). However, since the range of periods for stable alternans is very narrow (for example, 30 ms using parameter set 1 compared to 85 ms for set 2), spiral waves with abruptly steep APD restitutions, are more likely to break up by mechanism 1 than by mechanism 2. One possible exception to this hypothesis could occur if a spiral were able to pin to an inhomogeneity of a size such that the period lay precisely in the alternans region. Similarly, breakup by mechanism 1 can occur in models with steep (but not abruptly steep) APD restitution curves whenever a large change in cycle length occurs.

*Non-steep APD Restitution Curves*

Without diminishing the significance of steep APD restitution curves as a breakup mechanism, especially since their relevance has been shown in a number of experiments[49 187 188], it is important to recognize that there are a number of other mechanisms that also can cause spiral wave breakup. In fact, one of the most widely used ionic models for cardiac simulations, the Luo-Rudy-I[101], breaks up with its original parameter settings despite its flat APD restitution. The LR-I model has a much faster (and more realistic) sodium conductance than its predecessor, the BR model[83], resulting in a smaller $DI_{min}$, (25 vs. 46 ms) and a relatively high $APD_{min}$, producing a fairly flat APD restitution. Even when the calcium is speeded up by two (to decrease the maximum APD from about 360 ms to a more physiological value of about 230ms), the restitution curve is flat, with slope less than one (Figure 40). Because of its high excitability, small refractory period, and large wavelength, the model follows linear core trajectories with sharp turns that can block wave propagation as they turn due to the Doppler shift in the frequency, as shown in Figure 21 in the discussion of mechanism 4.

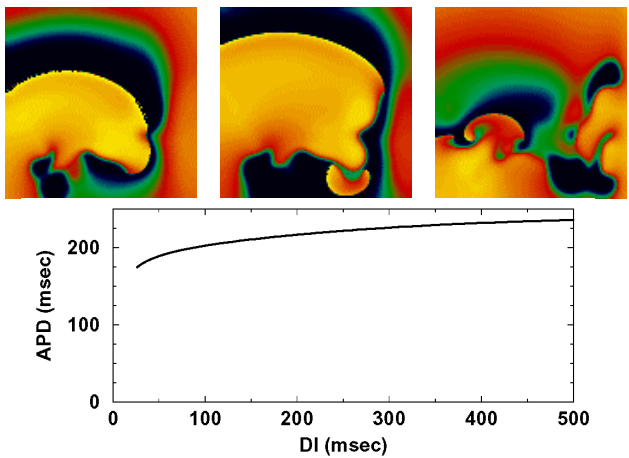

**Figure 40. Breakup in the LR-I model with speeded up calcium. The first two panels show the initial breakup as the spiral wave turns, blocks, and breaks. The third panel shows the voltage profile after**

**a few more wave breaks have occurred. Because of the large wavelength and the linear core, the breakup was transient, and eventually all waves left the area. The lower panel shows the APD restitution curve for the LR-I model with calcium speeded up by a factor of two, which has slope less than one everywhere. Similar breakup is obtained when the calcium dynamics are up to 2.8 times faster than with the original parameters[126].**

Other possibilities for spiral wave breakup in two dimensions that do not require steep APD restitutions are the cases of non-monotonically decreasing restitution curves, such as in the cases of biphasic APD (mechanism 5) and supernormal CV (mechanism 6) restitution curves. With a biphasic APD restitution curve, even when the magnitude of the slope never exceeds one, the prolongations of APD as the DI is decreased (over a certain range of DIs) can block subsequent waves, thereby causing breakup. Similarly, a supernormal CV restitution curve causes waves at short DIs to move more quickly than waves at long DIs. In this way, waves can stack and even block by collisions as spirals turn. Other mechanisms that do not require a steep or otherwise specifically-shaped APD or CV restitution curves to produce breakup are discussed below. Mechanisms not involving steep APD restitution may be relevant in the study of atrial fibrillation, where extensive experimental evidence has shown that the rate adaptation of atrial tissue diminishes or is eliminated after prolonged periods of fibrillation[189,190].

*Quantifying Restitution*

While details of restitution curves such as shape, steepness, and the $DI_{min}$ can be important in determining spiral wave stability and are a principal tool in classifying the various breakup mechanisms of this paper, we should mention two important issues regarding their measurements, particularly in experiments.

First, APD restitution can be measured either in isolated cells or in intact tissue. Although both are similar since they represent the same system, differences can arise due to cell coupling. For example, the maximum APD measured in an isolated cell and in a cell in a tissue preparation can vary substantially due to electrotonic effects, especially as a function of excitability. This effect has been seen in experiments, where the APD in isolated myocytes has been estimated to be 10 to 15 percent longer than in tissue[191], and can be observed readily in numerical simulations. In the same manner, the value of the minimum diastolic interval measured in an isolated cell depends on the strength and duration of the stimulus and is much shorter than the minimum diastolic interval for propagation obtained in tissue. Therefore, the shapes obtained from tissue and isolated cells may differ in the sense that for isolated cells, the $APD_{max}$ may be higher and the $DI_{min}$ smaller, allowing more variation in slope. As a result, in some extreme cases alternans can be seen in an isolated cell but not in tissue. Since arrhythmias form in tissue rather than in single cells and electrotonic effects certainly are present, we believe that the restitution obtained in tissue is the relevant curve for this analysis.

Second, the restitution relations in real cardiac tissue do not depend only on the previous diastolic interval, but in fact there is *memory* of previous activations[192]. That is, there is an adaptation of APD to variations in cycle lengths which is believed to be due, among other things, to changes in electrochemical gradients and permeability arising from differences in accumulation of intracellular sodium and calcium and extracellular potassium as well as the nonequilibrium values of the ionic currents at different rates of stimulation[53,129,193,194].

Because of memory, there are different protocols commonly used to measure APD restitutions, and indeed one difficulty in using APD restitution curves as a predictor of spiral wave instability is determining an effective and correct measurement protocol. The so-called *steady-state restitution* curve is obtained by pacing the cardiac preparation at a fixed cycle length for a large number of beats (until the preparation reaches a steady state), then the last DI and APD are recorded at that cycle length and provide a point on the APD restitution curve. Repeating the same protocol for various cycle lengths and measuring one point for each cycle length constructs the full curve. Another protocol measures the *S1-S2 restitution at a given cycle length*. Here, a train of more than 20 stimuli at a fixed cycle length (S1) is used to set the tissue memory to that particular cycle length. Once a steady state is reached, a single premature stimulus (S2) is applied. The APD of the premature stimulus and the preceding DI provide a point on the restitution curve. Repeating the S1 train at the same cycle length and varying the time of the premature stimulus S2 provides a full restitution curve. A variant of the S1-S2 protocol uses a third stimulus S3[132 192] (or more) designed to reach shorter DIs and APDs than are accessible only by using successively closer stimuli. While the steady state protocol yields one APD curve, the S1-S2 produces an entire family of restitution curves[53 195], all of which have various slopes and shapes, bringing into question which restitution is the relevant one. Furthermore, the steady-state and S1-S2 protocols are limited in that they can be used only in quiescent tissue and not during an arrhythmic episode, which calls into question the relevance of these restitution curves to fibrillation initiation. During an arrhythmia, the only option for measuring restitution is the so-called *dynamic* protocol[196 187], in which voltage traces from many sites are used to obtain DI, APD pairs, which are then plotted together. A similar approach can be taken in the absence of an arrhythmia by introducing stimuli at random intervals and measuring all DI, APD pairs.

While the steady state and S1-S2 restitutions produce relatively clean curves, the dynamic protocol typically yields a cloud of points rather than a clearly discernible restitution curve. The cloud is due in part to the fact that measuring DIs and APDs from an extended system during an arrhythmia can include loading effects from curved fronts as well as double potentials with short APDs resulting from recordings close to the core of reentrant waves[15 72].

Currently it is believed that the dynamic restitution is the most useful predictor of spiral wave behavior, since it is measured under the conditions of arrhythmia. One possible method of extracting useful information from the cloud of points obtained is to plot the restitution curve using a density plot. If the DIs and APDs are binned, the restitution curve can take on a third dimension by plotting the number of points falling into the given APD, DI bin as a height. In this way, although outlying points remain in the plot, preference is given to the APD, DI pairs that appear most often, and a structure can emerge as points measured close to reentrant waves and due to loading effects are reduced to the background. Figure 41 A shows a density plot example obtained from five seconds of simulated fibrillation (using parameter set 6, mechanism 4, shown in Figure 18). The data were collected from all sites throughout the full five seconds and produced a highly scattered distribution. To construct the density plot, the DIs and APDs were rounded to the nearest 0.5 ms, thereby creating DI and APD bins. The number of DI, APD pairs obtained for each bin value were counted and plotted for that DI and APD with a color to represent the frequency of occurrence for that DI, APD pair. Those bins visited between two and 500 times are shown as the black "dust" and include 17 percent of the total data. (Those bins visited only once, constituting 8 percent of the total data, are not shown.) Bins visited a larger number of times are color-coded as follows: 501 to 1000 times, red, containing 12 percent of the data; 1001 to 2000 times, green, with 6 percent of the data; 2001 to 2500 times, blue, with 6 percent of the data; and greater than 2500 times (up to a maximum of more than 38,000), yellow, consisting of 51 percent of the data. The values shown in the figure take on an increasingly discernible structure as less-visited DI, APD pairs are excluded. For comparison, the model's APD restitution curve obtained from plane waves is shown in black, and it can be seen that the density plot clusters around this curve more and more tightly at those DI, APD pairs visited most frequently. In particular, the small APDs produced by curved fronts and wave tips are eliminated, and the minimum APD and DI of the denser clusters match those of the restitution curve.

      While this example uses data from a simulation, rather than experimental data, we believe a similar approach may be useful in clearing up cloud-like dynamic restitution curves obtained experimentally. Figure 41B shows the normalized power spectrum corresponding to the data in A, which resembles those observed experimentally[16 197]. A dominant frequency of about 10 Hz is present (even in the absence of a dominant spiral wave), along with a secondary peak at half the value of the first (corresponding to the frequency of the 2:1 block region). These regions are visited preferentially in the restitution density plot, as can be seen by the clustering of DI, APD pairs visited most often (colored yellow) around the cycle lengths corresponding to the two frequency peaks (10 Hz ~ 100 ms, 5 Hz ~ 200 ms). Despite the width of the large peak and the presence of a wide range of other frequencies in smaller amounts, use of the density plot allows a restitution curve structure to emerge. We anticipate that a similar procedure should help to clarify a useful restitution curve from experimental data.

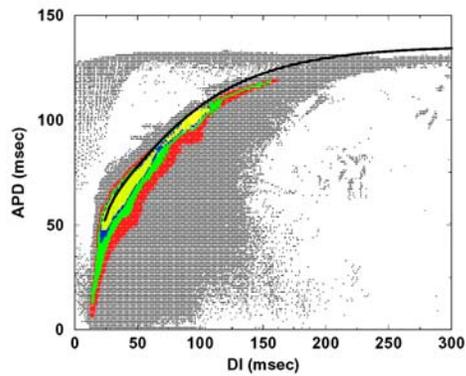

A

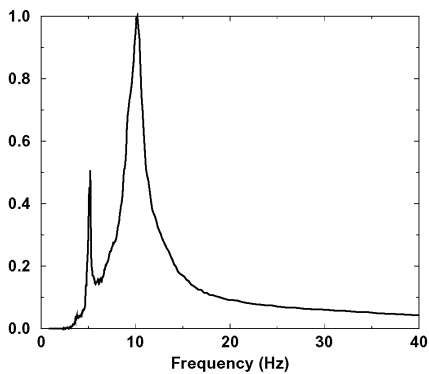

B

Figure 41. (A) Density plot of restitution. DI, APD pairs are gathered from all sites over a five-second simulation and grouped into 0.5 ms wide bins. A point is plotted for each DI, APD pair whose bin was visited at least twice during the simulation, representing 92 percent of the total data collected. The points appearing as a diffuse black "dust" were visited up to 500 times and represent 17 percent of the data. Bins visited more often were color-coded: up to 1000 visits, red; up to 2000 visits, green; up to 2500 visits, blue; more than 2500 visits, yellow. Those DI, APD pairs visited most often cluster closely around the model's APD restitution curve, shown in black. (B) Power spectrum of frequencies obtained during the same simulation. The two yellow regions in the density plot, with cycle lengths near 100 ms and 200 ms, correspond to the peaks in the spectrum of 10 Hz and 5 Hz, respectively.

The CV restitution curve[198] has not been studied as broadly as its APD counterpart, mostly because its measurement becomes complicated since fiber orientation and three-dimensional effects distort propagating fronts. Therefore, it still is largely unknown how CV restitution depends on previous activations. Although it has been shown[199] that maximum longitudinal and transverse CVs are not significantly affected by basic cycle length in normal tissue, it is unknown whether the shape of the curve may change. Furthermore, during ischemia cellular coupling and thus upstroke velocity change with cycle length[182 200], so that at least the maximum conduction velocity becomes a function of cycle length, and memory effects on CV may become important.

**The Role of Initial Conditions**

Initial conditions, in many cases, are also important determinants of spiral wave stability. A striking example of this is shown in the case of steep APD restitution curves that have a second region with slope less than one at short DIs, such as the 1962 Noble model[99] and the Fox et al.[115] models. When such regions exist, periodic pacing at short or long cycle lengths where the slope is less than one results in stable waves, while in between there is a region of cycle lengths where the slope is greater than one and thus alternans develops and conduction block may occur. Therefore, spiral waves with long periods falling in the region of slope less than one with long DIs are stable, while spirals with short periods falling in the other region of slope less than one can be either stable or unstable depending on initial conditions and tissue size, as shown in Figure 15 A and B. Where panel A shows a complex spatiotemporal pattern due to continuous wave breaks, panel B shows a stable spiral wave. In both cases, all parameters are the same, and the only difference was the initial conditions.

While the previous example illustrates the effect of initial conditions on the onset of breakup, it requires specific initial conditions; however, other mechanisms require only a simple spatial gradient in DI's to produce conduction blocks. An example is the rate-dependent bistability and hysteresis of APD (mechanism 3), in which the passing of a previous wave can lead to a gradient of recovery, causing parts of the tissue to be on the 1:1 branch while others are on the 2:1 branch and resulting in a wave break, as shown in Figure 17.

Discordant alternans-induced breakup (mechanism 2) also can depend on initial conditions, again to a lesser extent than in the first example. When a spiral wave is initiated in uniform tissue, the breakup typically takes many beats to occur, as the oscillations grow slowly due to conduction velocity restitution before developing conduction block. However, discordant alternans can be generated immediately following a gradient of recovery[48]; therefore, if the spiral wave is initiated in tissue with an existing gradient of refractoriness, the oscillations can grow faster and breakup can occur rather quickly.

The biphasic APD restitution curve (mechanism 5) also is sensitive to initial conditions, as seen in the cobweb diagram of Figure 23. Because of the presence of a region with negative slope in the middle of an otherwise positively sloped curve, different initial conditions for the same period can produce both stable and unstable solutions for a given frequency. Another example is negative tension (mechanism 8), discussed below, in which the difference in obtaining a stable scroll wave or turbulence depends on initial conditions. While a straight scroll wave will be stable, any small deviation from that will result in breakup.

**The Role of Thickness (2D vs. 3D)**

Even though restitution relations and initial conditions are very important in explaining 2D conduction blocks, structural effects manifested in 3D can have relevant consequences in the destabilization of three-dimensional reentrant waves (scroll waves) and their vortex filaments. In section VI three different mechanisms for spiral wave breakup in 3D are discussed. The first, negative tension (mechanism 8), develops as the excitability of the medium is decreased, which can occur physiologically when cardiac tissue is deprived of oxygen and becomes ischemic. In such cases any small perturbation applied to the vortex filament of a scroll wave grows. Vortices can then, after many rotations, elongate, curve and twist intramurally, until they collide with a boundary and break, producing a new vortex and scroll wave. This process can repeat itself, leading to multiple scroll waves and turbulence. Similarly, in twist-induced breakup (mechanism 9), vortices elongate and break at boundaries. However, unlike negative tension, this mechanism occurs in the high excitability limit, physiologically corresponding to healthy tissue, and the breakup requires many fewer rotations. It is important to note that while mechanism 8 can arise in completely homogenous tissue, mechanism 9 is a consequence of the natural anisotropic fiber rotation found in cardiac tissue. Likewise, mechanism 10 depends on fiber rotation and scroll breakup further depends on discrete cell effects, which grow more pronounced as the tissue becomes ischemic.

We note that while ischemia is a complicated condition that induces numerous physiological alterations in cardiac tissue, some of its most important effects are reductions in excitability and cell coupling. The reduced excitability may possibly shift the dynamical state of the tissue into the negative tension regime, thereby activating that breakup mechanism. In addition, the poor cell coupling may induce wave breakup in a manner like the coarse discretization mechanisms presented earlier. The twist instability mechanism, on the other hand, becomes less important in this parameter regime.

Although mechanism 10 in principle can occur in 2D tissue but is amplified in 3D by anisotropic fiber rotation, mechanisms 8 and 9 are purely three-dimensional and they need a minimum thickness to develop. Vortex lines are constrained to be locally normal to any boundary at which they attach (sealed boundaries that conserve current), and in very thin layers this requirement can prevent elongation in both mechanisms. Mechanism 9, in addition, has a thickness limit as a function of fiber rotation rate[40 126], below which the twist induced is not enough to elongate vortex filaments substantially.

It is important to note that whereas in 2D conduction blocks are a requirement for breakup, in 3D they are not necessary even when they occur, since breakup can be produced purely by topological changes in the vortex filaments, which have no counterpart in 2D. Once multiple vortices are created, the complex behavior and interaction between vortices can render the system more turbulent. For example, conduction blocks between waves can suddenly generate intramural vortex rings that can eventually contract and disappear (see Figure 42A), or they can expand and even fuse with other existing vortices (see Figure 42B) if they have the same phase. Inversely, in other cases, vortex rings can be generated not by conduction block but by elongation and vortex self-pinching as shown in Figure 42C. Even though equations of motion have been derived for scroll waves in various regimes and under certain conditions[38 45 171 201 202], the dynamics of their interactions remain largely unknown.

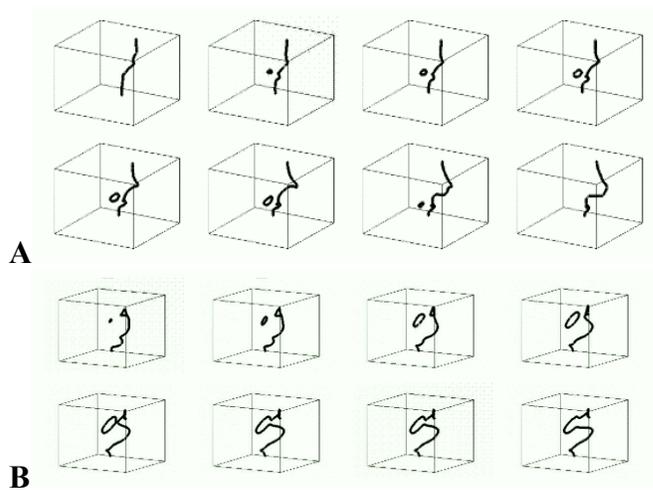

A

B

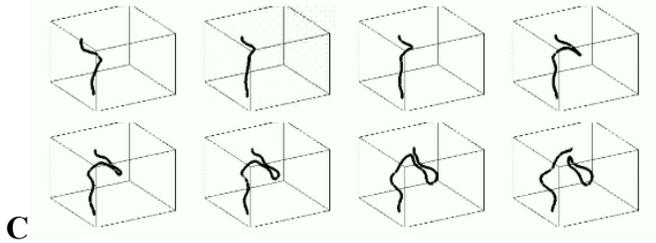
C

Figure 42. Vortex interactions. (A) A vortex ring is created and collapses again without touching the transmural filament. (B) A vortex ring is created and fuses with an existing transmural filament because its phase matches the filament phase at the fusion site. (C) A vortex ring is pinched off from a transumral filament as it elongates.

**The Role of Spiral Tip Trajectories**

As noted in Sections IV-VI the type of tip trajectory can be a crucial determinant of spiral wave stability for certain mechanisms. In mechanisms 1 and 2, any tip trajectory will produce breakup as long as the period of rotation lies in the region of slope less than one. On the other hand, mechanism 4 requires not only a relatively flat APD restitution curve and a small $DI_{min}$, but also a highly meandering or sharply turning tip trajectory, so that the waves emitted by the moving spiral tip will have a Doppler shift (as shown in Figure 19) and see shorter cycle lengths, resulting in conduction block (Figure 20 and Figure 21) whenever the shift lies below the period for propagation.

In the presence of periodic boundary conditions (mechanism 7), the type of spiral wave trajectory also is important. When the wavelength of a spiral is comparable to the length of the tissue with periodic boundaries, drift can be induced on the spiral wave by the interaction of self-generated incoming waves with the spiral tip, as illustrated in Figure 29. Spiral wave drift along the ventricles, which can be produced by periodicity or by other means[203], has been shown to produce fibrillation-like signals in the ECG [7,204]. Furthermore, breakup into multiple waves also can occur due to the periodic boundary effects as the movement of a hypermeandering spiral wave can cause complex gradients in repolarization, which can lead to conduction blocks as the spiral wave breaks, as shown in Figure 30.

The type of tip trajectory is also important in the 3D breakup mechanism of filament twist instability (mechanism 9), where sharp pivot turns in the trajectory are required. The rotational anisotropy of the medium causes the turn to take place at slightly different times across layers, producing a phase lag. Each time the pivot turn occurs, twist builds up in the filament, which the filament tries to eliminate through elongation (rather than by diffusion[75][176]). When the turn is sufficiently sharp, enough twist can accumulate to destabilize the filament, which elongates so much that it eventually hits the boundaries and breaks in two. This instability decreases and has little effect for tip trajectories with mild pivot turns and circular cores even at high degrees of rotational anisotropy[40][84][126].

**Limitations**

A number of limitations of the present study should be mentioned. First, this manuscript has presented a set of mechanisms that can occur in homogeneous tissue, while in reality, cardiac muscle contains a variety of intrinsic inhomogeneities. Nevertheless, this survey can serve as a foundation and framework for the analysis of mechanisms that can occur with or without heterogeneities. Many of the mechanisms operative in homogeneous tissue should continue to occur in heterogeneous tissue, but some may be enhanced or suppressed, while other new mechanisms certainly occur due to specific heterogeneities. For example, differences in cell dynamics have been found between the left and right ventricles, which may be important in arrhythmia development[10]. Likewise, different cell types exist along the thickness of the ventricular wall (epicardial, M, and endocardial cells), whose dynamics at large pacing cycles can vary greatly[205 206 207] and may destabilize waves under certain conditions[208 209]. Furthermore, natural dispersion in APD throughout the tissue may facilitate the formation of conduction blocks[210] or allow the development of sproing when the spiral period is shorter at one end of the vortex line, leading to breakup of scroll waves[75 176].

In addition, we do not describe effects of geometric structure that can contribute to arrhythmogenesis, such as curvature effects[141]; nonuniform fiber orientation producing drift[203]; abrupt changes in fiber direction at sites like the papillary muscle insertions in the ventricles[211], which provide anchoring sites for reentrant waves; and path structures for reentry, such as the pectinate muscles[73 212], the pulmonary veins[41], and the Purkinje system[213]. Furthermore, we have neglected memory effects[192 194 214 215], triggered activity including early afterdepolarizations[216], and localized heterogeneities such as scars[217 218 219] and ischemic regions[220], which have been shown to promote reentry and breakup. However, the analysis and effects of these and other types of native or illness-induced homogeneities on the breakup of scroll waves is beyond the scope of this paper.

## Conclusions

This paper has shown and classified a number of different mechanisms for spiral wave breakup in 2D and 3D tissues on the basis of the conditions necessary for their occurrence in cardiac tissue. APD and CV restitution curves can produce various types of breakup due to their steepness or due to their specific shapes, although other mechanisms exist even when APD and CV restitution curves are relatively flat. In some cases, initial conditions can play a crucial role in determining whether or not breakup will develop. The trajectory of the spiral wave tip also can be important for some breakup mechanisms to occur, as repolarization differences due to tip meander can lead to conduction block, and differences in pivot turns in 3D tissue with rotational anisotropy can destabilize transmural filaments. Finally, although spirals can break by a number of mechanisms in 2D, fully 3D tissue is needed for certain mechanisms to develop. All of the mechanisms discussed in this paper can operate in fully homogeneous and isotropic tissue, except for the twist- and discreteness-induced mechanisms, which require rotational anisotropy.

Which breakup mechanism or mechanisms are responsible for the transition from tachycardia to fibrillation in the human heart is still unclear. Experimental evidence in various animal models can be interpreted to support several mechanisms, including steep APD restitution[49,196,213] and tip trajectories[10,213], as well as one[7,8,9,10] or many spirals[10,11,12,13,14,15,16]. However, our intent here is to give a survey of possible arrhythmogenic mechanisms, any number of which may be present in a given patient or preparation. Numerous questions about how these mechanisms may operate deserve further study, including whether one or several mechanisms tend to underlie fibrillation, how fibrillation caused by various mechanisms may be differentiated, which mechanisms may be associated with other types of heart disease like ischemia and heart failure, and how to provide effective treatment to prevent fibrillation.

The last question in particular takes on more importance in light of the fact that multiple mechanisms may be capable of inducing fibrillation; therefore, pharmacotherapy studies designed to develop new drugs that target one breakup mechanism need to ensure that they do not activate, facilitate, or enhance other mechanisms that may exist. For example, flattening restitution can prevent mechanisms 1 and 2, but we can speculate that breakup may still persist or resume if by doing so a small refractory period and large wavelengths are produced, thus facilitating mechanism 4. Similarly, lowering the sodium conductance may suppress breakup by mechanisms 7 and 9 as the tip trajectory is changed, but could in principle enhance 8 and 10 if the sodium change is extreme. It appears that the best way to suppress all of the mechanisms discussed here is conversion of the tip trajectory to a large circular core, a conclusion similarly reached by Samie et al.[221][222] and Efimov et al.[62] As mentioned in Section II, the tip trajectory can be made circular and increased in size by enlarging the excitability gap or by decreasing the sodium or calcium conductances. However, decreasing the sodium conductance, as suggested in Ref. 62, may have arrhythmic effects due to mechanims 8 and 10 if the excitability is too low. Therefore, reducing the calcium conductance, as in Ref. 221, may be a good approach (with or without a concomitant small decrease in the sodium conductance), but the effects of low calcium on contraction need to be addressed.

To validate or dispute such approaches to arrhythmia prevention, additional studies of the dynamics of these and other breakup mechanisms in realistic cardiac geometries including variations in cell types is needed. In some cases, the parameter regimes in which the breakup occurs may be widened, making the breakup easier to occur. In other cases, structural effects may vary the roles of a breakup mechanism and conceivably new breakup mechanisms will emerge.

## Acknowledgements

We acknowledge support from the US National Science Foundation, the Mike and Louise Stein Philanthropic Fund, the Rosalyn S. Yalow Foundation for Medical Research, the Medtronic Foundation, the Guidant Foundation, the CR Bard Foundation, and J & J Biosense. This work also was supported by a grant of supercomputer time from the National Center for Supercomputing Alliance (Pittsburgh Supercomputing Center). We thank A. Karma and R.A. Gray for valuable discussions.

## Appendix

In this appendix, we describe the simplified ionic model introduced in Section III and used through this manuscript (unless specifically noted otherwise) to simulate cardiac electrical dynamics. The dynamics of the transmembrane potential $V_m$ is governed by the cable equation $\partial_t V(\mathbf{x},t) = \nabla \cdot (D \nabla V) - ( I_{fi}(V,v) + I_{so}(V) + I_{si}(V,w) ) / C_m$, where the ionic currents determine cell dynamics, $C_m$ represents the membrane capacitance (set to 1 µF/cm$^2$ ), and the diffusion tensor $D$ defines tissue structure and anisotropy described in Sec. VI. Except for the last two mechanisms, all simulations are isotropic, so that $D$ is a diagonal matrix whose off-diagonal elements are 0 and whose diagonal elements are 0.001 cm$^2$/ms. This is the most commonly used value in the literature, but it assumes a surface to volume ratio of 5000/cm, corresponding to a fairly small cell radius of around 4 µm[94]. The two gate-variables of the model, $v$ and $w$, follow first order equations in time:

$\partial_t v(\mathbf{x},t) = (1-p)(1-v)/\tau_v^-(V) - pv/\tau_v^+(V)$

$\partial_t w(\mathbf{x},t) = (1-p)(1-w)/\tau_w^-(V) - pw/\tau_w^+(V)$,

where $\tau_v^-(V) = (1-q)\,\tau_{v1}^-(V) + q\,\tau_{v2}^-(V)$ and

$$p = \begin{cases} 1 & \text{if } V \geq V_c \\ 0 & \text{if } V < V_c \end{cases} \qquad q = \begin{cases} 1 & \text{if } V \geq V_v \\ 0 & \text{if } V < V_v \end{cases}$$

The two gate variables and the maximum upstroke value in the transmembrane potential vary from 0 to 1. Therfore, when comparing with other models or experiments, $V_m$ needs to be rescaled, as shown in Figure 3. The three currents are given by the following equations:

$I_{fi}(V,v) = -vp(V-V_c)*(1-V)/\tau_d$,

$I_{so}(V) = V(1-p)/\tau_o + p/\tau_r$, and

$I_{si}(V,w) = -w(1 + \tanh(k(V-V_c^{si})))/(2\tau_{si})$..

Although we refer to this model as the 3V-SIM because it consists of 3 variables, there are two variations we commonly use, one with two variables (obtained by eliminating gate variable $w$ and the $I_{si}$ current), which produces simple flat APD restitution models as in set 2 and 7. The second variation, as mentioned in Section III, replaces the steady-state function $d_\infty(V)$ given by $(1+ \tanh(k(V-V_c^{si})))$ with a gate variable $d$, which is used to reproduce more accurately the AP shapes of other models[43,66], as shown in Figure 3A and B.

In section IV an extra current is added in order to obtain a biphasic APD restitution curve. This current is used in conjunction with parameter set 8 and an extra variable denoted here as $y$ and obeying the following equations:

$I_{biph} = pp(y)*0.355/\tau_r$,

where $pp(y)=1$ for $y>0.1$ and $pp(y)=0$ otherwise, and

$\partial_t y(\mathbf{x},t) = (1-p)(1-y)/400 - py/25$.

The 13 model parameters used in the various examples are given in the following table. Further description of model parameters and their functions can be found in Ref. 85.

| Parameter | Set 1 | Set 2 | Set 3 | Set 4 | Set 5 | Set 6 | Set 7 | Set 8 | Set 9 | Set 10 |
|---|---|---|---|---|---|---|---|---|---|---|
| $\tau_v^+$ | 3.33 | 10 | 3.33 | 3.33 | 3.33 | 3.33 | 10 | 13.03 | 3.33 | 10 |
| $\tau_{v1}^-$ | 19.6 | 10 | 19.6 | 15.6 | 12 | 9 | 7 | 19.6 | 15 | 8 |
| $\tau_{v2}^-$ | 1000 | 10 | 1250 | 5 | 2 | 8 | 7 | 1250 | 2 | 10 |
| $\tau_w^+$ | 667 | -- | 870 | 350 | 1000 | 250 | -- | 800 | 670 | 1000 |
| $\tau_w^-$ | 11 | -- | 41 | 80 | 100 | 30 | -- | 40 | 61 | 65 |
| $\tau_d$ | 0.25 | 0.25 | 0.25 | 0.407 | 0.362 | 0.39 | 0.25 | 0.45 | 0.25 | 0.115 |
| $\tau_0$ | 8.3 | 10 | 12.5 | 9 | 5 | 9 | 12 | 12.5 | 12.5 | 12.5 |
| $\tau_r$ | 50 | 190 | 33.33 | 34 | 33.33 | 33.33 | 100 | 33.25 | 29 | 80 |
| $\tau_{si}$ | 45 | -- | 29 | 26.5 | 29 | 29 | -- | 29 | 29 | 77 |
| $K$ | 10 | -- | 10 | 15 | 15 | 15 | -- | 10 | 10 | 10 |
| $V_c^{si}$ | 0.85 | -- | 0.85 | 0.45 | 0.70 | 0.50 | -- | 0.85 | 0.45 | 0.85 |
| $V_c$ | 0.13 | 0.13 | 0.13 | 0.15 | 0.13 | 0.13 | 0.13 | 0.13 | 0.13 | 0.13 |
| $V_v$ | 0.055 | -- | 0.04 | 0.04 | 0.04 | 0.04 | -- | 0.04 | 0.05 | 0.025 |

**Table 1. Parameter values used for the 3V-SIM to produce the simulations included in this study.**

All simulations were performed using an exact integration scheme for the gate variables and a Crank-Nicolson scheme for the voltage as described in Ref. 40, thus allowing larger time steps compared to the forward Euler scheme. However, it is important to mention that even though all simulations (except for mechanism 10) are resolved, in some circumstances, such as in the low excitability limit, the tip trajectory regimes as a function of parameters may vary slightly as the time constants for integration are decreased further. That is, a small percentage shift in $\tau_d$ (excitability) may be needed to obtain the same results as the time constants for integration are varied. However, the results are valid at the resolutions given throughout this paper.

For the periodic boundary cases, where a cylinder is approximated by making the horizontal direction of a 2D domain periodic, the solution of the tridiagonal matrix in the Crank-Nicolson scheme acquires two additional entries at the corners and becomes cyclic tridiagonal. We solve this matrix by adding corrections to the Thomas algorithm (used in the zero flux boundary conditions) based on the perturbation correction Sherman-Morrison method. Therefore, a tridiagonal matrix of the form

$A_i^- V_{i-1}^1 + A_i^0 V_i^1 + A_i^+ V_{i+1}^1 = F_i(V_i^{t+\Delta t/2})$

$A_i^- V_{i-1}^2 + A_i^0 V_i^2 + A_i^+ V_{i+1}^2 = 0$

$A_i^- V_{i-1}^3 + A_i^0 V_i^3 + A_i^+ V_{i+1}^3 = 0,$

where $V_0^1 = V_n^1 = V_n^2 = V_0^3 = 0$ and $V_1^2 = V_n^3 = 1$, has a solution of the form

$V_i^{t+\Delta t} = V_i^1 + (V_n^1(1 - V_1^3) + V_1^1 V_n^3) V_i^2 + V_1^1(1 - V_n^2) + V_n^1 V_1^2) V_i^{3)}/[(1 - V_n^2)(1 - V_1^3) - V_1^2 V_n^3].$

For the 1D reductions of target waves in the description of mechanism 2 and the 2D reductions of scroll waves in the description of mechanism 8, where radial coordinates are used, the value at r=0 is obtained by approximating the solution of the transmembrane potential to a polynomial; therefore, the radial Laplacian at that point can be approximated by $4(V_{1,j} - V_{0,j})/\Delta x^2$.

# References

[1] American Heart Association, 2001 Heart and Stroke Statistical Update (American Heart Association, Dallas, 2000).

[2] D.P. Zipes and H.J.J. Wellens, "Sudden Cardiac Death," Circ. **98**, 2334-2351 (1998).

[3] D. Scherf and A. Schott, *Extrasystoles and allied arrhythmias* (Grune and Stratton, New York, 1953).

[4] T. Sano and T. Sawanobori, "Mechanism initiating ventricular fibrillation demonstrated in cultured ventricular muscle tissue," Circ. Res. **26**, 201 (1962).

[5] M. Haïssaguerre, P. Jaïs, D.C. Shah, A. Takahashi, M. Hocini, G. Quinious, S. Garrigue, A. Le Mouroux, P. Le Métayer, and J. Clémenty, "Spontaneous initiation of atrial fibrillation by ectopic beats originating in the pulmonary veins," N. Engl. J. Med. **338**, 659-666 (1998).

[6] J.L. Lin, L.P. Lai, Y.Z. Tseng, W.P. Lien, and S.K.S. Huang, "Global Distribution of Atrial Ectopic Foci Triggering Recurrence of Atrial Tachyarrhythmia After Electrical Cardioversion of Long-Standing Atrial Fibrillation," Journal of the American College of Cardiology **37**, 904-910 (2001).

[7] R.A. Gray, J. Jalife, A. Panfilov, W.T. Baxter, C. Cabo, J.M. Davidenko, and A.M. Pertsov, "Nonstationary Vortexlike Reentrant Activity as a Mechanism of Polymorphic Ventricular Tachycardia in the Isolated Rabbit Heart," Circ. **91,** 2454-2469 (1995).

[8] J.M. Davidenko, A.M. Pertsov, R. Salomonsz, W.T. Baxter and J. Jalife, "Stationary and Drifting Spiral Waves of Excitation in Isolated Cardiac Muscle," Nature **355**, 349-351 (1992).

[9] M.J. Janse, F.J.G. Wilms-Schopman and R. Coronel, "Ventricular fibrillation is not always due to multiple wavelet re-entry," J. Cardiovasc. Electrophysiol. **6**, 512-521 (1995).

[10] F.H. Samie, O. Berenfeld, J. Anumonwo, S.F. Mironov, S. Udassi, J. Beaumont, S. Taffet, A.M. Pertsov, and J. Jalife, "Rectification of the Background Potassium Current: A Determinant of Rotor Dynamics in Ventricular Fibrillation," Circulation Research **89**, 1216-1223 (2001).

[11] P.V. Bayly, B.H. KenKnight, J.M. Rogers, E.E. Johnson, R.E. Ideker, W.M. Smith, "Spatial organization, predictability, and determinism in ventricular fibrillation," Chaos **8**, 103-115 (1998).

[12] Y.H. Kim, M. Yashima, T.J. Wu, R. Doshi, P.S. Chen, and H. Karagueuzian, "Mechanism of Procainamide-Induced Prevention of Spontaneous Wave Break During Ventricular Fibrillation: Insight Into the Maintenance of Fibrillation Wave Fronts," Circulation **100**, 666-674 (1999).

[13] F. Witkowski, L.J. Leon, P.A. Penkoske, W.R. Giles, M.L. Spano, W. L. Ditto, and A.T. Winfree, "Spatiotemporal evolution of ventricular fibrillation," Nature **392**, 78-82 (1998).

[14] G.P. Walcott, G.N. Kay, V.J. Plumb, W.M. Smith, J.M. Rogers, A.E. Epstein, and R.E. Ideker, "Endocardial Wave Front Organization During Ventricular Fibrillation in Humans," J. Am. Coll. Cardiol. **39**, 109-115 (2002).

[15] I.R. Efimov, V. Sidorov, Y. Cheng, and B. Wollenzier, "Evidence of Three-Dimensional Scroll Waves with Ribbon-Shaped Filament as a Mechanism of Ventricular

[9] M.J. Janse, F.J.G. Wilms-Schopman and R. Coronel, "Ventricular fibrillation is not always due to multiple wavelet re-entry," J. Cardiovasc. Electrophysiol. **6**, 512-521 (1995).

[10] F.H. Samie, O. Berenfeld, J. Anumonwo, S.F. Mironov, S. Udassi, J. Beaumont, S. Taffet, A.M. Pertsov, and J. Jalife, "Rectification of the Background Potassium Current: A Determinant of Rotor Dynamics in Ventricular Fibrillation," Circulation Research **89**, 1216-1223 (2001).

[11] P.V. Bayly, B.H. KenKnight, J.M. Rogers, E.E. Johnson, R.E. Ideker, W.M. Smith, "Spatial organization, predictability, and determinism in ventricular fibrillation," Chaos **8**, 103-115 (1998).

[12] Y.H. Kim, M. Yashima, T.J. Wu, R. Doshi, P.S. Chen, and H. Karagueuzian, "Mechanism of Procainamide-Induced Prevention of Spontaneous Wave Break During Ventricular Fibrillation: Insight Into the Maintenance of Fibrillation Wave Fronts," Circulation **100**, 666-674 (1999).

[13] F. Witkowski, L.J. Leon, P.A. Penkoske, W.R. Giles, M.L. Spano, W. L. Ditto, and A.T. Winfree, "Spatiotemporal evolution of ventricular fibrillation," Nature **392**, 78-82 (1998).

[14] G.P. Walcott, G.N. Kay, V.J. Plumb, W.M. Smith, J.M. Rogers, A.E. Epstein, and R.E. Ideker, "Endocardial Wave Front Organization During Ventricular Fibrillation in Humans," J. Am. Coll. Cardiol. **39**, 109-115 (2002).

[15] I.R. Efimov, V. Sidorov, Y. Cheng, and B. Wollenzier, "Evidence of Three-Dimensional Scroll Waves with Ribbon-Shaped Filament as a Mechanism of Ventricular

Tachycardia in the Isolated Rabbit Heart," J. Cardiovasc. Electrophysiol. **10**, 1452-1462 (1999).

[16] R.A. Gray, A.M. Pertsov and J. Jalife, "Spatial and temporal organization during cardiac fibrillation," Nature **392**, 75-78 (1998).

[17] Cardiac Arrhythmia Suppression Trial (CAST) Investigators, "Preliminary report: effect of encainide and flecainide on mortality in a randomized trial of arrhythmia suppression after myocardial infarction," N. Engl. J. Med. **321**, 406-412 (1989).

[18] Cardiac Arrhythmia Suppression Trial (CAST) II Investigators, "Effect of the antiarrhythmic agent moricizine on survival after myocardial infarction," N. Engl. J. Med. **327**, 227-233 (1992).

[19] A. Waldo, A.J. Camm, H. deRuyter, P.L. Friedman, D.J. MacNeil, J.F. Pauls, B. Pitt, C.M. Pratt, P.J. Schwartz, and E.P. Veltri, "Effect of d-sotalol on mortality in patients with left ventricular dysfunction after recent and remote myocardial infarction," Lancet **348**, 7-12 (1996).

[20] B.C. Hill, A.J. Hunt, and K.R. Courtney, "Reentrant Tachycardia in a Thin Layer of Ventricular Subepicardium: Effects of d-Sotalol and Lidocaine," Journal of Cardiovascular Pharmacology **16**, 871-880 (1990).

[21] R.E. Ideker, T.N. Chattipakorn, R.A. Gray, "Defibrillation mechanisms: the parable of the blind men and the elephant," J Cardiovasc Electrophysiol. **11**, 1008-13 (2000).

[22] R.A. Gray and J. Jalife, "Ventricular fibrillation and atrial fibrillation are two different beasts," Chaos **8**, 65-78 (1998).

[23] *Optical mapping of Cardiac Excitation and Arrhythmias,* edited by D.S. Rosenbaum and J. Jalife (Futura Publishing Co., New York, 2001).

[24] W.T. Baxter, S.F. Mironov, A.V. Zaitsev, J. Jalife, and A.M. Pertsov, "Visualizing excitation waves inside cardiac muscle using transillumination," Biophys. J. **80,** 516-530 (2001).

[25] D.A. Hooks, I.J. LeGrice, J.D. Harvey, and B.H. Smaill, "Intramural Multisite Recording of Transmembrane Potential in the Heart," Biophys. J. **81**, 2671-2680 (2001).

[26] J.M. Rogers, S.B. Melnick, and J. Huang, "Fiberglass needle electrodes for transmural cardiac mapping," submitted.

[27] N. Winer and A. Rosenblueth, "The Mathematical Formulation of the Problem of Conduction of Impulses in a Network of Connected Excitable Elements, Specifically in Cardiac Muscle," Arch. Inst. Cardiol. Mex. **16**, 205-265 (1946).

[28] G.R. Mines, "On dynamic equilibrium in the heart," J. Physiol. **46**, 349-383 (1913).

[29] W.E. Garrey, "The nature of fibrillatory contraction of the heart, its relation to tissue mass and form," Am. J. Physiol. **33**, 397-414 (1914).

[30] J.A. McWilliam, "Fibrillar contraction of the heart," J. Physiol. **8**, 296-310, (1887).

[31] G.K. Moe, "On the multiple avelet hypothesis of atrial fibrillation," Arch. Int. Pharmacodyn Ther. **140**, 183-188 (1962).

[32] G.K. Moe, W.C. Rheinboldt and J.A. Abildskov, "A computer model of atrial fibrillation," Am. Heart J. **67**, 200-220 (1964).

[33] M. Courtemanche, "Complex spiral wave dynamics in a spatially distributed ionic model of cardiac electrical activity," Chaos **6**, 579-600 (1996).

[34] A. Karma, "Electrical alternans and spiral wave breakup in cardiac tissue," Chaos **4**, 461-472 (1994).

[35] M. Bär and M. Eiswirth, "Turbulence due to spiral breakup in a continuous excitable medium," Phys. Rev. E **48**, R1635-R1637 (1993).

[36] A.F.M. Maree and A.V. Panfilov, "Spiral Breakup in Excitable Tissue due to Lateral Instability," Phys. Rev. Lett **78**, 1819-1822 (1997).

[37] A. Giaquinta, S. Boccaletti, and F.T. Arecchi, "Superexcitability induced spiral breakup in excitable systems," Int. Journ. of Bifurcation and Chaos **6**, 1753-1759 (1996).

[38] V.N. Biktashev, A.V. Holden, and H. Zhang, "Tension of organizing filaments of scoll waves," Phil. Trans. R. Soc. Lond. A **347**, 611-630 (1994).

[39] A.V. Panfilov and J.P. Keener, "Re-entry in three-dimensional Fitzhugh-Nagumo medium with rotational anisotropy," Physica D **84**, 545-552 (1995).

[40] F.H. Fenton and A. Karma, "Vortex dynamics in three-dimensional continuous myocardium with fiber rotation: Filament instability and fibrillation," Chaos **8**, 20-47 (1998).

[41] E.J. Vigmond, R. Ruckdeschel, and N. Trayanova, "Reentry in a Morphologically Realistic Atrial Model," J. Cardiovasc. Electrophysiol. **12**, 1046-1054 (2001).

[42] N. Virag, J.M. Vesin, and L. Kappenberger, "A computer model of cardiac electrical activity for the simulation of arrhythmias, Pacing Clinical Electrophysiol." **21** (Part II), 2366-2371 (1998).

[43] E.M. Cherry, S.J. Evans, H.Hastings and F.H. Fenton, in preparation. (add the naspe abstract in May)

[44] M. Courtemanche, L. Glass, and J.P. Keener, "Instabilities of a propagating pulse in a ring of excitable media," Phys. Rev. Lett. **70**, 2182-2185 (1993).

[45] J.J. Tyson and J.P. Keener, "Singular perturbation theory of traveling waves in an excitable medium," Physica D **32**, 327-361 (1988).

[46] H. Ito and L. Glass, "Theory of reentrant excitation in a ring of cardiac tissue," Physica D **56B**, 84-106 (1992).

[47] A. Karma, H. Levine, and X. Zou, "Theory of pulse instabilities in electrophysiological models of excitable tissues," Physica D **73**, 113-127 (1994).

[48] M. Watanabe, F. Fenton, S. Evans, H. Hastings, and A. Karma, "Mechanisms for Discordant Alternans," J. Cardiovasc. Electrophysiol. **12**, 196-206 (2001).

[49] A. Garfinkel, Y.H. Kim, O. Voroshilovsky, Z. Qu, J.R. Kil, M.H. Lee, H.S. Karagueuzian, J.N. Weiss, and P.S. Chen, "Preventing ventricular fibrillation by flattening cardiac restitution," Proceedings of the National Academy of Sciences **97,** 6061-6066 (2000).

[50] C. Delgado, B. Steinhaus, M. Delmar, D.R. Chialvo, and J. Jalife, "Directional differences in excitabilities and margin of safety for propagation in sheep ventricular epicardial muscle," Circ. Res. **67**, 97-110 (1990).

[51] T. Watanabe, P.M. Rautaharju, and T.F. McDonald, "Ventricular Action Potentials, Ventricular Extracellular Potential, and the ECG of Guinea Pig," Circ. Res. **57**, 362-373 (1995).
107

[52] C. Antzelevitch, S. Sicouri, S.H. Litovsky, A. Lukas, S.C. Krishnan, J.M. Di Diego, G.A. Gintant, and D.W. Liu, "Heterogeneity within the Ventricular Wall," Circ. Res. **69**, 1427-1449 (1991).

[53] M.R. Boyett and B.R. Jewell, "A study of the factors responsible for rate-dependent shortening of the action potential in mammalian ventricular muscle," J. Physiol. (Lond) **285**, 359-380 (1978).

[54] J.B. Nolasco and R.W. Dahlen, "A graphic method for the study of alternation in cardiac action potentials," J. App. Physiol. **25**, 191-196 (1968).

[55] W. Quan and Y. Rudy, "Unidirectional block and reentry of cardiac excitation: a model study," Circ. Res. **66**, 367-382 (1990).

[56] D.R. Chialvo, D.C. Michaels, and J. Jalife, "Supernormal excitability as a mechanism of chaotic dynamics of activation in cardiac Purkinje fibers," Circ. Res. **66**, 525-545 (1990).

[57] V.I. Krinsky and I.R. Efimov, "Vortices with linear cores in mathematical models of excitable media," Physica A **188**, 55-60 (1992).

[58] V.I. Krinsky, I.R. Efimov and J. Jalife "Vortices with linear cores in excitable media," Proc. Roy. Soc. London A **437**, 645-655 (1992).

[59] D.T. Kim, Y. Kwan, J.J. Lee, T. Ikeda, T. Uchida, K. Kamjoo, Y.H. Kim, J.J.C. Ong, C.A. Athill, T.J. Wu, L. Czer, H.S. Karagueuzian, and P.S. Chen, "Patterns of spiral tip motion in cardiac tissue," Chaos **8**, 137-148 (1998).

[60] A.T. Winfree, "Varieties of spiral wave behavior in excitable media," Chaos **1**, 303-333 (1991).

[61] D. Barkley, "Euclidean symmetry and the dynamics of rotating spiral waves," Phys. Rev. Lett. **72**, 164-167 (1994).

[62] I.R. Efimov, V.I. Krinsky, and J. Jalife, "Dynamics of rotating vortices in the Beeler-Reuter model of cardiac tissue," Chaos, Solitons and Fractals **5**, 513-526 (1995).

[63] A.T. Winfree, "Heart muscle as a reaction-diffusion medium: the roles of electric potential diffusion, activation front curvature and anisotropy," Journal of Bifurcation and Chaos **7**, 487-526 (1997).

[64] A.S. Mikhailov and V.S. Zykov, "Kinematical theory of spiral waves in excitable media: comparison with numerical simulations," Phys. D. **52,** 379-397 (1991).

[65] J. Beaumont, N. Davidenko, J.M. Davidenko, and J. Jalife, "Spiral Waves in Two-Dimensional Models of Ventricular Muscle: Formation of a Stationary Core," Biophysical Journal **75**, 1-14 (1998).

[66] F. Fenton, "Effects of restitution and activation shape on the dynamics of cell models," in preparation.

[67] V. Hakim, A. Karma, "Spiral wave in excitable media: the large core limit," Phys. Rev. Lett. **79,** 665-668 (1997).

[68] V. Hakim and A. Karma, "Theory of spiral wave dynamics in weakly excitable media: Asymptotic reduction to a kinematic model and applications," Phys. Rev. E **60**, 5073-5105 (1999).

[69] D. Barkley, "Linear Stability Analysis of Rotating Spiral Waves in Excitable Media," Phys. Rev. Lett. **68**, 2090-2093 (1992).

[70] G. Li, Q. Ouyang, V.V. Petrov, and H.L. Swinney, "Transition from Simple Rotating Chemical Spirals to Meandering and Traveling Spirals," Phys. Rev. Lett. **77**, 2105-2108 (1996).

[71] Z. Qu, J. Kil, F. Xie, A. Garfinkel, and J.N. Weiss, "Scroll wave dynamics in a three-dimensional cardiac tissue model: Roles of restitution, thickness, and fiber rotation," Biophysical Journal **78**, 2761-2775 (2000).

[72] Y.H. Kim, A. Garfinkel, T. Ikeda, T.J. Wu, C.A. Athill, J.N. Weiss, H.S. Karagueuzian, and P.S. Chen, "Spatiotemporal complexity of ventricular fibrillation revealed by tissue mass reduction in isolated swine right ventricle," J. Clin. Invest. **100**, 2486-2500 (1997).

[73] T.J. Wu, M. Yashima, F. Xie, C.A. Athill, Y.H. Kim, M.C. Fishbein, Z. Qu, A. Garfinkel, J.N. Weiss, H.S. Karagueuzian, P.S. Chen, "Role of pectinate muscle bundles in the generation and maintenance of intra-atrial reentry," Circ. Res. **83**, 448-462 (1998).

[74] A. Garfinkel, P.S. Chen, D.O. Walter, H.S. Karagueuzian, B. Kogan, S.J. Evans, M. Karpoukhin, C. Hwang, T. Uchida, M Gotoh, O. Nwasokwa, and P. Sager, "Quasiperiodicity and chaos in cardiac fibrillation," J. Clin. Invest. **99**, 305-314 (1997).

[75] C. Henze, E. Lugosi, and A.T. Winfree, "Helical organizing centers in excitable media," Can. J. Phys. **68**, 683-710 (1989).

[76] H. Zhang and A.V. Holden, "Chaotic meander of spiral waves in the Fitzhugh-Nagumo system," Chaos, Solitons and Fractals **5**, 661-670 (1995).

[77] D. Barkley, M. Kness, and L.S. Tuckerman, "Spiral wave dynamics in a simple model of excitable media: the transition from simple to compound rotation," Phys. Rev. A. **42**, 2489-2492 (1990).

[78] W. Jahnke, W.E. Skaggs, and A.T. Winfree, "Chemical vortex dynamics in the Belousov-Zhabotinsky reaction and in the two-variable Oregonator model," J. Phys. Chem **93**, 740-749 (1989).

[79] A.N. Iyer and R.Gray, "An experimentalist's approach to accurate localization of Phase Singularities during Reentry," Annals of Biomedical Eng. **29**, 47-59 (2001).

[80] R.A. Gray, private communication (2002).

[81] J. Beaumont, N. Davidenko, A. Goodwin, J.M. Davidenko, and J. Jalife, "Dynamics of Cardiac Excitation During Vortex-like Reentry," in preparation.

[82] R. FitzHugh, "Impulses and Physiological States in Theoretical Models of Nerve Membrane," Biophys. J. **1**, 445-466 (1961).

[83] G.W. Beeler and H. Reuter, "Reconstruction of the action potential of ventricular myocardial fibres," J. Physiol. **268**, 177-210 (1977).

[84] F. Fenton and A. Karma "Fiber-Rotation-Induced vortex turbulence in thick myocardium," Physical Review Letters **81**, 481-484 (1998).

[85] F. Fenton, "Theoretical Investigation of Spiral and Scroll Wave Instabilities Underlying Cardiac Fibrillation," Ph.D. thesis, Northeastern University, Boston, MA 02115 (1999).

[86] I. Banville and R. Gray "Effect of action potential duration and conduction velocity restitution and their spatial dispersion on alternans and the stability of arrhythmias," submitted.

[87] M. Courtemanche, R.J. Ramirez, and S. Nattel, "Ionic mechanisms underlying human atrial action potential properties: insights form a mathematical model," Am. J. Phys. **275**, H301-H321 (1998).

[88] M. Courtemanche and A.T. Winfree, "Re-entrant rotating waves in a Beeler-Reuter based model of two-dimensional cardiac electrical activity," International Journal of Bifurcation and Chaos **1**, 431-444 (1991).

[89] M.G. Fishler and N.V. Thakor, "A massively parallel computer model of propagation through a two-dimensional cardiac syncytium," Pacing Clin. Electrophysiol. **14**, 1694-1699 (1991).

[90] A.V. Holden and A.V. Panfilov, "Spatiotemporal chaos in a model of cardiac electrical activity," International Journal of Bifurcation and Chaos **1**, 219-225 (1991).

[91] J.B. Nolasco and R.W. Dahlen, "A graphic method for the study of alternation of cardiac action potentials," Journal of Applied Physiology **25**, 191-196 (1968).

[92] M.R. Guevara, G. Ward, A. Shrier, and L. Glass, "Electrical alternans and period doubling bifurcations," Computers in Cardiology, 167-170 (1984).

[93] A. Karma, "Spiral Breakup in Model Equations of Action Potential Propagation in Cardiac Tissue," Phys. Rev. Lett. **71**, 1103-1106 (1993).

[94] F.H. Fenton, E.M. Cherry, H.M. Hastings, and S.J. Evans, "Real-time simulations of excitable media: JAVA as a scientific language and as a wrapper for C and FORTRAN programs," BioSystems **64**, 73-96 (2002), with applets in http://arrhythmia.hofstra.edu.

[95] J.J. Fox, M. L. Riccio, F. Hua, E. Bodenschatz, and R.F. Gilmour, Jr., "Spatiotemporal Transition to Conduction Block in Canine Ventricle," Circ Res **90**, 289-296 (2002).

[96] M. Courtemanche, J.P. Keener, and L. Glass, "A delay equation representation of pulse circulation in a ring in excitable media," SIAM J. Appl. Math. **56**, 119-142 (1996).

[97] F.H. Fenton, A. Karma, H.M. Hastings and S.J. Evans, "Transition from Ventricular Tachycardia to Ventricular Fibrillation as a Function of Tissue Characteristics," IEEE Chicago 2000, World Congress on Medical Physics and Biomedical Engineering, CD-ROM, paper no. 5617-90379 (2000).

[98] M.C. Strain and H.S. Greenside, "Size-dependent transition to high-dimensional chaotic dynamics in a two-dimensional excitable medium," Phys. Rev. Lett. **80**, 2306-2309 (1998).

[99] D. Noble, "A modification of the Hodgkin-Huxley equations application to Purkinje fibre action and pace-maker potentials," J. Physiol. **160**, 317-352 (1962).

[100] H. Zhang and N. Patel, "Spiral wave breakdown in an excitable medium model of cardiac tissue," Chaos, Solitons and Fractals **5**, 635-643 (1995).

[101] C.H. Luo and Y. Rudy, "A model of the ventricular cardiac action potential: depolarization, repolarization, and their interaction," Circ. Res. **68**, 1501-1526 (1991).

[102] Z. Qu, J.N. Weiss, and A. Garfinkel, "Cardiac electrical restitution properties and stability of reentrant spiral waves: a simulation study," Am. J. Physiol. **276**, H269-H283 (1999).

[103] Z.Qu, A. Garfinkel, P.S. Chen, and J.N. Weiss, "Mechanisms of discordant alternans and induction of reentry in simulated cardiac tissue," Circ. **102**, 1664-1670 (2000).

[104] L.H. Frame and M.B. Simson, "Oscillations of conduction, action potential duration, and refractoriness: A mechanism for spontaneous termination of reentrant tachycardias," Circulation **78**, 1277-1287 (1988).

[105] J.M. Pastore, S.D. Girouard, K.R. Laurita, F.G. Akar, and D.S. Rosenbaum, "Mechanism linking T-wave alternans to the genesis of cardiac fibrillation," Circ. Res. **99**, 1385-1394 (1999).

[106] J.M. Pastore and D.S. Rosenbaum, "Role of structural barriers in the mechanism of alternans-induced reentry," Circ. Res. **87**, 1157-1163 (2000).

[107] D.S. Rosenbaum, L.E. Jackson, J.M. Smith, H. Garan, J.N. Ruskin, and R.J. Cohen, "Electrical alternans and vulnerability to ventricular arrhythmias," N. Engl. J. Med. **330**, 235-241 (1994).

[108] F. Fenton, "Numerical Simulations of Cardiac Dynamics: What can we learn from simple and complex models?", Computers in Cardiology (IEEE) **27**, 251-254 (2000).

[109] B.Echebarria and A. Karma, "Instability and Spatiotemporal dynamics of alternans in paced cardiac tissue," submitted (preprint at LANL cond-mat/0111552).

[110] M.Kay and R.Gray, "The effect of activation wave front curvature on action potential duration restitution," Fellowship award winners' presentation of research, NASPE 2001.

[111] P. Comtois and A. Vinet, "Curvature effects on activation speed and repolarization in an ionic model of cardiac myocytes," Phys. Rev. E **4**, 4619-4628 (1999).

[112] V.G. Fast and A.G. Kleber, "Role of wavefront curvature in propagation of cardiac impulse," Cardiovascular Research **33**, 258-271 (1997).

[113] J.P. Keener, "An Eikonal-Curvature Equation for Action Potential Propagation in Myocardium," J. Math. Biol. **29**, 629-651 (1991).

[114] B.Echebarrria and A. Karma, "Control of Alternans in spatially extended cardiac tissue," this issue of Chaos.

[115] J.J. Fox, J.L. McHarg and R.F. Gilmour Jr., "Ionic mechanism of electrical alternans," Am. J. Physiol **282,** H516-H530 (2002).

[116] G. M. Hall, S. Bahar and D.J. Gauthier, "Prevelence of rate-dependent behavior in cardiac muscle," Phys. Rev. Lett. **82**, 2995-2998 (1999).

[117] P. Lorente and J. Davidenko, "Hysteresis phenomena in excitable cardiac tissue," Ann. N.Y. Acad. Sci. **591**, 109-127 (1990).

[118] R.A. Oliver, G.M. Hall, S. Bahar, W. Krassowska, P.D. Wolf, Ellen G. Dixon-Tulloch, D.J. Gauthier, "Existence of bistability and correlation with arrhythmogenesis in paced sheep atria," J. Cardiovasc. Electrophys. **11**, 797-805 (2000).

[119] P. Lorente, C. Delgado, M. Delmar, D. Henzel and J. Jalife, "Hysteresis in excitability of isolated guinea pig ventricular myocytes," Circ. Res. **69**, 1301-1315 (1991).

[120] A.R. Yehia, D. Jeandupeux, F.Alonso and M.R. Guevara, "Hystersis and bistability in the direct transition from 1:1 to 2:1 rhythm in periodically driven single ventricular cells," Chaos **9**, 916-931 (1999).

[121] S. Bahar, "Reentrant waves induced by local bistabilities in a cardiac model," in Proceedings of the 5th Experimental Chaos Conference, edited by M. Ding, W.L. Ditto, L.M. Pecora, and M.L. Spano (World Scientific, 2001), 215-222.

[122] V.G. Fast and A.M. Pertsov, "Drift of a vortex in the myocardium," Biophysics **35**, 489-494 (1990).

[123] L.J. Leon, F.A. Roberge, and A. Vinet, "Simulation of two-dimensional anisotropic cardiac reentry: effects of the wavelength on the reentry characteristics," Annals of Biomedical Engineering **22**, 592-609 (1994).

[124] A.M. Pertsov, J.M. Davidenko, R. Salomonsz, W.T. Baxter, and J. Jalife, "Spiral waves of excitation underlie reentrant activity in isolated cardiac muscle," Circ. Res. **72**, 631-650 (1993).

[125] M. Bär, M. Hildebrand, M. Eiswirth, M. Falcke, H. Engel, and M. Neufeld, "Chemical turbulence and standing waves in a surface reaction model: The influence of global coupling and wave instabilities," Chaos **4**, 499-508 (1994).

[126] W.J. Rappel, "Filament instability and rotational tissue anisotropy: A numerical study using detailed cardiac models," Chaos **11**, 71-80 (2001).

[127] P. Szigligeti, T. Banyasz, J. Magyar, G.Y. Szigeti, Z. Papp, A. Varro, and P.P. Nanasi, "Intracellular calcium and electrical restitution in mammalian cardiac cells," Acta Physiol. Scand. **163**, 139-147 (1998).

[128] S.M. Horner, D.J. Dick, C.F. Murphy, and M.J. Lab, "Cycle Length Dependence of the Electrophysiological Effects of Increased Load on the Myocardium," Circ. **94**, 1131-1136 (1996).

[129] M.R. Franz, C.D. Swerdlow, L.B. Liem, and J. Schaefer, "Cycle Length Dependence of Human Action Potential Duration in Vivo," J. Clin. Invest. **82**, 972-979 (1988).

[130] M. Watanabe, N.F. Otani, R.F. Gilmour, "Biphasic restitution of action potential duration and complex dynamics in ventricular myocardium," Circ. Res. **76**, 915-921 (1995).

[131] Z.Qu, J. Weiss and A. Garfinkel, "Spatiotemporal chaos in a simulated rings of cardiac cells," Phys. Rev. Lett. **78**, 1387-1390 (1997).

[132] Y. Kobayashi, W. Peters, S.S. Khan, W.J. Mandel, and H.S. Karagueuzian, "Cellular mechanisms of differential action potential duration restitution in canine ventricular muscle cells during single versus double premature stimuli," Circ. **86**, 955-967 (1992).

[133] D.R. Chialvo, D.C. Michaels, and J. Jalife, "Supernormal Excitability as a Mechanism of Chaotic Dynamics of Activation in Cardiac Purkinje Fibers," Circ. Res. **66**, 525-545 (1990).

[134] J.W. Buchanan Jr., T Saito and L.S. Gettes, "The effects of antiarrhythmic drugs, stimulation frequency, and potassium-induced resting membrane potential changes on conduction velocity and $dV/dt_{max}$ in guinea pig myocardium," Circ. Res. **56**, 696-703 (1985).

[135] K. Endresen, J.P. Amlie, K. Forfang, S. Simonsen, and O. Jensen, "Monophasic action potentials in patients with coronary artery disease: reproducibility and electrical restitution and conduction at different stimulation rates," Cardiovascular Research **21**, 696-702 (1987).

[136] K. Endresen and J.P. Amlie, "Electrical Restitution and Conduction Intervals of Ventricular Premature Beats in Man: Influence of Heart Rate," Pacing Clin. Electrophysiol. **12**, 1347-1354 (1989).

[137] N. Manz, S.C. Muller, and O. Steinbock, "Anomalous Dispersion of Chemical Waves in a Homogeneously Catalyzed Reaction System," J. Phys. Chem. A **104**, 5895-5897 (2000).

[138] C.T. Hamik, N. Manz, and O. Steinbock, "Anomalous Dispersion and Attractive Pulse Interaction in the 1,4-Cyclohexanedione Belousov-Zhabotinsky Reaction," J. Phys. Chem. A **105**, 6144-6153 (2001).

[139] C. Fisch, "Electrocardiographic Manifestations of Exit Block and Supernormal and Concealed Conduction," in *Cardiac Electrophysiology From Cell to Bedside*, **3**rd edition, edited by D. P. Zipes and J. Jalife (W.B. Saunders Co., Philadelphia, 2000), p.185-690.

[140] J. Maselko and K. Showalter, "Chemical waves on spherical surfaces," Nature **339**, 609-611 (1989).

[141] J.M. Rogers, "Wavefront Fragmentation Due to Ventricular Geometry in a Model of the Rabbit Heart," Chaos (THIS FOCUS ISSUE) (2002).

[142] Y.A. Yermakova, V.I. Krinskii, A.V. Panfilov, and A.M. Pertsov, "Interaction of Helical and Flat Periodic Autowaves in an Active Medium," Biophysics **31**, 348-354 (1986).

[143] F.H. Fenton, S.J. Evans, H.M. Hastings and A. Karma, "Transition from Ventricular Tachycardia to Ventricular Fibrillation as a Function of Tissue Characteristics in a Computer Model," Europace **1**, Supplement D, 109P/10 (2000).

[144] D.M. Harrild and C.S. Henriquez, "A computer model of normal conduction in the human atria," Circ. Res. **87**, e25-e36 (2000).

[145] S.M. Dillon, M.A. Allessie, P.C. Ursell, and A.L. Wit, "Influence of anisotropic tissue structure on reentrant circuits in the epicardial border zone of subacute canine infarcts," Circ. Res. **63**, 182-206 (1988).

[146] K.M. Kavanagh, J.S Kabas, D.L. Rollins, S.B. Melnick, W.M. Smith, and R.E. Ideker, "High-current stimuli to the spared epicardium of a large infarct induced ventricular tachycardia," Circ. **85**, 680-698 (1992).

[147] D.P. Zipes, J. Fischer, R.M. King, A. Nicoll, and W.W. Jolly, "Termination of ventricular fibrillation in dogs by depolarizing a critical amount of myocardium," Am. J. Cardiol. **36**, 37-44 (1975).

[148] M.A. Allessie, M.J. Schalij, C.J. Kirchoff, L. Boersma, M. Huybers, and J. Hollen, "Experimental electrophysiology and arrhythmogenicity, anisotropy and ventricular tachycardia," Eur. Heart J. **10**, supplement E, 2-8 (1989).

[149] G. Breithardt, M. Borggrefe, A. Martinez-Rubio, and T. Budde, "Pathophysiological mechanisms of ventricular tachyarrhythmias," Eur. Heart J. **10**, supplement E, 9-18 (1989).

[150] M.J. Schalij, W.J. Lammers, P.L. Rensma and M.A. Allessie, "Anisotropic conduction and reentry in perfused epicardium of rabbit left ventricle," Am. J. Physiol. **263**, H1466-H1478 (1992).

[151] A.T. Winfree, "Electrical turbulence in three-dimensional heart muscle," Science **266**, 1003-1006 (1994).

[152] A.T. Winfree, "Rotors, fibrillation, and dimensionality," in *Computational Biology of the Heart*, edited by A.V. Panfilov and A.V. Holden (John Wiley and Sons Ltd., Chichester, 1997), p. 101-135.

[153] D. Vaidya, G.E. Morley, F.H. Samie, and J. Jalife, "Reentry and fibrillation in the mouse heart, a challenge to the critical mass hypothesis," Circ. Res. **85**, 174-181 (1999).

[154] A.V. Panfilov and P. Hogeweg, "Scroll breakup in a three-dimensional excitable medium," Physical Review E **53**, 1740-1743 (1996).

[155] A.T. Winfree, "Scroll-shaped waves of chemical activity in three dimensions," Science **181**, 937-938 (1973).

[156] P.S. Chen, P.D. Wolf, E.G. Dixon, N.D. Danieley, D.W. Frazier, W.M. Smith, and R.E. Ideker, "Mechanism of ventricular vulnerability to single premature stimuli in open-chest dogs," Circ. Res. **62**, 1191-1209 (1988).

[157] D.W. Frazier, P.D. Wolf, J.M. Wharton, A.S.L. Tang, W.M. Smith, and R.E. Ideker, "Stimulus-induced critical point: mechanism for the electrical initiation of reentry in normal canine myocardium," J. Clin. Invest. **83**, 1039-1052 (1989).

[158] M. Vinson, A. Pertsov, and J. Jalife, "Anchoring of vortex filaments in 3D excitable media," Physica D **72**, 119-134 (1993).

[159] A.S. Mikhailov, A.V. Panfilov, and A.N. Rudenko, "Twisted scroll waves in active three-dimensional media," Phys. Lett. A **109**, 246-250 (1985).

[160] A.M. Pertsov, R.R. Aliev, and V.I. Kriinsky, "Three-dimensional twisted vortices in an excitable chemical medium," Nature **345**, 419-421 (1990).

[161] B.J. Welsh, J. Gomatam, and A.E. Burgess, "Three-dimensional chemical waves in the Belousov-Zhabotinskii reaction," Nature **304**, 611-614 (1983).

[162] M. Vinson, S. Mironov, S. Mulvey, and A. Pertsov, "Control of spatial orientation and lifetime of scroll rings in excitable media," Nature **386**, 477-480 (1997).

[163] A.T. Winfree, "Persistent tangled vortex rings in generic excitable media," Nature **371**, 233-236 (1994).

[164] A. Malevanets and R. Kapral, "Links, knots, and knotted labyrinths in bistable systems," Phys. Rev. Lett. **77**, 767-770 (1996).

[165] A.V. Panfilov and A.N. Rudenko, "Two regimes of the scroll ring drift in the three-dimensional active media," Physica D **28**, 215-218 (1987).

[166] A.V. Panfilov and A.V. Holden, "Computer simulation of re-entry sources in myocardium in two and three dimensions," J. Theor. Biol. **161**, 271-285 (1993).

[167] P.J. Nandapurkar and A.T. Winfree, "Dynamical stability of untwisted scroll rings in excitable media," Physica D **35**, 277-288 (1998).

[168] M. Courtemanche, W. Skaggs, and A. T. Winfree, "Stable 3-dimensional action-potential circulation in the FitzHugh-Nagumo model," Physica D **41**, 173-182 (1990).

[169] Y.A. Yermakova and A.M. Pertsov, "Interaction of Rotating Spiral Waves with a Boundary," Biophysics **31**, 932-940 (1986).

[170] J.P. Keener, "The dynamics of three-dimensional scroll waves in excitable media," Physica D **31**, 269-276 (1988).

[171] J.P. Keener and J.J. Tyson, "The dynamics of scroll waves in excitable media," SIAM Review **34**, 1-39 (1992).

[172] C. Thomas, "The muscular architecture of the ventricles of hog and dog hearts," Am. J. Anat. **101**, 17 (1957).

[173] D. Streeter, "Gross morphology and fiber geometry in the heart," in *Handbook of Physiology,* edited by R. Berne (American Physiological Society, Bethesda, MD, 1979), Vol 1, Section 2, p. 61-112.

[174] C.S. Peskin, "Fiber architecture of the leftvventricular wall: An asymptotic analysis," Commun. Pur. Appl. Math. **42**, 79-113 (1989).

[175] D.D. Streeter Jr, H.M. Spotnitz, D.P. Patel, J. Ross, and E.H. Sonnenblick, "Fiber orientation in the canine left ventricle during diastole and systole," Circ. Res., **24**, 339-347 (1969).

[176] S. Mironov, M. Vinson, S. Mulvey, and A. Pertsov, "Destabilization of three-dimensional rotating chemical waves in an inhomogeneous BZ reaction," J. Physiol. (London) **100**, 1975-1983 (1996).

[177] M.S. Spach, W.T. Miller 3rd, D.B. Geselowitz, R.C. Barr, J. Kootsey, and E.A. Johnson, "The discontinuous nature of propagation in normal canine cardiac muscle: evidence for recurrent discontinuities of intracellular resistance that affect the membrane currents," Circ. Res. **48,** 39-54 (1981).

[178] L. Clerc, "Directional Differences of Impulse Spread in Trabecular Muscle from Mammalian Heart," J. Physiol. **255**, 334-346 (1976).

[179] L.J. Leon and F.A. Roberge, "Directional characteristics of action potential propagation in cardiac muscle," Circ. Res. **69**, 378-395 (1991).

[180] M.S. Spach and J.F. Heidlage, "A multidimensional model of cellular effects on the spread of electrotonic currents and on propagating action potentials," in *High-Performance Computing in Biomedical Research*, edited by T.C. Pilkington, B. Loftis, J.F. Thompson, S.L.-Y. Woo, T.C. Palmer, and T.F. Budinger (CRC Press, Boca Raton, Fla., 1993), p.289-317.

[181] A.G. Kleber, C.B. Riegger, and M.J. Janse, "Electrical uncoupling and increase of extracellular resistance after induction of ischemia in isolated, arterially perfused rabbit papillary muscle," Circ. Res. **61**, 271-279 (1987).

[182] J.R. Harper, T.A. Johnson, C.L. Engle, D.G. Martin, W. Fleet, and L.S. Gettes, "Effect of Rate Changes in Conduction Velocity and Extracellular Potassium Concentration During Acute Ischemia in the In Situ Pig Heart," J. Cardiovasc. Electrophysiol. **4**, 661-671 (1993).

[183] A.V. Panfilov and J.P. Keener, "Generation of reentry in anisotropic myocardium," J. Cardiovasc. Electrophysiol. **4**, 412-421 (1993).

[184] J.P. Keener, "Propagation and its failure in coupled systems of discrete excitable cells," SIAM J. Appl. Math. **47**, 556-572 (1987).

[185] J.P. Keener, "On the formation of circulating patterns of excitation in anisotropic excitable media," J. Math. Biol. **26**, 41-56 (1988).

[186] J. P. Keener, "The effects of discrete gap junction coupling on propagation in myocardium," J. Theor. Biol. **148**, 49-82 (1991).

[187] M.L. Riccio, M.L. Koller, and R.F. Gilmour, Jr., "Electrical restitution and spatiotemporal organization during ventricular fibrillation," Circ. Res. **84**, 955-963 (1999).

[188] C. Omichi, S. Zhou, M.H. Lee, A. Naik, C.M. Chang, A. Garfinkel, J.N. Weiss, S.F. Lin, H.S. Karagueuzian, P.S. Chen, "Effects of amiodarone on wave front dynamics during ventricular fibrillation in isolated swine right ventricle," Am. J. Physiol. **282**, H1063-H1070 (2002).

[189] M.C.E.F. Wijffels, C.J.H.J. Kirchhof, R. Dorland, and M.A. Allessie, "Atrial Fibrillation Begets Atrial Fibrillation," Circ. **92**, 1954-1968 (1995).

[190] S. Nattel, D. Li, and L. Yue, "Basic Mechanisms of Atrial Fibrillation—Very New Insights into Very Old Ideas," Annu. Rev. Physiol. **62**, 51-77 (2000).

[191] T. Watanabe, P.M. Rautaharju, and T.F. McDonald, "Ventricular Action Potentials, Ventricular Extracellular Potentials, and the ECG of Guinea Pig," Circ. Res. **57**, 362-373 (1985).

[192] R.F. Gilmour Jr., N. F. Ontani, and M.A. Watanabe, "Memory and complex dynamics in cardiac Purkinje fibers,"Am. J. Physiol. **272,** H1826-H1832 (1997).

[193] T.J. Hund and Y. Rudy, "Determinants of Excitability in Cardiac Myocytes: Mechanistic Investigation of Memory Effect," Biophys. J. **79**, 3095-3104 (2000).

[194] M.A. Watanabe and M.L. Koller, "Mathematical analysis of dynamics of cardiac memory and accommodation: theory and experiment," Am. J. Physiol. Heart Circ. Physiol. **282**, H1534-H1547 (2002).

[195] V. Elharrar and B. Surawicz, "Cycle length effect on restitution of action potential duration in dog cardiac fibers," Am. J. Physiol. **244**, H782-H792 (1983).

[196] M.L. Koller, M.R. Riccio, and R.F. Gilmour, Jr., "Dynamic restitution of action potential duration during electrical alternans and ventricular fibrillation," Am. J. Physiol. **275**, H1635-H1642 (1998).

[197] A.V. Zaitsev, O. Berenfeld, S.F. Mironov, J. Jalife, and A.M. Pertsov, "Distribution of excitation frequencies on the epicardial and endocardial surfaces of fibrillating ventricular wall in the sheep heart," Circ. Res. **86**, 408-417 (2000).

[198] J.L.R.M. Smeets, M.A. Allessie, W.J.E.P. Lammers, F.I.M. Bonke, and J. Hollen, "The wavelength of the cardiac impulse and reentrant arrhythmias in isolated rabbit atrium," Circ. Res. **58**, 96-108 (1986).

[199] M.J. Reiter, M. Landers, Z. Zetelaki, C.J.H. Kirchhof, and M.A. Allessie, "Electrophysiological effects of acute dilatation in the isolated rabbit heart," Circ. **96**, 4050-4056 (1997).

[200] Y. Hiramatsu, J.W. Buchanan, Jr., S.B. Knisley, G.G. Koch, S. Kropp, and L.S. Gettes, "Influence of rate-dependent cellular uncoupling on conduction change during simulated ischemia in guinea pig papillary muscles: effect of verapamil," Circ. Res. **65**, 95-102 (1989).

[201] D. Margerit and D. Barkley, "Selection of Twisted Scroll Waves in Three-Dimensional Excitable Media," Phys. Rev. Lett. **86**, 175-178 (2001).

[202] S. Setayeshgar and A.J. Bernoff, "Scroll Waves in the Presence of Slowly Varying Anisotropy with Application to the Heart," Phys. Rev. Lett. **88**, 028101 (2002).

[203] J.M. Rogers and A.D. McCulloch, "Nonuniform muscle fiber orientation causes spiral wave drift in a finite element model of cardiac action potential propagation," J. Cardiovasc. Electrophys. **5**, 496-509 (1994).

[204] R.A. Gray, J. Jalife, A.V. Panfilov, W.T. Baxter, C. Cabo, J.M. Davidenko and A.M. Pertsov, "Mechanisms of cardiac fibrillation," Science **270**, 1222-12225 (1995).

[205] P.M. Tande, E. Mortensen, and H. Refsum, "Rate-dependent differences in dog epi- and endocardial monophasic action potential configuration in vivo," Am. J. Physiol. **261**, H1387-H1391 (1991).

[206] E.P. Anyukhovsky, E.A. Sosunov, and M.R. Rosen, "Regional Differences in Electrophysiological Properties of Epicardium, Midmyocardium, and Endocardium," Circ. **94**, 1981-1988 (1996).

[207] C. Antzelevitch, W. Shimizu, G.X. Yan, S. Sicouri, J. Weissenburger, V.V. Nesterenko, A. Burashnikov, J. Di Diego, J. Saffitz, and G.P. Thomas, "The M Cell: Its Contribution to the ECG and to Normal and Abnormal Electrical Function of the Heart," J. Cardiovasc. Electrophysiol. **10**, 1124-1152 (1999).

[208] A.W. Cates and A.E. Pollard, "A Model Study of Intramural Dispersion of Action Potential Duration in the Canine Pulmonary Conus, Annals of Biomedical Engineering," **26**, 567-576 (1998).

[209] J. Kil, Z. Qu, J.N. Weiss, and A. Garfinkel, "Electrophysiological Transmural Heterogeneity and Scroll Wave Stability in Simulated 3D Cardiac Tissue," in preparation.

[210] K.J. Sampson and C.S. Henriquez, "Simulation and prediction of functional block in the presence of structural and ionic heterogeneity," Am. J. Physiol. **281**, H2597-H2603 (2001).

[211] M. Valderrabano, M.H. Lee, T. Ohara, A.C. Lai, M.C. Fishbein, S.F. Lin, H.S. Karagueuzian, and P.S. Chen, "Dynamics of intramural and transmural reentry during ventricular fibrillation in isolated swine ventricles," Circ. Res. **88**, 839-848 (2001).

[212] R.A. Gray, A.M. Pertsov, and J. Jalife, "Incomplete Reentry and Epicardial Breakthrough Patterns During Atrial Fibrillation in the Sheep Heart," Circ. **94**, 2649-2661 (1996).

[213] F. Ouyang, R. Cappato, S. Ernst, M. Goya, M. Volkmer, J. Hebe, M. Antz, T. Vogtmann, A. Schaumann, P. Fotuhi, M. Hoffmann-Riem, and K.-H. Kuck, "Electroanatomic Substrate of Idiopathic Left Ventricular Tachycardia: Unidirectional Block and Macroreentry Within the Purkinje Network," Circulation **105**, 462-469 (2002).

[214] F. Fenton, S. Evans, and H. Hastings, "Memory in an excitable medium; A mechanism for spiral wave breakup in the low excitable limit," Physical Review Letters **83**, 3964-3967 (1999).

[215] T.J. Hund and Y. Rudy, "Determinants of Excitability in Cardiac Myocytes: Mechanistic Investigation of Memory Effect," Biophys. J. **79**, 3095-3104 (2000).

[216] P.C. Viswanathan, Y. Rudy, "Pause Induced Early Afterdepolarizations in the Long QT Syndrome: A Simulation Study," Cardiovascular Research **42**, 530-542 (1999).

[217] A.V. Panfilov and J.P. Keener, "Effects of high frequency stimulation on cardiac tissue with an inexcitable obstacle," J. Theor. Biol. **163**, 439-448 (1993).

[218] J.M. Starobin, Y.I. Zilberter, E.M. Rusnak, and C.F. Starmer, "Wavelet formation in excitable cardiac tissue: The role of wavefront obstacle interactions in initiating high-frequency fibrillatory-like arrhythmias," Biophys. J. **70**, 581-594 (1996).

[219] C. Cabo, A.M. Pertsov, J.M. Davidenko, and J. Jalife, "Electrical turbulence as a result of the critical curvature for propagation in cardiac tissue," Chaos **8**, 116-126 (1998).

[220] H. Arce, A. Xu, H. Gonzalez, and M.R. Guevara, "Alternans and higher-order rhythms in an ionic model of a sheet of ischemic ventricular muscle," Chaos **10**, 411-426 (2000).

[221] F.H. Samie, R. Mandapti, R.A. Gray, Y. Watanabe, C. Zuur, J. Beaumont, and J. Jalife, "A mechanism of transition from ventricular fibrillation to tachycardia: effect of calcium channel blockade on the dynamics of rotating waves," Circ. Res. **86**, 684-691 (2000).

[222] F.J. Samie and J. Jalife, "Mechanisms underlying ventricular tachycardia and its transition to ventricular fibrillation in the structurally normal heart," Cardiovascular Research **50**, 242-250 (2001).